\documentclass[11pt,a4paper]{article}
\pdfoutput=1
\usepackage{jheppub}
\usepackage{afterpage}
\usepackage[figureposition=bottom,font=small,labelfont=bf,labelsep=period,skip=5pt,tableposition=top]{caption}
\usepackage{ecolumn}
\usepackage{sidecap}\sidecaptionvpos{figure}{t}
\usepackage{array}
\usepackage{dsfont}
\usepackage{slashed}
\usepackage{mathtools}
\usepackage[shortskips,subtag]{methtools}
\usepackage{mathrsfs}
\usepackage[squaren,Gray]{SIunits}
\usepackage{booktabs}
\usepackage{dcolumn}
\usepackage{multirow}
\usepackage{bbold}
\usepackage{widetable}
\usepackage{placeins}
\usepackage{makerobust}
\MakeRobustCommand{\vphantom}
\newcommand{\overbar}[1]{\mkern 1.5mu\overline{\mkern-1.5mu#1\mkern-1.5mu}\mkern 1.5mu}

\newcommand{\MSbar}{\overbar{\text{MS}}}
\newcommand{\vect}[1]{\ensuremath{\boldsymbol{#1}}}
\renewcommand*{\cdot}{\mathbin{\mbox{\textperiodcentered}}}
\newcommand{\RI}{\text{RI}^\prime\mkern-3mu\raisebox{1pt}{$/$}\text{SMOM}}
\newcommand{\RInoS}{\text{RI}^\prime\mkern-3mu\raisebox{1pt}{$/$}\text{MOM}}
\newcommand{\SU}[1]{\ensuremath{\text{SU}(#1)}}
\def\M{M}
\def\m{m}
\def\etaa{{\eta^8}}
\DeclareFontEncoding{LS1}{}{}
\DeclareFontSubstitution{LS1}{stix}{m}{n}
\DeclareSymbolFont{stix}{LS1}{stix}{m}{it}
\DeclareMathAccent{\leftarrowaccent}{\mathalpha}{stix}{"91}
\DeclareMathAccent{\rightarrowaccent}{\mathalpha}{stix}{"92}
\DeclareMathAccent{\leftrightarrowaccent}{\mathalpha}{stix}{"95}
\renewcommand{\overleftarrow}[1]{\leftarrowaccent{#1}}
\renewcommand{\overrightarrow}[1]{\rightarrowaccent{#1}}
\renewcommand{\overleftrightarrow}[1]{\leftrightarrowaccent{#1}}
\title{Light-cone distribution amplitudes of pseudoscalar mesons from lattice QCD}
\collaborationImg{\includegraphics{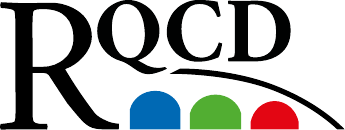}}
\author[a,b]{Gunnar~S.~Bali,}
\author[a]{Vladimir~M.~Braun,}
\author[a]{Simon~B{\"u}rger,}
\author[a]{Meinulf~G{\"o}ckeler,}
\author[a]{Michael~Gruber,}
\author[a]{Fabian~Hutzler,}
\author[c]{Piotr~Korcyl,}
\author[a]{Andreas~Sch\"afer,}
\author[d]{Andr\'e~Sternbeck,}
\author[a]{and Philipp~Wein}
\affiliation[a]{Institut f{\"u}r Theoretische Physik, Universit{\"a}t Regensburg, 93040 Regensburg, Germany}
\affiliation[b]{Department of Theoretical Physics, Tata Institute of Fundamental Research,\\Homi Bhabha Road, Mumbai 400005, India.}
\affiliation[c]{Marian Smoluchowski Institute of Physics, Jagiellonian University,\\ul.\ \L ojasiewicza 11, 30-348 Krak\'ow, Poland}
\affiliation[d]{Theoretisch-Physikalisches Institut, Friedrich-Schiller-Universit{\"a}t Jena, 07743 Jena, Germany}
\emailAdd{gunnar.bali@ur.de}
\emailAdd{vladimir.braun@ur.de}
\emailAdd{simon.buerger@ur.de}
\emailAdd{meinulf.goeckeler@ur.de}
\emailAdd{michael1.gruber@ur.de}
\emailAdd{fabian.hutzler@ur.de}
\emailAdd{piotr.korcyl@uj.edu.pl}
\emailAdd{andreas.schaefer@ur.de}
\emailAdd{andre.sternbeck@uni-jena.de}
\emailAdd{philipp.wein@ur.de}
\abstract{We present the first lattice determination of the two lowest Gegenbauer moments of the leading-twist pion and kaon light-cone distribution amplitudes with full control of all errors. The calculation is carried out on $35$ different CLS ensembles with $N_f=2+1$ flavors of dynamical Wilson-clover fermions. These cover a multitude of pion and kaon mass combinations (including the physical point) and $5$~different lattice spacings down to~$a=\unit{0.039}{\femto\meter}$. The momentum smearing technique and a new operator basis are employed to reduce statistical fluctuations and to improve the overlap with the ground states. The results are obtained from a combined chiral and continuum limit extrapolation that includes three separate trajectories in the quark mass plane.\par%
The present arXiv version (v3) includes an Addendum where we update the results using the recently calculated three-loop matching factors for the conversion from the $\RI$ to the $\MSbar$ scheme. We find $a_2^\pi=0.116^{+19}_{-20}$ for the pion, $a_1^K=0.0525^{+31}_{-33}$ and $a_2^K=0.106^{+15}_{-16}$ for the kaon. We also include the previous values, which were obtained with two-loop matching.}%
\begin{document}
\maketitle%
\flushbottom%
\section{Introduction}%
Hadron light-cone distribution amplitudes (LCDAs) have been introduced four decades ago~\cite{Radyushkin:1977gp,Chernyak:1977as,Lepage:1979za,Lepage:1979zb,Efremov:1978rn,Efremov:1979qk,Lepage:1980fj} in the context of the QCD description of hard exclusive reactions. The LCDAs are scale-dependent nonperturbative functions that can be interpreted as quantum-mechanical amplitudes. Within this article we will use the term ``LCDAs'' synonymous with the leading-twist LCDAs. The latter describe the distribution of the longitudinal momentum amongst the quarks in the leading Fock state contribution of a hadron wave function at small transverse parton separations. The pion LCDA is both the simplest LCDA and also the most important one in phenomenological applications. Unsurprisingly, it has received the most attention in the literature. Its precise knowledge is becoming increasingly relevant in flavor physics (where weak decays, such as $B\to \pi \ell\nu_\ell$, $B\to\pi\pi$, etc., are providing information on the Cabibbo--Kobayashi--Maskawa matrix), in two-photon hard reactions (like $\gamma^\ast\to \gamma\pi$ or $\gamma\gamma\to\pi\pi$), and --- as a tool to access the flavor separation in the nucleon generalized parton distributions --- in hard exclusive electro-production ($e N\to e N\pi$) with Bjorken kinematics.\par
\begin{SCfigure}[1][t]%
\centering%
\includegraphics[width=0.54\textwidth]{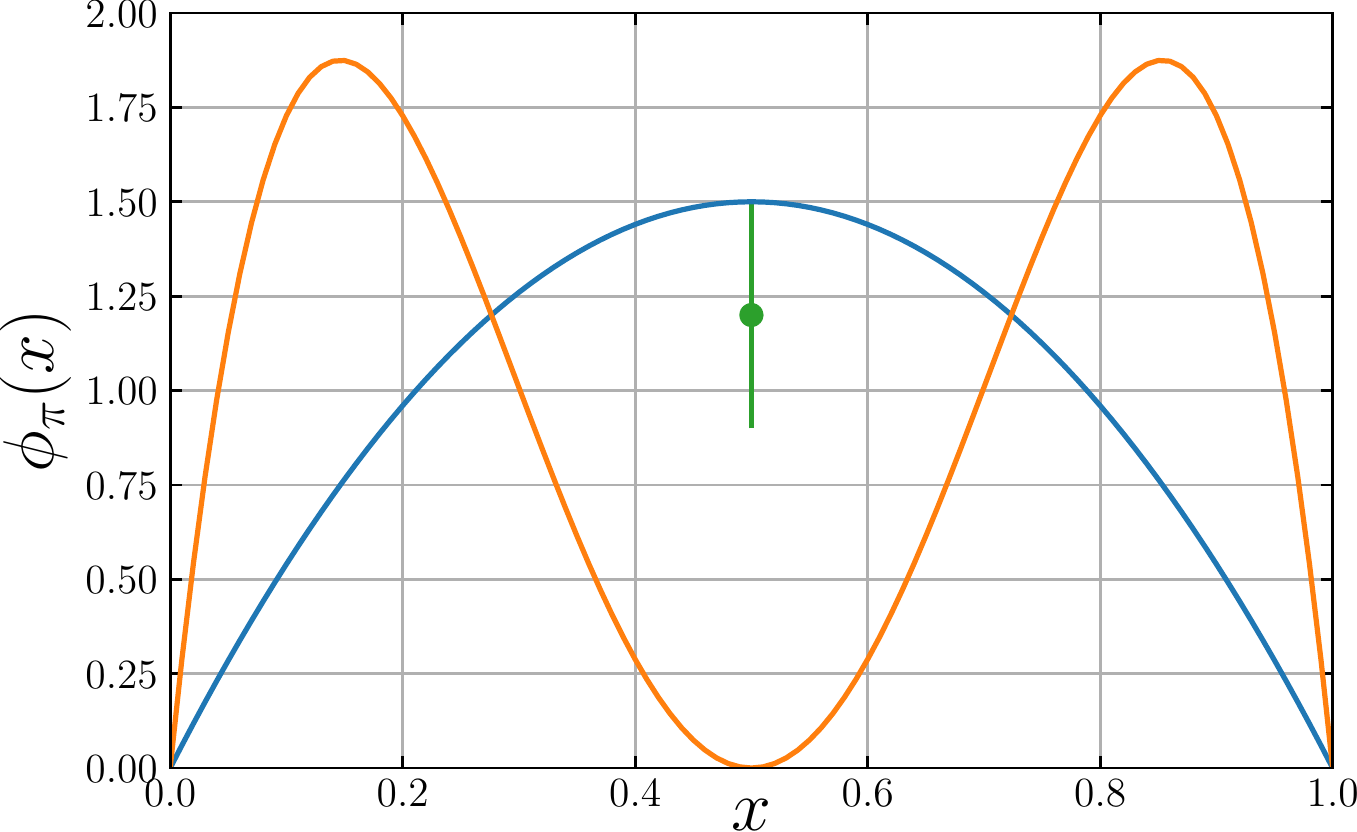}%
\caption{Models for the pion LCDA: the blue line shows the asymptotic shape corresponding to the limit $\mu\to\infty$, while the orange line depicts the CZ model~\cite{Chernyak:1981zz,Chernyak:1983ej} for $\mu=\unit{0.5}{\giga\electronvolt}$. The green point shows the QCD light-cone sum rule result~\cite{Braun:1988qv} for the mid-point at $\mu=\unit{1}{\giga\electronvolt}$.\label{fig_DAs}}%
\end{SCfigure}

%
Theoretical attempts to predict the shape of the pion LCDA $\phi_{\pi}(x,\mu^2)$ as a function of the longitudinal momentum fraction $x$ at a scale $\mu$ have a long history. The discussion was shaped for many years by the famous paper by Chernyak and Zhitnitsky (CZ)~\cite{Chernyak:1981zz} who calculated the second moment in~$x$ of the pion LCDA using QCD sum rules~\cite{Shifman:1978bx} and found a number much larger than the result expected at asymptotically large scales. Based on this calculation, CZ proposed a particular model for the pion LCDA at low scales, known as the CZ~model. Assuming the validity of perturbative QCD factorization, this model allowed for a consistent description of all experimental data on hard exclusive processes that were available at that time~\cite{Chernyak:1983ej}. In figure~\ref{fig_DAs} we compare the asymptotic LCDA \mbox{$\phi_\pi(x,\mu^2)\xrightarrow{\mu\to\infty}6x(1-x)$}~\cite{Efremov:1978rn,Lepage:1979zb} with the CZ model. The latter corresponds to a double-peaked distribution, where one of the constituents is most likely to carry a small ($\sim0.15$) and the other one a large ($\sim0.85$) fraction of the longitudinal pion momentum.\par%
The CZ model received some criticism. On the one hand, the validity of collinear factorization in hard exclusive reactions at relatively low momentum transfer was questioned \cite{Isgur:1988iw,Radyushkin:1990te} and the role of a competing ``soft'' or ``end-point'' mechanism was emphasized. In particular it was shown~\cite{Nesterenko:1982gc,Radyushkin:1990te} that the data on the pion form factor at $Q^2\sim\unit{1\text{--}3}{\giga\electronvolt^2}$ could be described by the soft contribution alone, without any ``hard'' corrections. On the other hand, it was argued that the QCD sum rules employed in ref.~\cite{Chernyak:1981zz} were not reliable as they may suffer from large contributions from operators of higher dimension. A model for such higher-order contributions using the concept of nonlocal vacuum condensates~\cite{Mikhailov:1991pt} yielded a much smaller value of the second moment than the CZ model, see ref.~\cite{Bakulev:2001pa} for a state-of-the-art study. Finally, the explicit calculation~\cite{Braun:1988qv} of the value of the pion LCDA at the mid-point $x=\tfrac{1}{2}$, using an at that time novel method, the light-cone sum rule (LCSR) technique, gave a rather large number, see~figure~\ref{fig_DAs}, inconsistent with the pronounced ``dip'' of the CZ model. Using the LCSR approach it was also shown for many examples, see, e.g., refs.~\cite{Braun:1994ij,Braun:1999uj,Khodjamirian:2000ds,Ball:2004ye,Bakulev:2009ib,Agaev:2010aq,Khodjamirian:2011ub,Bakulev:2012nh,Mikhailov:2016klg}, that the CZ model leads to very large soft contributions to hard reactions, which contradict the data. Nevertheless, the paradigm ``asymptotic-like LCDA versus CZ-like LCDA'' continues to be the preferred language of many model studies.\par%
A new wave of interest in the pion LCDA was inspired by the BaBar measurement~\cite{Aubert:2009mc} of the pion transition form factor $\gamma\gamma^\ast \to \pi$ that indicates very strong scaling violations up to the highest virtualities $Q^2\sim \unit{40}{\giga\electronvolt^2}$ available. In order to explain this behavior, an unconventional ``constant'' shape of the pion LCDA was proposed~\cite{Polyakov:2009je,Radyushkin:2009zg}, which triggered further discussion, see, e.g., ref.~\cite{Agaev:2010aq}. Although the similar Belle experiment~\cite{Uehara:2012ag} does not suggest strong scaling violations, the problem is far from being resolved and this measurement will be repeated by Belle~II at the upgraded SuperKEKB accelerator at KEK~\cite{Kou:2018nap} with a much improved projected precision. Motivated by these experimental needs and in the absence of a convincing first-principles calculation, the pion LCDA continues to attract a lot of attention. In the last few years several new calculations appeared, most notably using techniques based on Dyson--Schwinger equations (DSE)~\cite{Chang:2013pq}. A short overview of several existing models and their distinctive features can be found in ref.~\cite{Stefanis:2015qha}. For further models see, e.g., refs.~\cite{RuizArriola:2002bp,RuizArriola:2006jge}.\par%
Within the past 10--20 years lattice QCD has firmly established itself as the method of choice for nonperturbative calculations in QCD, as it has the potential to provide quantitative results with full control over all sources of uncertainty. The problem that we address here, however, is not simple. Lattice calculations of moments of the pion LCDA were proposed more than 30 years ago~\cite{Kronfeld:1984zv,Martinelli:1987si}. First pioneering studies were carried out within the quenched approximation in refs.~\cite{Martinelli:1987si,DeGrand:1987vy,DelDebbio:1999mq,DelDebbio:2002mq} and with $N_f=2$ Wilson fermions in ref.~\cite{Daniel:1990ah}. The first modern calculations were performed more than a decade ago by the QCDSF/UKQCD collaboration using $N_f=2$ nonperturbatively improved Wilson fermions~\cite{Braun:2006dg} and somewhat later by RBC/UKQCD~\cite{Arthur:2010xf} as part of their $N_f = 2+1$ domain-wall fermion phenomenology program. More recently, the study of ref.~\cite{Braun:2006dg} was extended in ref.~\cite{Braun:2015axa} to a larger set of lattice ensembles with different volumes, lattice spacings, and pion masses down to $m_\pi=\unit{150}{\mega\electronvolt}$, also implementing several technical improvements. In this way the errors due to the chiral extrapolation could be brought under control but still no controlled continuum limit extrapolation could be carried out.\par%
In this paper we close this last gap and present results of the first lattice calculation of the two lowest moments of the pion and kaon light-cone distribution amplitudes with full control of all systematic errors. This progress has become possible by the CLS (Coordinated Lattice Simulations) community effort~\cite{Bruno:2014jqa} aiming at the production of very fine lattices using open boundary conditions in time and further algorithmic improvements to reduce the autocorrelations within the Monte-Carlo time-series. (Autocorrelations increase as the continuum limit is approached.) The calculation reported in this work has been carried out on $35$~ensembles (see appendix~\ref{app_ens} for details) using $N_f=2+1$ flavors of nonperturbatively improved Wilson (clover) fermions with pion masses down to the physical point, employing $5$ different lattice spacings down to $a=\unit{0.039}{\femto\meter}$. In addition, we use the momentum smearing technique~\cite{Bali:2016lva}, which enables us to reduce statistical fluctuations by improving the overlap of the meson interpolating field with the ground state. Employing this technique, first results for the second moment of the pion LCDA for a single lattice spacing were reported in ref.~\cite{Bali:2017ude}. Since then we have enlarged the operator basis (cf.\ also ref.~\cite{Bali:2018qat}) and added four lattice spacings as well as other quark mass combinations. The results are then obtained pursuing combined chiral and continuum limit extrapolations, utilizing data from three separate trajectories in the quark mass plane. As a by-product we also obtain the continuum limit quark mass dependence of the LCDA moments. A similar determination of the wave function normalization constants and the first LCDA moments of the lowest-lying baryon octet can be found in the companion article~\cite{Bali:2019ecy}.\par%
This article is organized as follows. In section~\ref{sec_general} we first introduce LCDAs as well as the operators and correlators used in our analysis. Next, the renormalization of the lattice matrix elements is explained. This includes two steps: nonperturbative renormalization in the $\RI$ (or $\RInoS$) scheme and perturbative conversion from this scheme to the $\MSbar$ scheme. In section~\ref{sec:lattice} we describe the set of gauge ensembles employed. Subsequently, we detail the analysis of the correlation functions (including the specific choice of operators and external momenta) and extract the relevant matrix elements from the lattice. We also provide the extrapolation formulae for the quark mass and lattice spacing dependence. In section~\ref{sec_results}, we present our results for the LCDA moments and assess the error budget, before we discuss our findings and confront these with values from the literature in section~\ref{sec_discussion}. Finally, the addendum in section~\ref{sec_addendum} provides updated values using the recently calculated three-loop matching factors for the conversion from the $\RI$ to the $\MSbar$ scheme~\cite{Kniehl:2020sgo,Kniehl:2020nhw}.\par%
\section{General formalism\label{sec_general}}%
\subsection{Continuum definitions}%
Each pseudoscalar meson has only one independent leading-twist LCDA,~$\phi_M$, which can be defined via a meson-to-vacuum matrix element of a renormalized nonlocal quark-antiquark light-ray operator,
\begin{align} \label{eq_Bethe}
\langle 0 | \bar{q}(z_2 n) [z_2n,z_1n] \slashed{n} \gamma _5 u(z_1 n) | M (p) \rangle =   i f_M (p\cdot n) \smash[t]{\int _0 ^1}\! \mathrm{d}x\, e^{-i(z_1x+z_2(1-x))p\cdot n} \phi_{M}  (x,\mu^2)\,,
\end{align}%
where we consider the pion~($M=\pi^+$) with~$\bar{q}=\bar{d}$ and the kaon~($M=K^+$) with~$\bar{q}=\bar{s}$. Here, $z_{1,2}$ are real numbers, $n^\mu$ is an auxiliary light-like ($n^2=0$) vector, and $|M(p)\rangle$ represents the ground state meson $M$ with on-shell momentum $p^2=m_M^2$. The light-like Wilson line connecting the quark fields, $[z_2n,z_1n]$, is inserted to secure gauge invariance. The scale dependence of~$\phi_M$ is indicated by the argument~$\mu^2$. We denote the quark masses as $m_q$.\par%
Neglecting the isospin breaking due to electromagnetic effects and nondegenerate light quark masses (by setting $m_u=m_d\equiv m_\ell$), the LCDAs of all (charged and neutral) pions are trivially related such that it is sufficient to consider only one representative; the same holds for the kaons.
The decay constant $f_M$ appearing in eq.~\eqref{eq_Bethe} can be obtained as the matrix element of a local operator,
\begin{align}
\langle 0 | \bar{q}(0)\gamma _0 \gamma _5 u(0) | M^+(p) \rangle = if_M p_0\,,
\end{align}
and has the value $f_\pi\approx\unit{130}{\mega\electronvolt}$~\cite{Tanabashi:2018oca} for the pion and  $f_K\approx\unit{156}{\mega\electronvolt}$~\cite{Aoki:2016frl} for the kaon.%
\par%
Within eq.~\eqref{eq_Bethe} a fraction~$x$ of the longitudinal meson momentum is carried by the $u$~quark, while the $\bar{q}$~antiquark carries the remaining fraction~$1-x$. The difference of the momentum fractions is usually denoted as
\begin{align}
 \xi = x-(1-x) = 2x-1\,.
\end{align}
The complete information on the LCDA can be encoded in a set of moments. One such set is defined by%
\begin{align}\label{eq_moments}%
 \langle\xi^n\rangle_{\!M}(\mu^2) = \int_0^1 \!\!\mathrm{d}x\, (2x-1)^n \phi_M(x,\mu^2) \,.
\end{align}%
Another possible set of moments is%
\begin{align}%
 a_n^M\!(\mu^2) &= \frac{2(2n+3)}{3(n+1)(n+2)} \int_0^1 \!\!\mathrm{d}x\, C^{3/2}_n(2x-1)\, \phi_M(x,\mu^2)\,,
\end{align}%
where $C_n^{3/2}(\xi)$ are Gegenbauer polynomials, which correspond to irreducible representations of the collinear conformal group $\text{SL}(2,\mathds{R})$. Both sets, the $\xi$-moments $\langle \xi^n\rangle$ and the
Gegenbauer moments $a_n^M$, are related by a simple linear transformation, cf.\ eqs.~\subeqref{eq_renormalized_moments1}{b} and~\subeqref{eq_renormalized_moments2}{b} below.\footnote{Note that the second $\xi$-moment is given by a matrix element of an operator that contains two derivatives, which, in the case of parton distributions, would be relevant for the determination of the third Mellin moment.} Since the Gegenbauer polynomials form a complete set of functions, the LCDAs can be expanded as%
\begin{align}%
 \phi_M(x,\mu^2) &= 6x(1-x) \biggl[1+\smashoperator{\sum_{n=1}^\infty} a_n^M\!(\mu^2) C^{3/2}_n(2x-1)\biggr]\,,
\end{align}%
where the coefficients~$a_n^M$ are renormalized multiplicatively in leading logarithmic order as a consequence of conformal symmetry~\cite{Braun:2003rp}. Due to C-parity, all odd moments of the pion, i.e., $\langle\xi^n\rangle_{\pi}$ and  $a_n^\pi$ for $n=1,3,\ldots$\,, vanish in the limit of exact isospin symmetry. Higher-order contributions in the Gegenbauer expansion are suppressed at large scales, since the anomalous dimensions of~$a_n^M$ increase with~$n$~\cite{Efremov:1978rn}. Hence, in the asymptotic limit $\mu\rightarrow\infty$ only the leading term survives,%
\begin{align}
\phi _M  (x,\mu^2\to\infty ) = \phi^{\rm{as}}(x)=6x(1-x)\,,
\end{align}
which is usually referred to as the asymptotic LCDA. From here on we will suppress the explicit scale dependence of the DAs and their moments in the notation. Our lattice results will be given at the fixed scale $\mu=\unit{2}{\giga\electronvolt}$ in the $\MSbar$~scheme with three active flavors.\par%
\subsection{Lattice definitions}
From now on we will work in Euclidean spacetime and follow the conventions of ref.~\cite{Braun:2015axa}. The renormalized light-ray operator on the left-hand side of eq.~\eqref{eq_Bethe} generates renormalized local operators. This means that the moments~\eqref{eq_moments} of the LCDAs can be expressed in terms of matrix elements of local operators that can be evaluated using lattice QCD. In order to calculate the first and second moments of the pseudoscalar LCDAs we define the bare lattice operators%
\begin{subequations}\label{eq_operators}%
\begin{align}
\mathcal{P}(x) &= \bar{q}(x) \gamma_5 u(x)\,,\label{eq_operators_a}\\
\mathcal{A}_\rho(x) &= \bar{q}(x) \gamma_\rho\gamma_5 u(x)\,,\label{eq_operators_b}\\
\mathcal{O}^-_{\rho\mu}(x) &= \bar{q}(x) \bigl[\overrightarrow{D}_{(\mu}-\overleftarrow{D}_{(\mu}\bigr] \gamma_{\rho)}\gamma_5 u(x)\,,\label{eq_operators_c}\\
\mathcal{O}^-_{\rho\mu\nu}(x) &= \bar{q}(x) \bigl[\overrightarrow{D}_{(\mu}\overrightarrow{D}_{\mathstrut\nu}-2\overleftarrow{D}_{(\mu}\overrightarrow{D}_{\mathstrut\nu}+\overleftarrow{D}_{(\mu}\overleftarrow{D}_{\mathstrut\nu}\bigr] \gamma_{\rho)}\gamma_5 u(x)\,,\label{eq_operators_d}\\
\mathcal{O}^+_{\rho\mu\nu}(x) &= \bar{q}(x) \bigl[\overrightarrow{D}_{(\mu}\overrightarrow{D}_{\mathstrut\nu}+2\overleftarrow{D}_{(\mu}\overrightarrow{D}_{\mathstrut\nu}+\overleftarrow{D}_{(\mu}\overleftarrow{D}_{\mathstrut\nu}\bigr] \gamma_{\rho)}\gamma_5 u(x)\,,\label{eq_operators_e}
\end{align}%
\end{subequations}%
where the covariant derivative $D_\mu$ is discretized symmetrically. To obtain a leading-twist projection we symmetrize over all Lorentz indices and subtract all traces. This procedure is indicated by parentheses, e.g., $\mathcal{O}_{(\mu \nu)} = \frac{1}{2} \left( \mathcal{O}_{\mu \nu}+\mathcal{O}_{\nu \mu} \right)-\frac{1}{4}\delta _{\mu \nu} \mathcal{O}_{\lambda \lambda}$. In principle, one could also consider an operator $\mathcal{O}^+ _{\rho \mu}$, replacing the minus sign in eq.~\eqref{eq_operators_c} by a plus sign. However, as $\mathcal{O}^+_{\rho \mu}$ differs in C-parity from $\mathcal{O}^-_{\rho \mu}$, these two operators cannot mix with each other so that  $\mathcal{O}^+ _{\rho \mu}$ is irrelevant for our calculation. In contrast, the operator $\mathcal{O}^+_{\rho\mu\nu}$ has the same C-parity as $\smash{\mathcal{O}^-_{\rho\mu\nu}}$ and must be taken into account. Introducing the shorthand notation $\overleftrightarrow{D}_\mu = \overrightarrow{D}_\mu - \overleftarrow{D} _\mu$, the operator $\mathcal{O}^- _{\rho\mu\nu}$ can also be written as $ \bar{q}(x)  \overleftrightarrow{D}_{(\mu} \overleftrightarrow{D}_{\mathstrut\nu}  \gamma _{\rho )} \gamma _5 u(x)$ in the continuum.\par%
On a hypercubic lattice, the continuous $\mathrm{O}(4)$ symmetry is reduced to its discrete $\mathrm{H}(4)$ subgroup. This symmetry breaking can in principle induce mixing of the operators of interest with lower-dimensional operators accompanied by coefficient functions that diverge with a power of~$1/a$. For the first two $\xi$-moments this mixing can be avoided by selecting lattice operators that belong to a suitable irreducible representation of the hypercubic group $\mathrm{H}(4)$~\cite{Arthur:2010xf,Braun:2006dg}. For the calculation of the first moment we use the operators $\mathcal{O}_{4\mu}^-$, while for the second moments we choose $\mathcal{O}_{\rho\mu\nu}^\pm$ with all three indices different, see also section~\ref{sec_renorm}.\par%
In order to extract the desired moments we use two-point correlation functions of the operators with an interpolating current,%
\begin{subequations}%
\begin{align}%
C_\rho (t,\vect{p}) &= a^3 \sum _{\vect{x}} e^{-i\vect{p}\cdot\vect{x}} \langle \mathcal{A}_\rho (\vect{x},t)\mathcal{P}_{\vect{p}}^\dagger(0) \rangle\,, \\
C_{\rho \mu}^{-} (t,\vect{p}) &= a^3 \sum _{\vect{x}} e^{-i\vect{p}\cdot\vect{x}} \langle \mathcal{O}_{\rho \mu}^- (\vect{x},t)\mathcal{P}_{\vect{p}}^\dagger(0) \rangle\,,\\
C_{\rho \mu \nu}^\pm (t,\vect{p}) &= a^3 \sum _{\vect{x}} e^{-i\vect{p}\cdot\vect{x}} \langle \mathcal{O}_{\rho \mu \nu}^\pm (\vect{x},t)\mathcal{P}_{\vect{p}}^\dagger(0) \rangle\,,
\end{align}%
\end{subequations}%
where the index $\vect{p}$ indicates that the quarks appearing within the interpolator~\eqref{eq_operators_a} have been momentum smeared~\cite{Bali:2016lva,Bali:2017ude} (employing APE smeared~\cite{Falcioni:1984ei} spatial gauge transporters) to optimize the overlap with the ground state. The smearing parameters are not only adjusted according to the momentum but also optimized with respect to lattice spacing and quark mass. The ground state will dominate for sufficiently large values of the source-sink separation~$t$. In this limit, neglecting effects from the temporal boundaries, one obtains
\begin{align}
  C _{\mathcal{O}} (t,\vect{p}) &= \frac{1}{2E} \langle 0 | \mathcal{O}(0) | M(p) \rangle \langle M(p) | \mathcal{P}_{\vect{p}}^\dagger(0) | 0\rangle e^{-Et}\,,
\end{align}
with the ground state energy~$E=\sqrt{\smash[b]{m_M^2+\vect{p}^2}}$. For ensembles with open boundaries in time we place the source and sink within a window where the exponentially suppressed boundary effects can be neglected and translational invariance in time is restored within statistical accuracy. Regarding ensembles with the conventional anti-periodic fermionic boundary conditions in time, one should include a second exponential, $e^{-Et}\mapsto e^{-Et}+\tau_\mathcal{O}\tau_{\mathcal{P}}e^{-E(T-t)}$, where the sign factors $\tau_\mathcal{O},\tau_\mathcal{P}$ represent the transformation properties of~$\mathcal{O}$ and~$\mathcal{P}$ under time reversal.\par
For the extraction of the first moment we consider the ratios%
\begin{subequations}\label{eq_ratio1}%
\begin{align}%
R_{1,a}^{-}&=\frac{i}{3}\sum _{j=1}^3\frac{1}{p_j}\frac{C^{-} _{4 j}(t,\vect{p})}{C_4 (t,\vect{p})}  \,, &
R_{1,b}^{-}&=\frac{4E}{3E^2+\vect{p}^2}\frac{C^{-} _{44}(t,\vect{p})}{C_4 (t,\vect{p})}  \,.\subtag{2}
\end{align}%
\end{subequations}%
Similarly, for the required matrix elements for the second moment we consider%
\begin{subequations}\label{eq_ratio2}%
\begin{align}%
R_{2,a_1}^\pm&=-\frac{1}{3}\sum _{\substack{i,j=1\\i<j}}^3\frac{1}{p_i p_j}\frac{C^\pm _{4 i j}(t,\vect{p})}{C_4 (t,\vect{p})}\,, &
R_{2,a_2}^\pm&=\frac{1}{3}\sum_{i=1}^3\frac{p_i}{p_1 p_2 p_3}\frac{C^\pm _{123}(t,\vect{p})}{C_i (t,\vect{p})}\,.\subtag{2}
\end{align}%
\end{subequations}%
In contrast to the ratios~\eqref{eq_ratio1}, the two ratios defined in eqs.~\eqref{eq_ratio2} transform according to the same irreducible representation of~$\mathrm{H}(4)$ and will give the same result $R_{2}^\pm=R_{2,a_1}^\pm=R_{2,a_2}^\pm$ (in the limit $t\rightarrow\infty$, $p_j\ll a^{-1}$). However, $R_{2,a_1}^\pm$ and $R_{2,a_2}^\pm$ are affected differently by excited states, cf.\ section~\ref{sec_correlator_analysis}.\par%
\subsection{Renormalization procedure\label{sec_renorm}}%
The lattice operators have to be renormalized to obtain matrix elements in the $\MSbar$ scheme. As mentioned above, the continuous Euclidean $\mathrm{O}(4)$ symmetry is reduced to that of its finite hypercubic subgroup $\mathrm{H}(4)$ on the lattice. Therefore, symmetry imposes much weaker constraints on the mixing of operators under renormalization. In order to avoid mixing as far as possible, in particular mixing with lower-dimensional operators, we use operators from suitably chosen multiplets that transform according to irreducible representations of~$\mathrm{H}(4)$ and possess a definite C-parity. In the case of the operators \eqref{eq_operators_c} with one derivative we consider two multiplets transforming according to nonequivalent representations: one, labeled $a$, consisting of the six operators $\mathcal{O}^- _{\rho \mu}$ with $1 \leq \mu < \rho \leq 4$ and another one, labeled $b$, consisting of $\mathcal{O}^- _{44}$ and two further linear combinations of $\mathcal{O}^- _{11}$, $\mathcal{O}^- _{22}$, $\mathcal{O}^- _{33}$, $\mathcal{O}^- _{44}$. These do not mix with any other operators.\par%
The operators~\eqref{eq_operators_d} and~\eqref{eq_operators_e} with two derivatives have equal C-parity and behave identically under both continuum and lattice spacetime transformations. Hence, they will necessarily mix with each other. We utilize the multiplets%
\begin{subequations}\label{eq_multiplets}%
\begin{alignat}{4}%
&\mathcal{O}^+_{423}\,, &\quad& \mathcal{O}^+_{413}\,, &\quad& \mathcal{O}^+_{412}\,, &\quad& \mathcal{O}^+_{123}\\
\shortintertext{and}
&\mathcal{O}^-_{423}\,, && \mathcal{O}^-_{413}\,, && \mathcal{O}^-_{412}\,, && \mathcal{O}^-_{123}\,,
\end{alignat}%
\end{subequations}%
which transform under $\mathrm{H}(4)$ according to one and the same four-dimensional irreducible representation. Their symmetry properties guarantee that they do not mix with any other operators.\par%
We determine the renormalization and mixing coefficients nonperturbatively on the lattice using the same $\RI$ scheme~\cite{Sturm:2009kb} as was used in ref.~\cite{Braun:2015axa}. For the coarser lattice spacings ($\beta=3.4, \, 3.46, \, 3.55$) we have ensembles with different quark mass values $m_\ell=m_s$ and (anti-)periodic boundary conditions in time at our disposal so that we can proceed in exactly the same way as in ref.~\cite{Braun:2015axa}, starting from Landau-gauge-fixed three-point functions%
\begin{align}%
\frac{a^{12}}{V} \sum_{x,y,z} e^{- i p \cdot x -i(q-p) \cdot z + i q \cdot y }
\langle d(x) \mathcal{O}(z) \bar{u} (y) \rangle \,,
\end{align}%
where $\mathcal{O}$ represents the operators from eqs.~\mbox{\eqref{eq_operators_b}--\eqref{eq_operators_e}} with an antiquark flavor $\bar{q}=\bar{d}$ that is mass-degenerate with the $u$~quark. However, a problem arises on the finer lattices. For $\beta=3.7$ and $3.85$ we are forced to work with open boundary conditions in time to reduce autocorrelations in the Monte-Carlo time-series~\cite{Luscher:2011kk,Luscher:2012av}. In this case we modify the computation of the required three-point functions in two respects: we place the momentum sources within a subvolume, keeping a sufficiently large distance from the boundaries in the time direction, and we restrict the (final) sum over $z$ to an even smaller volume inside this subvolume. The further analysis can then be performed as in the periodic case. A detailed discussion, including a justification of this method and a comparison with the results from periodic boundary conditions, will be the topic of a dedicated, forthcoming publication. The ensembles with symmetric quark masses (\mbox{$m_\ell=m_s$}) used for the calculation of the renormalization factors are detailed in table~\ref{table_sym}. Unfortunately, we could only afford to generate ensembles for two distinct values of \mbox{$m_\ell=m_s$} at $\beta=3.7$ and~$3.85$. In the other cases the mass dependence of the amputated three-point functions is rather mild, so that we are confident that this restriction does not significantly affect the reliability of the required chiral extrapolations.\par%
In the case of the first LCDA moment of the kaon it is also possible to carry out the renormalization via the $\RInoS$ scheme~\cite{Martinelli:1994ty,Chetyrkin:1999pq}, where even the three-loop matching to the $\MSbar$ scheme is available~\cite{Gracey:2003yr,Gracey:2003mr,Gracey:2011zn}. Therefore, we choose to present four distinct results: with one- and two-loop matching~\cite{Gracey:2011fb,Gracey:2011zg} via the $\RI$ scheme as well as with two- and three-loop matching using the $\RInoS$ scheme.\par%
The tiny statistical errors of the results are negligible in comparison to the systematic uncertainties. In order to estimate the latter we proceed similarly to ref.~\cite{Braun:2015axa} and perform a number of fits, varying one element of the analysis at a time. We carry out two independent determinations of the renormalization and mixing coefficients, namely with one-loop and two-loop truncations of the perturbative expansion of the conversion factors from the $\RI$ scheme to the $\MSbar$~scheme for use in NLO and NNLO calculations in perturbative QCD, respectively. In both cases we vary the initial scale~$\mu_1$ of the fit range and the number~$n_{\mathrm {disc}}$ of terms in the parametrization $A_1a^2\mu^2+\dots+A_{n_{\mathrm{disc}}}(a^2\mu^2)^{n_{\mathrm{disc}}}$ of the lattice artifacts. In order to take into account the uncertainties in the determination of the lattice spacing, the central values of~$1/a^2$ shown in table~\ref{tab_spacings} are multiplied by a factor~$\lambda^2_{\mathrm {scale}}= 1.03$. This value contains the scale uncertainty of~$8t_0^*=\mu_{\mathrm{ref}}^{*\,-2}$ given in ref.~\cite{Bruno:2017gxd} and the largest error of our determination of~$t_0^*/a^2$, added in quadrature. Finally, also $\Lambda_{\MSbar}^{(3)}=\unit{341(12)}{\mega\electronvolt}$~\cite{Bruno:2017gxd} is varied within its uncertainty. Thus, we end up with five types of fits; the different settings are compiled in table~\ref{tab_fits}.\par%
\begin{table}[tb]%
\newcommand*{\tabstrut}{\rule[0pt-\depthof{$_{\MSbar}^($}]{0sp}{\totalheightof{$_{\MSbar}^($}}}%
\begin{minipage}[t]{.36\textwidth}%
\centering
{\small\vspace{\baselineskip}}%
\caption{Lattice spacings.\label{tab_spacings}}%
\begin{widetable}{\textwidth}{cD{.}{.}{1.3}D{.}{.}{2.2}}%
\toprule
\tabstrut$\beta$ & \multicolumn{1}{c}{$a\,[\femto\meter]$} & \multicolumn{1}{c}{$1/a^2\,[\giga\electronvolt\squared]$}\\
\midrule
$3.40$ & 0.086 &  5.28\\
$3.46$ & 0.076 &  6.75\\
$3.55$ & 0.064 &  9.44\\
$3.70$ & 0.050 & 15.75\\
$3.85$ & 0.039 & 25.54\\
\bottomrule
\end{widetable}%
\end{minipage}%
\hfill
\begin{minipage}[t]{.55\textwidth}%
\centering
\caption{Fit choices regarding the determination of the renormalization factors.\vphantom{pg}\label{tab_fits}}%
\begin{widetable}{\textwidth}{cD{.}{.}{2.0}cD{.}{.}{1.2}c}%
\toprule
Fit & \multicolumn{1}{c}{$\mu_1^2\,[\giga\electronvolt\squared]$} & $n_{\mathrm {disc}}$ & \multicolumn{1}{c}{$\lambda^2_{\mathrm {scale}}$} & $\Lambda_{\MSbar}^{(3)}\,[\mega\electronvolt]$\\
\midrule
1 &  4 & $3$ & 1.0  & 341\\
2 & 10 & $3$ & 1.0  & 341\\
3 &  4 & $2$ & 1.0  & 341\\
4 &  4 & $3$ & 1.03 & 341\\
5 &  4 & $3$ & 1.0  & 353\\
\bottomrule
\end{widetable}%
\end{minipage}%
\end{table}%
We determine the LCDA moments separately for each of the resulting renormalization and mixing coefficients, thus generating a set of five values per renormalization scheme at a given loop order. In this way we obtain two sets of results for the second LCDA moments, one using the two-loop SMOM conversion factors and another one employing the one-loop SMOM conversion factors. As explained above, for the first moment of the kaon LCDA we even have four such sets of results, as we can also nonperturbatively convert the bare lattice results to the $\RInoS$ scheme instead and then utilize the two-loop or three-loop conversion factors between the $\RInoS$ and the $\MSbar$ schemes.\par%
In each set we take the results of fit~1 as our central values. Defining $\delta_i$, $i=2,3,4,5$, as the difference between the number based on fit~$i$ and the result based on fit~1, we estimate the systematic uncertainties due to the renormalization factors as $\sqrt{\smash[b]{\delta_2^2+\delta_3^2+\delta_4^2+\delta_5^2}}$. The dominant uncertainties are related to the low-momentum cut-off of our fit range ($\delta_2$), i.e., the scale dependence, and the parametrization of lattice artifacts~($\delta_3$). The former becomes smaller when going from one-loop to two-loop perturbative accuracy, while the latter uncertainty shrinks as the lattice spacing is reduced. The uncertainty induced by the scale setting~($\delta_4$) and the error of the strong coupling parameter~($\delta_5$) are negligible. Note that all figures in this article showing renormalized data are generated using the $\RI$ intermediate scheme with two-loop matching to the $\MSbar$~scheme.\par%
Finally, the renormalized first moments are related to the ratios defined in eqs.~\eqref{eq_ratio1} by%
\begin{subequations}\label{eq_renormalized_moments1}%
\begin{align}%
\langle\xi^1\rangle^{\MSbar} &= \zeta_aR_{1,a}^- = \zeta_bR_{1,b}^-\,, &
a_1^{\MSbar} &= \frac{5}{3} \langle\xi^1\rangle^{\MSbar}\,,\subtag{2}
\end{align}%
\end{subequations}%
while the second moments are related to the ratios~\eqref{eq_ratio2} via%
\begin{subequations}\label{eq_renormalized_moments2}%
\begin{align}%
\langle\xi^2\rangle^{\MSbar} &= \zeta_{11}R_2^- + \zeta_{12}R_2^+\,, &
a_2^{\MSbar} &= \frac{7}{12} \bigl[5\langle\xi^2\rangle^{\MSbar}-\langle\mathbb{1}^2\rangle^{\MSbar}\bigr]\,,\subtag{2}\\
\langle\mathbb{1}^2\rangle^{\MSbar} &= \zeta_{22}R_2^+\,.\label{eq_eins}
\end{align}%
\end{subequations}%
In the continuum $\langle\mathbb{1}^2\rangle^{\MSbar}=1$, while it can differ from unity on the lattice, see section~\ref{sec_game}. The~$\zeta$s denote ratios of the renormalization constants of the operators~\mbox{\eqref{eq_operators_c}--\eqref{eq_operators_e}} over the renormalization constant of the axialvector current~\eqref{eq_operators_b}, cf.\ ref.~\cite{Braun:2015axa}.  Henceforth, $\langle\xi^n\rangle$, $\langle\mathbb{1}^n\rangle$, and~$a_n$ are always implied to be renormalized in the $\MSbar$~scheme and we omit the superscript~$\MSbar$.\par%
\section{Details of the lattice analysis\label{sec:lattice}}%
\subsection{Lattice ensembles\label{sec_lattices}}%
We use lattice ensembles generated within the CLS effort~\cite{Bruno:2014jqa} employing $N_f=2+1$ flavors of nonperturbatively~$\mathcal{O}(a)$ improved Wilson fermions~\cite{Sheikholeslami:1985ij,Bulava:2013cta} combined with the tree-level Symanzik improved gauge action~\cite{Weisz:1982zw}. For details on the action and the simulation see ref.~\cite{Bruno:2014jqa}.\footnote{Some of the $m_\ell=m_s$ ensembles with (anti-)periodic boundary conditions in time have been generated by RQCD using the BQCD~code~\cite{Nakamura:2010qh}.} Since that publication more CLS simulation points have been added, see, e.g., ref.~\cite{Bali:2016umi}. An overview of the ensembles analyzed here is given in appendix~\ref{app_ens}. Most CLS ensembles use open boundary conditions in the time direction, which allows us to carry out simulations at very fine lattice spacings without facing the problem of topological charge freezing~\cite{Luscher:2012av,Luscher:2011kk}.
\par
Five values of the inverse coupling constant $\beta=6/g^2$ are realized, corresponding to lattice spacings ranging from $a=\unit{0.086}{\femto\meter}$ down to $a=\unit{0.039}{\femto\meter}$, see table~\ref{tab_spacings}. Here we set the scale using $\sqrt{\smash[b]{8t_0^*}}=\unit{0.413(6)}{\femto\meter}$~\cite{Bruno:2017gxd}, where~$t_0^*$ is defined in ref.~\cite{Bruno:2016plf} as the Wilson flow scale~$t_0$~\cite{Luscher:2010iy}, computed at a particular reference point in the quark mass plane. The numerical value was obtained by matching the average continuum limit pion and kaon decay constant $f_{\pi K}=(2f_K+f_\pi)/3$ to experiment~\cite{Bruno:2016plf}.\par
\begin{SCfigure}[1][t]%
\centering%
\includegraphics[width=0.5\textwidth]{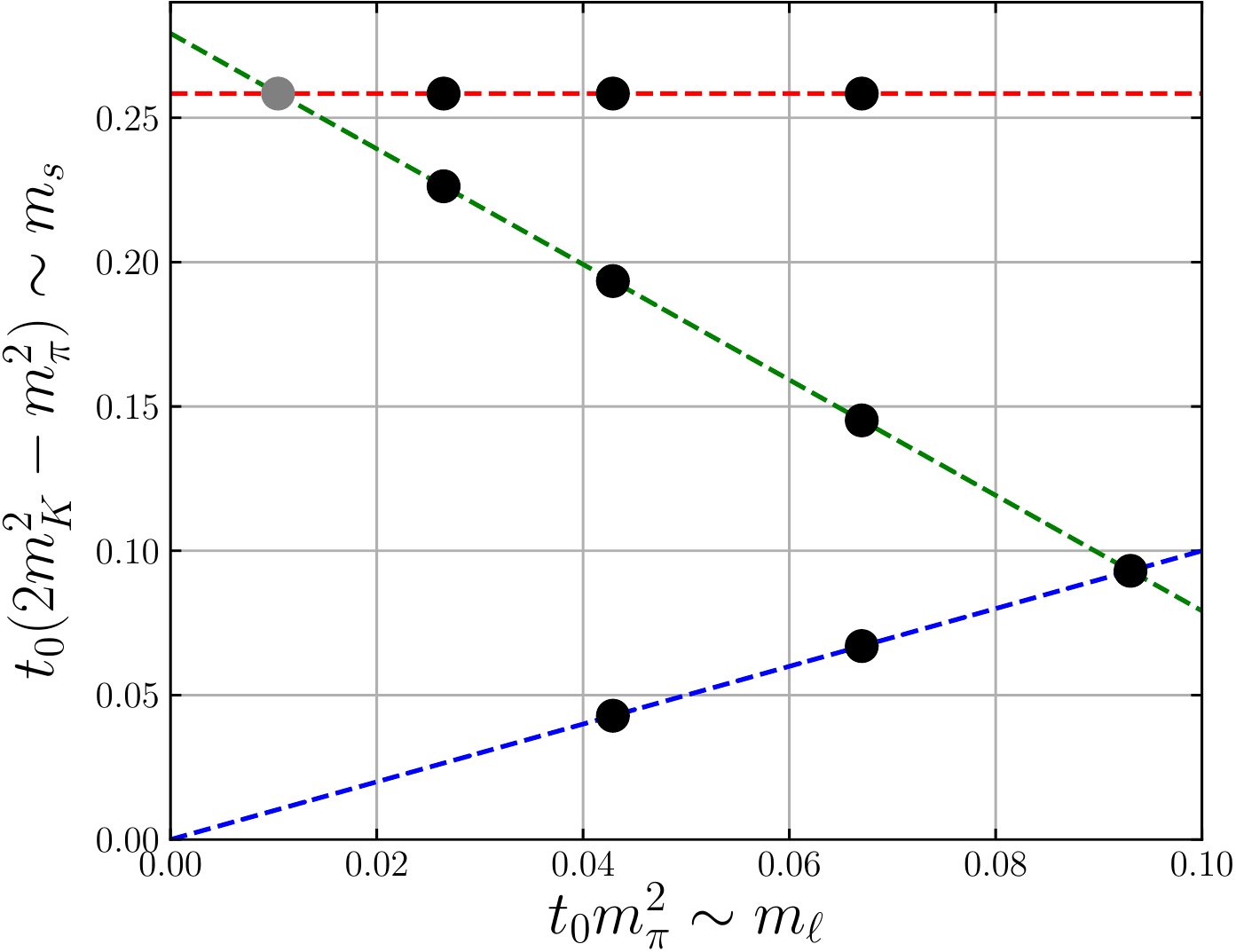}%
\caption{Schematic illustration of the mass trajectories of the lattice ensembles used in this study. Along the flavor symmetric line (blue) all pseudoscalar mesons have equal mass ($m_K^2=m_\pi^2$), which is equivalent to equal quark masses ($m_\ell=m_s$). The (green) line of the physical average quadratic meson mass ($2 m_K^2+m_\pi^2=\text{phys.}$) corresponds to an approximately physical mean quark mass ($2m_\ell+m_s\approx\text{phys.}$). The red line is defined by $2 m_K^2-m_\pi^2=\text{phys.}$ and indicates an approximately physical strange quark mass ($m_s\approx\text{phys.}$). The gray dot marks the physical point.\label{fig_trajectories}}%
\end{SCfigure}%
At each lattice spacing we have several points in the quark mass plane, along three trajectories: (a) along a nearly-physical fixed value of the trace of the mass matrix $\operatorname{Tr}\mathcal{M}\equiv m_u+m_d+m_s=2m_\ell+m_s=\text{phys.}$, (b) varying the light quark mass while trying to keep the renormalized strange quark mass $m_s$ constant at its physical value, and (c) along the ``symmetric'' line $m_\ell=m_s$, where light and strange quark masses are equal. The first two trajectories intersect close to the physical quark mass point. The locations of these three lines are shown in figure~\ref{fig_trajectories}. We determine the LCDA moments on various ensembles along these trajectories; our largest pion mass is about~$\unit{420}\mega\electronvolt$ and the smallest one is~$\unit{130}{\mega\electronvolt}$. Table~\ref{table_TrM} of appendix~\ref{app_ens} contains all lattices lying on line~(a) ($\operatorname{Tr}\mathcal{M}=\mathrm{constant}$). This line starts with a lattice at the flavor symmetric point and approaches the physical point, decreasing the light quark mass while simultaneously increasing the strange quark mass. Table~\ref{table_msc} contains all lattices lying on line (b) ($m_s\approx \mathrm{constant}$), where the strange quark mass is fixed to its physical value. This line starts with lattices that have unphysically large values of the $u$ and $d$~quark mass $m_\ell$ and approaches the physical point with decreasing light quark mass. Finally, table~\ref{table_sym} contains all lattices on the $\SU3$\nobreakdash-symmetric line where $m_\ell=m_s$. Along this line, which also includes the symmetric point of the $\operatorname{Tr}\mathcal{M}=\mathrm{constant}$ trajectory, all pseudoscalar mesons are members of a mass-degenerate \SU3 multiplet and their properties are related by symmetry.\par%
The spatial extents of the lattices used to determine the LCDA moments are always larger than~$\unit{2.4}{\femto\meter}$ and, with very few exceptions, larger than four times the inverse mass of the lightest pseudoscalar meson, see also tables~\mbox{\ref{table_TrM}--\ref{table_sym}}. For the pseudoscalar meson masses the expected corrections due to finite volume effects calculated at next-to-leading order in chiral perturbation theory (ChPT)~\cite{Gasser:1986vb,Gasser:1987zq} are smaller than half of their statistical errors. To this order the LCDAs are not affected by finite volume corrections at all since they are normalized with respect to the decay constant, see eq.~\eqref{eq_Bethe}. Therefore, it is well justified to neglect volume effects in our analysis.\par%
\subsection{Analysis of correlation functions\label{sec_correlator_analysis}}%
\begin{SCfigure}[1][t]%
\centering%
\includegraphics[width=0.55\textwidth]{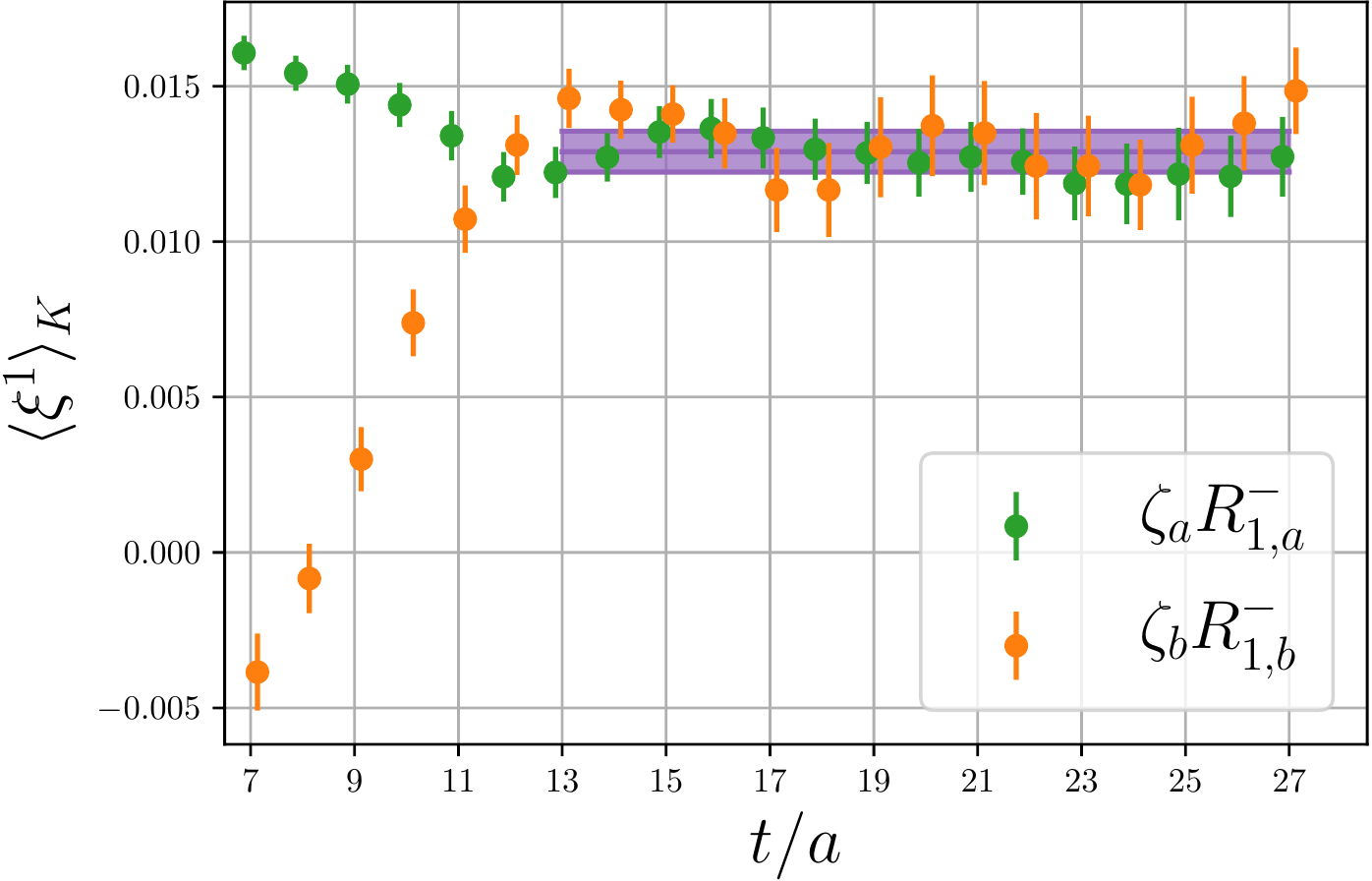}%
\caption{The ratios corresponding to the renormalized moment~$\langle\xi^1\rangle_{\!K}$ defined in eqs.~\eqref{eq_extract_Ra} and~\eqref{eq_extract_Rb} as a function of the time~$t$ in lattice units for the example of ensemble J501 with~$a=\unit{0.039}{\femto\meter}$. The result of a combined fit to both ratios is depicted in purple.\label{fig_a1}}%
\end{SCfigure}

\def\p{\wp}
Below we specify our choice of correlators and momentum directions. For the first moment we have operators from two different H(4) multiplets at our disposal (cf.\ eqs.~\eqref{eq_ratio1}). For the ratio in eq.~\subeqref{eq_ratio1}{a} we select the momenta $\vect{p}=(\pm 1,0,0)\p$, $\vect{p}=(0,\pm 1,0)\p$, and $\vect{p}=(0,0,\pm 1)\p$, where $\p=\frac{2\pi}{L}$. We then extract $R_{1,a}^{-}$ as a function of $t$ according to%
\begin{align} \label{eq_extract_Ra}
R_{1,a}^{-}&=\frac{i}{3\p}\sum _{j=1}^3\frac{\hat{p}_- C^{-} _{4 j}(t,\p\,\vect{e}_j)}{\hat{p}_+ C_4 (t,\p\,\vect{e}_j)}\,,
\end{align}%
where the forward/backward momentum averaging is performed by the operator $\hat{p}_\pm$:%
\begin{align}
 \hat{p}_\pm C(t,\vect{p}) = \tfrac{1}{2}\bigl( C(t,\vect{p})\pm C(t,-\vect{p}) \bigr)\,.
\end{align}%
For the ratio in eq.~\subeqref{eq_ratio1}{b} we may simply set $\vect{p}=\vect{0}$ to obtain%
\begin{align}\label{eq_extract_Rb}
 R_{1,b}^{-}&=\frac{4}{3\m_K}\frac{C^{-} _{44}(t,\vect{0})}{C_4 (t,\vect{0})}\,.
\end{align}%
We then renormalize the above ratios, multiplying by $\zeta_a$ and $\zeta_b$ according to eq.~\subeqref{eq_renormalized_moments1}{a}. Finally, $\langle\xi^1\rangle_{\!K}$ is obtained by carrying out a simultaneous fit to the plateau that is reached at large $t$-values as depicted in figure~\ref{fig_a1}.\par%
\begin{figure}[b]%
\centering%
\newlength{\figheight}\setlength{\figheight}{4.9cm}%
\includegraphics[height=\figheight,clip,trim=25 0 0 0]{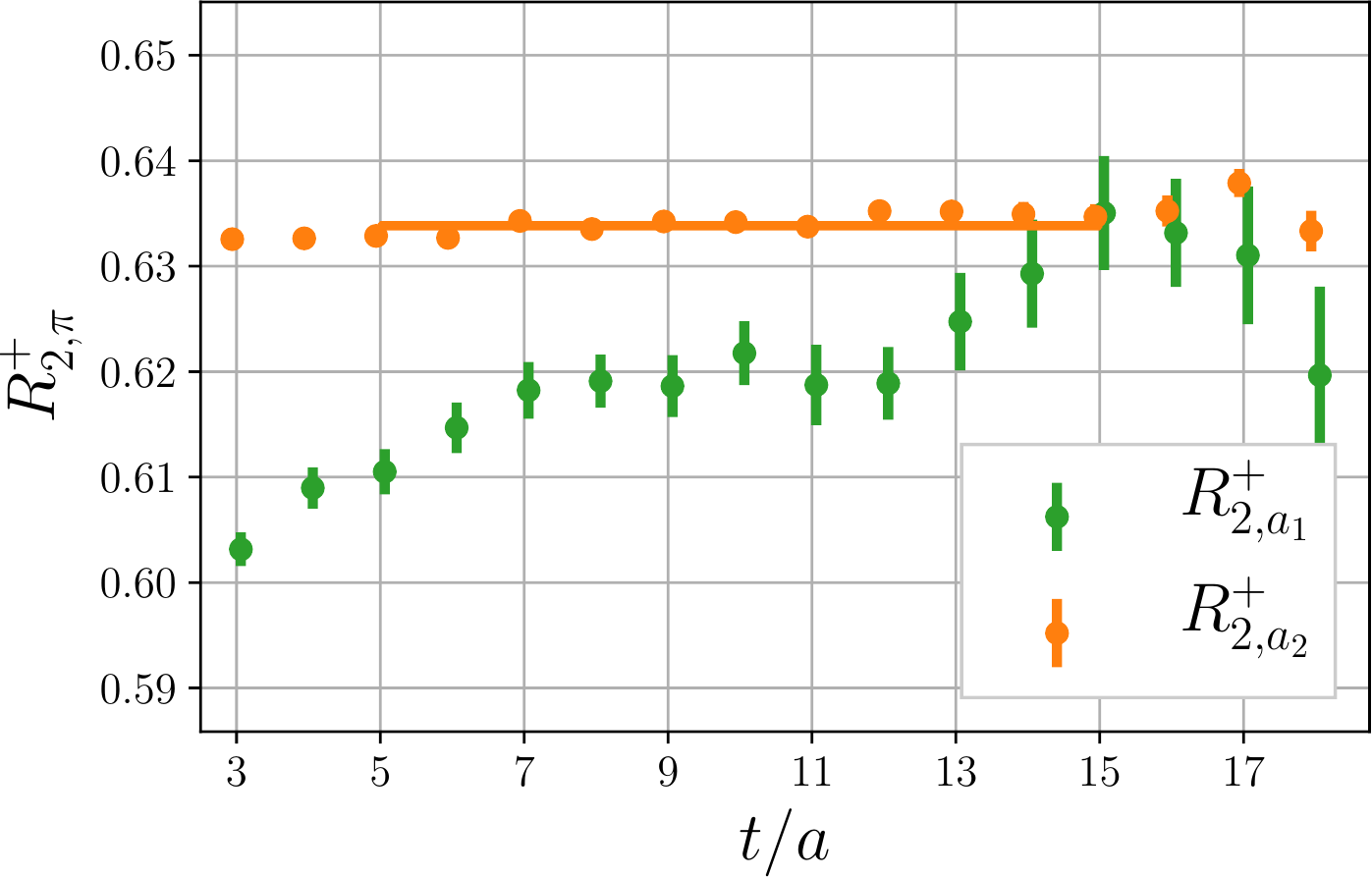}\hfill%
\includegraphics[height=\figheight]{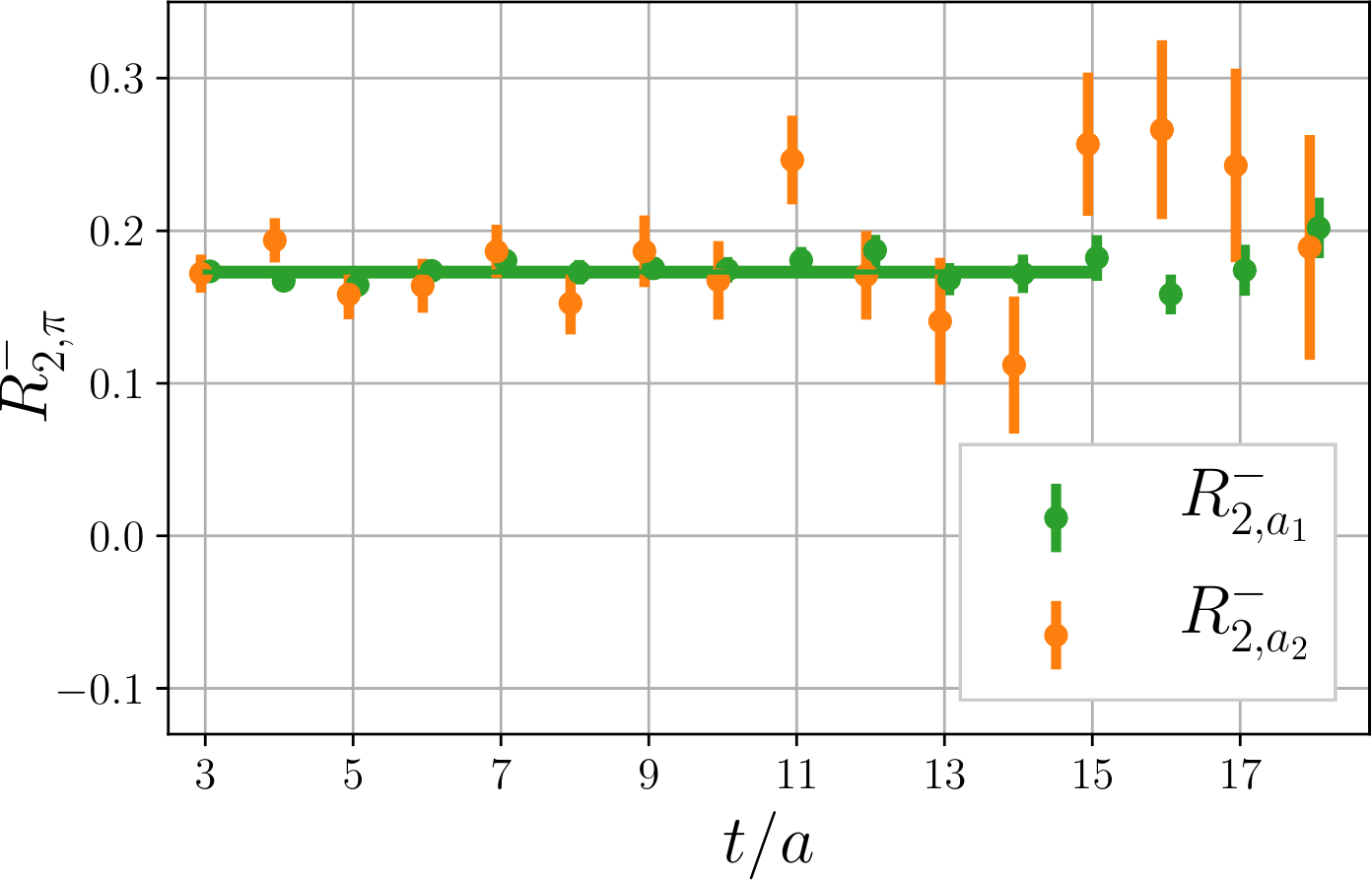}\llap{\textcolor{white}{\rule[.35\figheight]{.09\figheight}{.5\figheight}}\hspace{1.475\figheight}}%
\caption{\label{fig_R_plus} The ratios $R_{2,a_i}^{+}$ and $R_{2,a_i}^{-}$ (for the pion) defined in eqs.~\eqref{eq_ratio2} as functions of the lattice time $t$ with momentum $\vect{p}=(1,1,1)\p$ for the ensemble N203 ($a=\unit{0.064}{\femto\meter}$). Clearly, for the extraction of $R_2^+$~(left), the ratio $R_{2,a_2}^{+}$ is to be preferred to $R_{2,a_1}^{+}$, which suffers considerably from excited state effects and carries larger statistical errors. For the case of $R_2^-$~(right) neither data set seems to indicate any significant excited state contribution, but the statistical errors of $R_{2,a_1}^{-}$ are much smaller. The bands indicate the fit ranges and results.}%
\end{figure}

For the extraction of the second moments one needs at least two nonvanishing momentum components, cf. eqs.~\eqref{eq_ratio2}. We have already addressed the problem of the deterioration of the signal-to-noise ratio for increasing momenta $|\vect{p}|$ in our previous work~\cite{Bali:2017ude}, where we proposed to employ the momentum smearing technique (introduced in ref.~\cite{Bali:2016lva}) for all quark sources in order to improve the statistical error and to reduce contributions from excited states. The momentum smearing technique requires two inversions per momentum direction and in order to evaluate the full sum in eq.~\subeqref{eq_ratio2}{a} we performed six inversions to realize the momenta $\vect{p}=(1,1,0)\p$, $\vect{p}=(1,0,1)\p$, and $\vect{p}=(0,1,1)\p$ in ref.~\cite{Bali:2017ude}. In the present work we select the slightly higher momentum $\vect{p}=(1,1,1)\p$, which allows us to evaluate both eqs.~\subeqref{eq_ratio2}{a} and~\subeqref{eq_ratio2}{b}. This requires only two inversions in total. We compare the two ratios $R_{2,a_1}^{\pm}$ and $R_{2,a_2}^{\pm}$ for this momentum in figure~\ref{fig_R_plus}. We see that $R_{2,a_2}^{+}$ is by far superior for the extraction of $R^+ _2$, while $R_{2,a_1}^{-}$ is preferable for the determination of $R^- _2$. Since the operators $\mathcal{O}_{4ij}$ and $\mathcal{O}_{123}$ belong to the same $\mathrm{H}(4)$ multiplet, combining the results for $R_{2,a_2}^{+}$ and $R_{2,a_1}^{-}$ in order to obtain $\langle \xi ^2\rangle$ via eq.~\subeqref{eq_renormalized_moments2}{a} is allowed and does not require any additional considerations regarding the renormalization.\par%
As shown in~\cite{Bali:2017ude}, larger momenta can even improve the signal-to-noise ratio in certain situations. This is not the case here: the correlation functions with $\vect{p}=(1,1,1)\p$ have a slightly inferior signal-to-noise ratio compared to those using $\vect{p}=(1,1,0)\p$, cf.\ eq.~(27) of ref.~\cite{Bali:2017ude}. However, this choice enables us to obtain results for the whole operator multiplets in eqs.~\eqref{eq_multiplets} from a single momentum, which makes the calculation more efficient (roughly by a factor of four). That the additional ratio $R_{2,a_2}^{+}$ yields a much better ground state plateau (see the left panel of figure~\ref{fig_R_plus}) is an extra benefit.\par%
\subsection{Chiral extrapolation\label{sec_chpt}}
The CLS ensembles described in section~\ref{sec_lattices} (for more detail see appendix~\ref{app_ens}) enable us to perform a joint chiral and continuum limit extrapolation. As will be explained in section~\ref{sec_results}, both limits are well controlled, the latter due to the extended set of different lattice spacings at our disposal and the former due to the approach of the physical point along two distinct quark mass trajectories, with further constraints from the points along the symmetric line. The formulae for the chiral extrapolation of the first two LCDA $\xi$-moments of the lowest-lying pseudoscalar meson octet, i.e., the~$\pi$, the~$K$, and the $\etaa$~mesons,\footnote{The physical particles $\eta$ and $\eta^\prime$ are mixtures of the singlet $\eta^0$ meson and the octet $\etaa$ meson.} have been worked out in ref.~\cite{Chen:2003fp}. For the even moments~$\langle\xi^{2n}\rangle_{\!M}$ one obtains%
\begin{subequations}\label{eq_chpt_original}%
\begin{align}%
\langle\xi^{2n}\rangle_\pi &= \langle\xi^{2n}\rangle_0 + 2m_\ell\alpha^{(2n)} + (2m_\ell+m_s)\beta^{(2n)}\,,\\
\langle\xi^{2n}\rangle_{\!K} &= \langle\xi^{2n}\rangle_0 + (m_\ell+m_s)\alpha^{(2n)} + (2m_\ell+m_s)\beta^{(2n)}\,,\\
\langle\xi^{2n}\rangle_\etaa &= \langle\xi^{2n}\rangle_0+ \tfrac{2}{3}(m_\ell+2m_s)\alpha^{(2n)} + (2m_\ell+m_s)\beta^{(2n)}\,,
\end{align}%
\end{subequations}%
where $\alpha^{(2n)}$ and $\beta^{(2n)}$ are low energy constants (LECs). It is convenient to introduce the variables%
\begin{subequations}%
\begin{align}%
\bar{\m}^2 &= \frac{2m^2_K + m_{\pi}^2}{3}\approx 2B_0\frac{m_s+2m_\ell}{3}\,, &
\delta \m^2 &= m^2_K - m_{\pi}^2\approx B_0(m_s-m_\ell)\,, \subtag{2}
\end{align}%
\end{subequations}%
such that $\bar{\m}^2$ is approximately constant along the $\operatorname{Tr}\mathcal{M}=\mathrm{constant}$ trajectory, while $\delta \m^2$ vanishes for degenerate quark masses $m_\ell=m_s$. Here $B_0=|\langle \bar{u}u\rangle|/F_0^2\approx 2|\langle \bar{u}u\rangle|/f_{\pi}^2$ is the quark condensate parameter. Along the symmetric line the mesons have to form an exact \SU3 flavor octet with one and the same leading-twist
LCDA for the $\pi$, the $K$ and the $\eta^8$. This becomes evident when rewriting eqs.~\eqref{eq_chpt_original} in terms of the new variables:%
\begin{subequations}\label{eq_chpt}%
\begin{align}%
\langle\xi^{2n}\rangle_\pi &= \langle\xi^{2n}\rangle_0 + \bar{A}^{(2n)}\bar{\m}^2 - 2\delta\!A^{(2n)}\delta\m^2\,,\\
\langle\xi^{2n}\rangle_{\!K} &= \langle\xi^{2n}\rangle_0 + \bar{A}^{(2n)}\bar{\m}^2 + \phantom{2}\delta\!A^{(2n)}\delta\m^2\,,\\
\langle\xi^{2n}\rangle_\etaa &= \langle\xi^{2n}\rangle_0 + \bar{A}^{(2n)}\bar{\m}^2 + 2\delta\!A^{(2n)}\delta\m^2\,.\label{eq_eta}
\end{align}%
\end{subequations}%
Here, $\bar{A}^{(2n)}=\bigl( 2\alpha^{(2n)} + 3 \beta^{(2n)}\bigr)/(2B_0)$ and $\delta\!A^{(2n)}=\alpha^{(2n)}/(3B_0)$ are linear combinations of the LECs of eqs.~\eqref{eq_chpt_original}. Note that the breaking of $\SU3$ flavor symmetry is highly constrained as, to one-loop order in ChPT, we have only one independent symmetry breaking parameter $\delta\!A^{(2n)}$ per LCDA moment. This will allow us to infer the shape of the $\eta^8$ LCDA from the pion and kaon data.\par
In the limit of exact isospin symmetry, C-parity implies that the LCDAs of the pion and~$\etaa$ are even functions of~$\xi$. Therefore, the odd moments vanish. This also applies to the LCDA of the kaon in the limit of exact flavor symmetry $\delta\m^2=0$. Therefore, re-expressing the corresponding formulae of ref.~\cite{Chen:2003fp} in terms of the variables $\bar{\m}$ and $\delta\m$ gives for the odd moments
\begin{subequations}%
\begin{align}%
\langle\xi^{2n+1}\rangle_\pi &= 0\,,&
\langle\xi^{2n+1}\rangle_{\!K} &= \delta\!A^{(2n+1)}\delta\m^2\,,&
\langle\xi^{2n+1}\rangle_\etaa &= 0\,.\subtag{3}
\end{align}%
\end{subequations}%
\subsection{Discretization effects} \label{sec_discret}
For both LCDA moments we expect the leading-order discretization effects to be linear in $a$, as the corresponding operators, $\mathcal{O}^- _{\rho\mu}$ and $\mathcal{O}^- _{\rho\mu\nu}$, have not been $\mathcal{O}(a)$ improved.\footnote{We remark that $\mathcal{O}(a)$~effects are actually suppressed by one power of the coupling constant~$g^2$.} We make the ansatz
\begin{subequations}%
\begin{align}%
\langle\xi^1\rangle_{\!\M} &= \bigl(1+c_0^{(1)}a+\bar{c}^{(1)}\bar{\m}^2a+\delta c^{(1)}_\M\delta\m^2a\bigr) \times
\begin{dcases}
 0\,, & \M=\pi\,,\\
 \delta\!A^{(1)}\delta\m^2\,, & \M=K\,,
\end{dcases}\label{eq_extrapolate_xi1}\displaybreak[0]\\
\langle\xi^2\rangle_{\!\M} &= \bigl(1+c_0^{(2)}a+\bar{c}^{(2)}\bar{\m}^2a+\delta c^{(2)}_\M\delta\m^2a\bigr) \times
\begin{dcases}
 \langle\xi^2\rangle_0+\bar{A}^{(2)}\bar{\m}^2-2\delta\!A^{(2)}\delta\m^2\,, & \M=\pi\,,\\
 \langle\xi^2\rangle_0+\bar{A}^{(2)}\bar{\m}^2+\phantom{2}\delta\!A^{(2)}\delta\m^2\,, & \M=K\,,
\end{dcases}\label{eq_extrapolate_xi2}
\end{align}%
\end{subequations}%
where the chiral extrapolation formulae of section~\ref{sec_chpt} are combined with a linear pa\-ram\-e\-triza\-tion of discretization effects, including mass-dependent terms. The $\SU3$ flavor constraints will be violated by $\mathcal{O}(a)$~terms since our fermion formulation explicitly breaks chiral symmetry. Therefore, $\delta c^{(2)}_\pi$~and~$\delta c^{(2)}_K$ are independent parameters. Within this ansatz we require a total of four parameters to describe the lattice spacing and quark mass dependence of~$\langle\xi^1\rangle_{\!K}$, while seven parameters are needed for our joint extrapolation of~$\langle\xi^2\rangle_\pi$ and~$\langle\xi^2\rangle_{\!K}$ that also yields~$\langle\xi^2\rangle_\etaa$. We will see that all lattice data are well described by the above ans\"atze. Nevertheless, we will vary the parametrization to explore the systematics associated with the choice of this particular functional dependence.\par%
In the continuum, the remaining operator $\mathcal{O}^+_{\rho\mu\nu}$ can be written as the second derivative of the axialvector current, $\mathcal{O}^+_{\rho\mu\nu}(x)=\partial_{(\mu}\partial_{\mathstrut\nu}\mathcal{A}_{\rho )}(x)$. This is not the case on the lattice and the renormalization factors of~$\mathcal{O}^+$ and~$\mathcal{A}$ differ. However, in the continuum limit the renormalized lattice ratio should approach unity,%
\begin{align}\label{eq_eins_continuum}
\zeta_{22}R_2^+ &= \langle\mathbb{1}^2\rangle \xrightarrow{a\to0} 1\,,
\end{align}%
such that the continuum relation $a_2=\tfrac{7}{12}\bigl[5\langle\xi^2\rangle-1\bigr]$ is recovered from eq.~\subeqref{eq_renormalized_moments2}{b}. We employ a nonperturbatively $\mathcal{O}(a)$~improved fermion action and tree-level $\mathcal{O}(a)$~improved derivatives in our operators. Assuming small order~$a$ discretization effects in $\langle\mathbb1^2\rangle_{\!\M}$,%
\begin{align}\label{eq_eins_extra_no_offset}%
\langle\mathbb1^2\rangle_{\!\M} &= 1 + e_{0,2}^{(2)}a^2 + \bar{e}_2^{(2)}\bar{\m}^2a^2 + \delta e^{(2)}_{\!\M,2}\delta\m^2a^2\,
\end{align}%
should provide a sensible parametrization of the data. In the next section we will discuss and check this ansatz.\par%
\section{Extrapolation strategy and error budget\label{sec_results}}%
In the following we present our results for the first and the second $\xi$-moments and Gegenbauer moments of the leading-twist pseudoscalar meson distribution amplitudes. In addition to the results for the pion and the kaon, which are extracted directly from the lattice data, we infer the second moment of the $\etaa$~meson using eq.~\eqref{eq_eta} from the $\SU3$ symmetry breaking constraints obtained from ChPT in ref.~\cite{Chen:2003fp}. Previous lattice determinations of the Gegenbauer moments~\cite{Martinelli:1987si,DeGrand:1987vy,Daniel:1990ah,DelDebbio:1999mq,DelDebbio:2002mq,Braun:2006ci,Boyle:2006xq,Boyle:2006pw,Braun:2006dg,Donnellan:2007xr,Arthur:2010xf,Braun:2015lfa,Braun:2015axa,Bali:2017ude} lacked ensembles with lattice spacings smaller than~$\unit{0.06}{\femto\meter}$ and so far no controlled continuum limit extrapolation has been carried out. This is particularly problematic for the second moment~$a_2^M$, which mixes with~$\langle\mathbb{1}^2\rangle_{\!\M}$ under renormalization, see eqs.~\eqref{eq_renormalized_moments2}.\par%
\subsection{A game of ones\label{sec_game}}%
\begin{SCfigure}[1][tp]%
\centering%
\includegraphics[width=.485\textwidth]{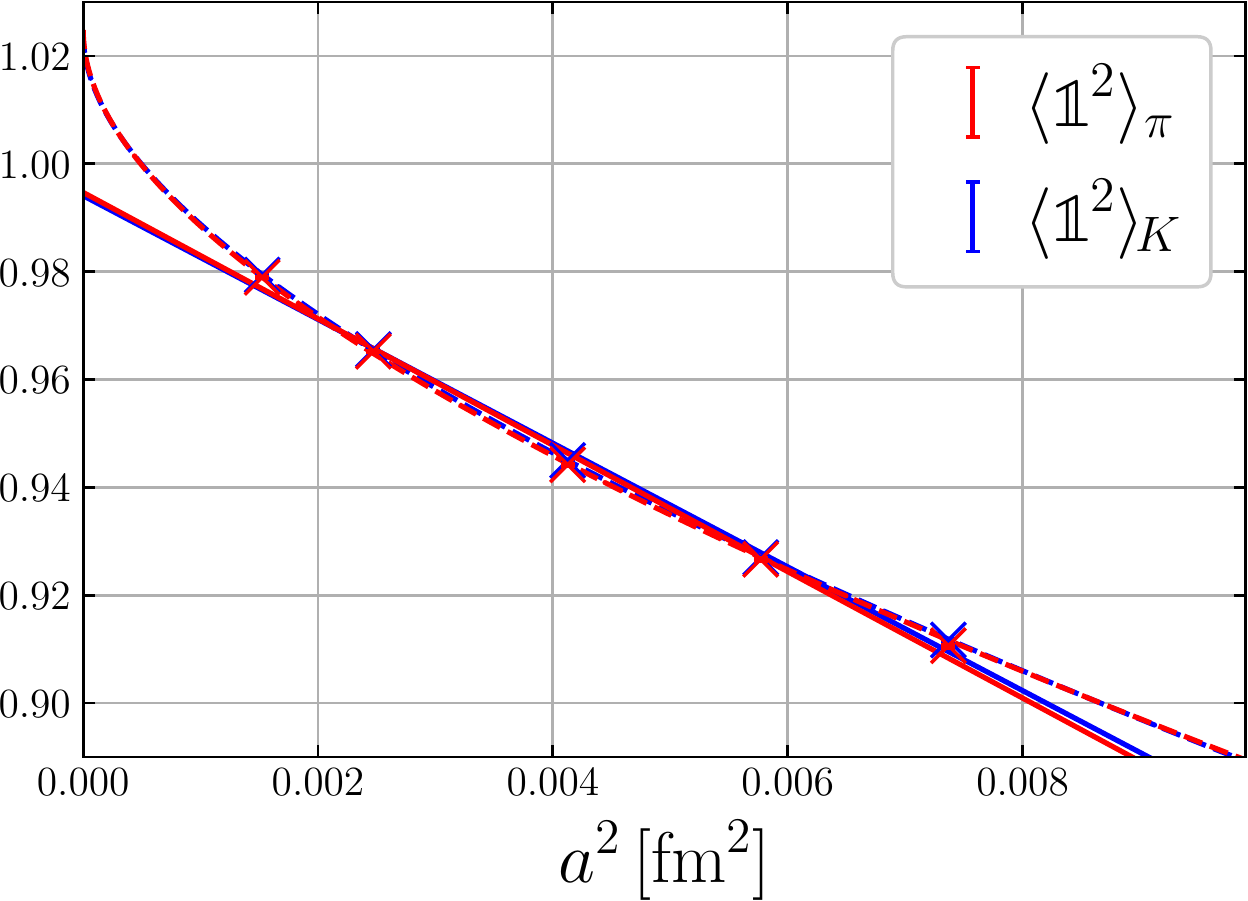}%
\caption{The quantity $\langle\mathbb{1}^2\rangle_{\!\M}$ as a function of the squared lattice spacing~$a^2$, plotted at the physical mass point. The solid lines represent the result of a global fit using eq.~\eqref{eq_eins_extra_with_offset}. The points shown have been obtained by translating all data points to the physical masses along the fitted function and then averaging measurements from the same lattice spacing. The dashed curves correspond to the alternative fit carried out to investigate linear terms as described in the main text.\label{fig_eins_a_world}}%
\end{SCfigure}%
The continuum limit~$\langle\mathbb{1}^2\rangle_{\!\M}\xrightarrow{a\to0}1$ is known. Using this value as a constraint and fitting our data, we find that the $a$ dependence is mostly quadratic and the possible linear contribution is small. This is consistent with expectations based on tree-level lattice perturbation theory, where linear terms vanish exactly.\par%
One can play another game, pretend that the continuum value of~$\langle\mathbb{1}^2\rangle_{\!\M}$ is not known, and try to determine it from the data. The quadratic fit ansatz%
\begin{align}\label{eq_eins_extra_with_offset}%
\langle\mathbb1^2\rangle_{\!\M} &= I_\M + e_{0,2}^{(2)}a^2 + \bar{e}_2^{(2)}\bar{\m}^2a^2 + \delta e^{(2)}_{\!\M,2}\delta\m^2a^2\,,
\end{align}%
using $I_M$ as a free parameter, gives a continuum limit value close to one with only $0.5\%$~deviation, see the solid line in figure~\ref{fig_eins_a_world}. This agreement is nontrivial (unrenormalized lattice values in the considered region of lattice spacings lie in the range \mbox{$0.59$--$0.68$}, see, e.g., the left panel of figure~\ref{fig_R_plus}) and can be viewed as confirmation of our calculation of the corresponding renormalization constant.\par%
However, without the constraint at $a=0$, the smallness of linear contributions in comparison to the quadratic $a$ dependence cannot be inferred from the data: an alternative fit including the additional linear terms~$e_0^{(2)}a$, $\bar{e}^{(2)}\bar{\m}^2a$, and~$\delta e^{(2)}_{\!\M}\delta\m^2a$ (dashed curve in figure~\ref{fig_eins_a_world}) leads to a continuum value that is about $2.5\%$ above unity. The difference can be viewed as a systematic uncertainty of the continuum extrapolation~(labeled~$a$ in the following), yielding the ``lattice values'' $I_\pi=0.9947^{+2}_{-2}(80)_r(301)_a$ and $I_K=0.9941^{+1}_{-2}(80)_r(300)_a$, where statistical errors are given by the sub-/superscript pair and the uncertainty due to the renormalization~($r$) is determined as described in section~\ref{sec_renorm}. To avoid misunderstanding: the values of $I_M$ (and the fits shown in figure~\ref{fig_eins_a_world}) are not used in the determination of the moments of meson LCDAs, to be discussed in the following sections. Their determination merely serves as a sanity check to strengthen the confidence in our renormalization procedure.\par%
In comparison to our previous work, see figure~3 of ref.~\cite{Braun:2015axa}, we achieve a much higher statistical precision for~$\langle\mathbb{1}^2\rangle_{\!\M}$, such that the statistical error now contributes by far the smallest uncertainty. This improvement in statistics is mostly due to employing the operator~$\mathcal{O}^+_{123}$ in the new method~\subeqref{eq_ratio2}{b}, compared to the old method involving the operators~$\mathcal{O}^+_{4ij}$, see also the left panel of figure~\ref{fig_R_plus}. Furthermore, it turns out that also the systematic uncertainties due to renormalization~($0.8\%$) and due to discretization effects~($3\%$) are quite small.\par%
\subsection{Extrapolation of the second LCDA moments}%
\begin{figure}[tp]%
\centering
\includegraphics[width=\textwidth]{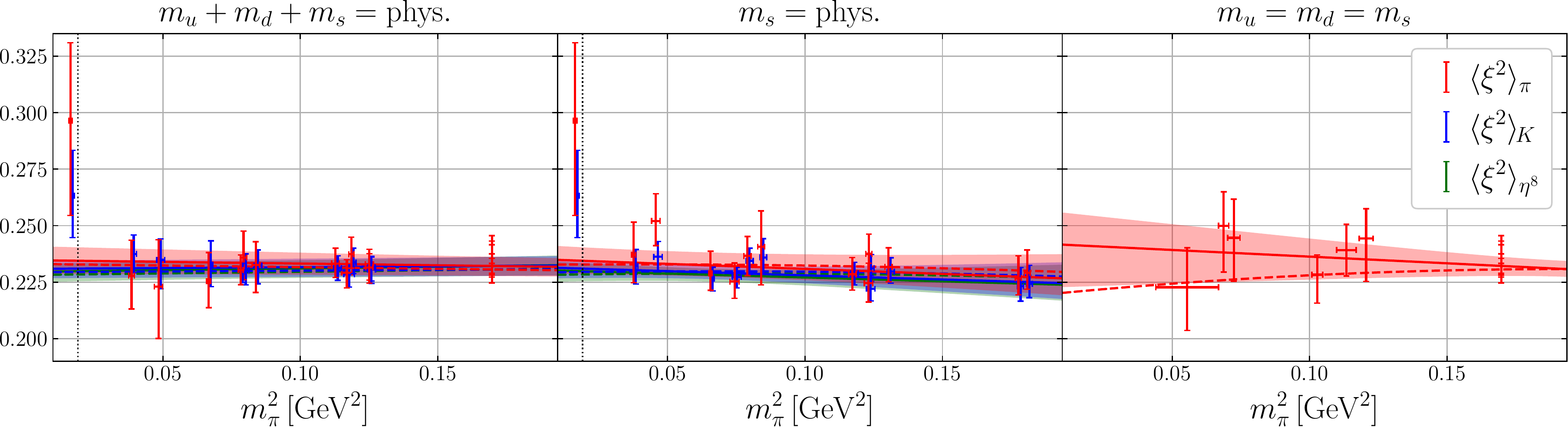}
\caption{Dependence of the moments~$\langle\xi^2\rangle_{\!\M}$ on the squared pion mass, plotted in the continuum limit. The points shown have been obtained by translating all data along the fitted function (keeping the masses fixed). The plots for the individual lattice spacings can be found in figure~\ref{fig_xi2} in appendix~\ref{app_plots}. The solid lines and shaded statistical error bands represent our main result. The dashed curves correspond to the mean value of an alternative fit (including a term of higher order in the masses) used to estimate the parametrization dependence as described in the main text.\label{fig_m_xi2}}%
\end{figure}%
For the extrapolation of the second moments~$\langle\xi^2\rangle_\pi$ and~$\langle\xi^2\rangle_{\!K}$ we use eq.~\eqref{eq_extrapolate_xi2}. We then insert the fitted LECs~$\langle\xi^2\rangle_0$, $\bar{A}^{(2)}$, and~$\delta\!A^{(2)}$ into eq.~\eqref{eq_eta} in order to obtain a prediction for~$\langle\xi^2\rangle_\etaa$ in the continuum. The combined extrapolation is shown in figure~\ref{fig_m_xi2} as a function of the pion mass and in figure~\ref{fig_a_xi2} as a function of the lattice spacing. Figure~\ref{fig_m_xi2} shows that the breaking of \SU3 flavor symmetry among these observables is rather small. Actually, within our errors, we find no differences between $\langle \xi^2 \rangle_\pi$, $\langle \xi^2 \rangle_{\!K}$, and $\langle \xi^2 \rangle_\etaa$. To estimate the systematic uncertainty due to the choice of the parametrization of the mass dependence we perform an alternative fit by including the additional term~$\bar{A}_2^{(2)}\bar{m}^4$, i.e., allowing for one extra parameter.\footnote{One could also try terms proportional to~$\bar{m}^2\delta m^2$ or~$\delta m^4$, but these introduce one new fit parameter for each meson instead of just one additional parameter in total, leading to overfitting.} This fit is indicated by the dashed line in figure~\ref{fig_m_xi2} and we take the difference with respect to the mean value of our main fit as the corresponding error. It can be seen that the second moments of the pseudoscalar LCDAs depend only mildly on the quark masses. In contrast, the discretization effects are quite sizable and amount to a correction of roughly $10\%$ from our largest lattice spacing of~$a=\unit{0.086}{\femto\meter}$ to the continuum, as shown in figure~\ref{fig_a_xi2}. To estimate the systematics of the $a$ dependence we again perform an alternative fit, this time adding the term~$c_{0,2}^{(2)}a^2$, indicated by the dashed line in figure~\ref{fig_a_xi2}. For our final results shown in table~\ref{table_results} we take the difference between this fit and our main fit as the estimate of the systematic error due to the continuum extrapolation.\par%
We have checked that other methods to estimate this systematic error lead to compatible results, e.g., omitting the data from the coarsest lattice spacing. Another possibility is to consider continuum extrapolations for two lattice observables that have the same continuum limit. To this end, we compare the second Gegenbauer moment $a_2^M$ defined in terms of $\langle\xi^2\rangle_{\!M}$ via the continuum theory relation%
\def\figheight{4.813cm}%
\begin{figure}[tp]%
\centering
\includegraphics[height=\figheight]{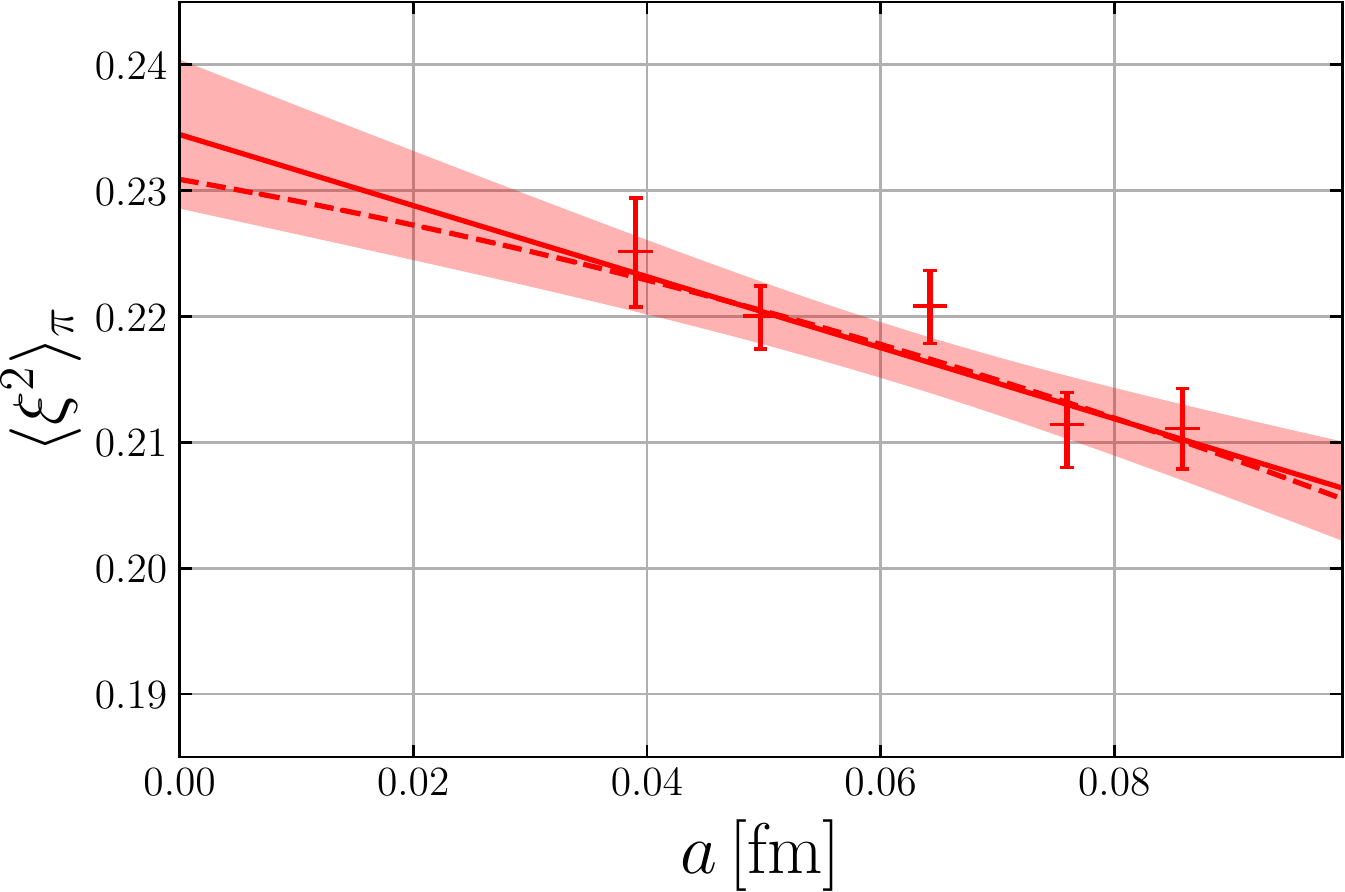}\hfill%
\includegraphics[height=\figheight]{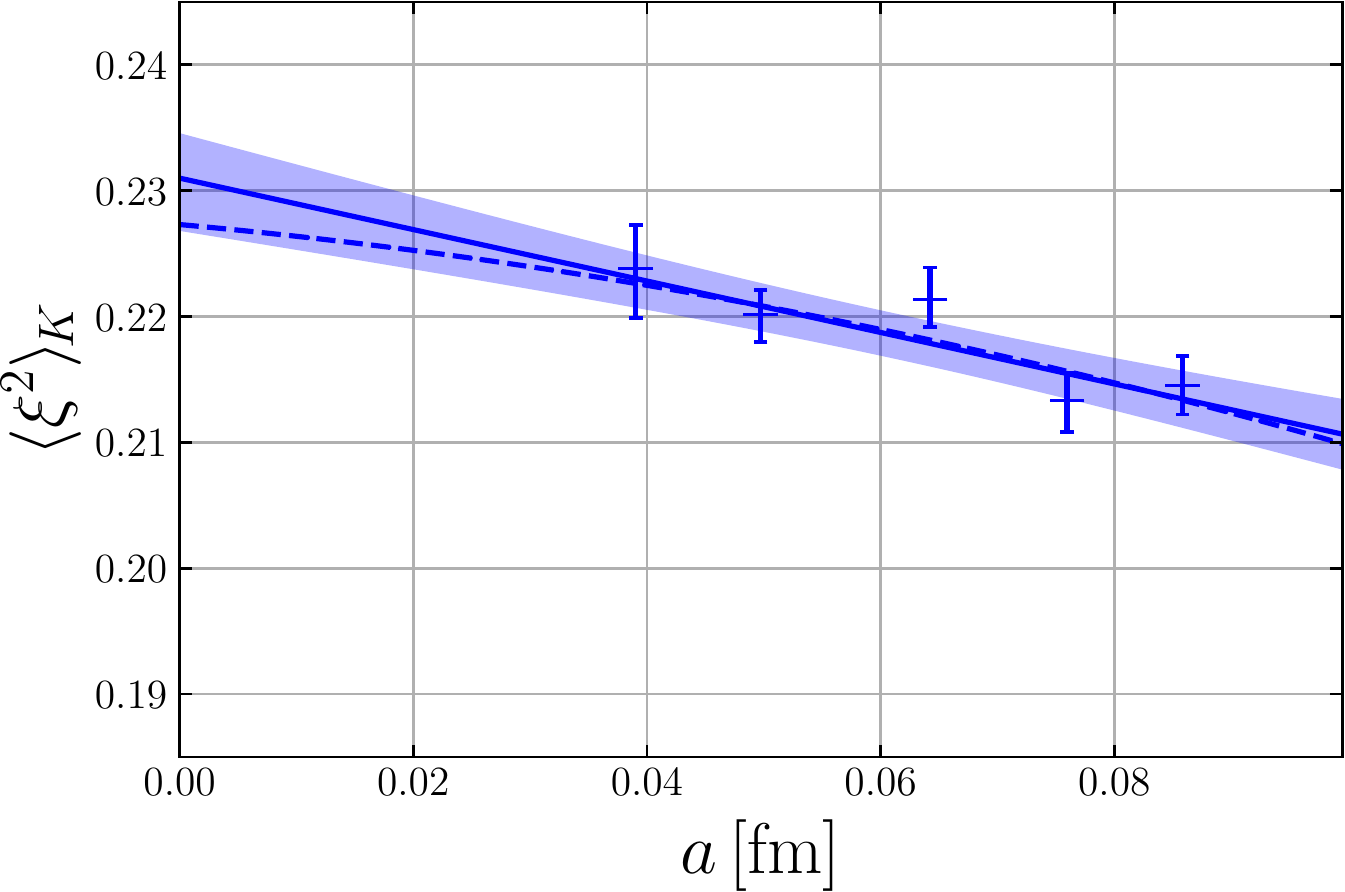}%
\caption{Dependence of the moments~$\langle\xi^2\rangle_{\!\M}$ on the lattice spacing~$a$, plotted at physical quark masses. The points shown have been obtained by translating all data along the fitted function (keeping the lattice spacing fixed) and then averaging measurements with the same~$a$. The plots for the individual trajectories can be found in figure~\ref{fig_xi2} in appendix~\ref{app_plots}. The solid lines and shaded statistical error bands represent our main result. The dashed curves correspond to the mean value of an alternative fit (including a term proportional to~$a^2$) used to estimate the parametrization dependence as described in the main text.\label{fig_a_xi2}}%
\end{figure}%
\begin{align}%
\label{eq_a2}
a_2^M &= \frac{7}{12} \bigl[5\langle\xi^2\rangle_{\!M}-1\bigr]\,,
\end{align}%
with the definition%
\begin{align}%
a_2^M &= \frac{7}{12} \bigl[5\langle\xi^2\rangle_{\!M}-\langle\mathbb{1}^2\rangle_{\!M}\bigr]\,, \label{eq_a2_otherlat}
\end{align}%
which is natural at a finite lattice spacing. As argued in section~\ref{sec_game}, the difference between these two quantities should be mainly due to $\mathcal O(a^2)$ effects. A comparison is shown in figure~\ref{fig_Braun} for the pion (left) and kaon (right). In both cases we perform a linear extrapolation in the lattice spacing. The difference in the continuum compares reasonably well to the estimates for $\mathcal{O}(a^2)$ effects obtained from the procedure explained above.\par%
\def\figheight{4.813cm}%
\begin{figure}[t]%
\centering
\includegraphics[height=\figheight]{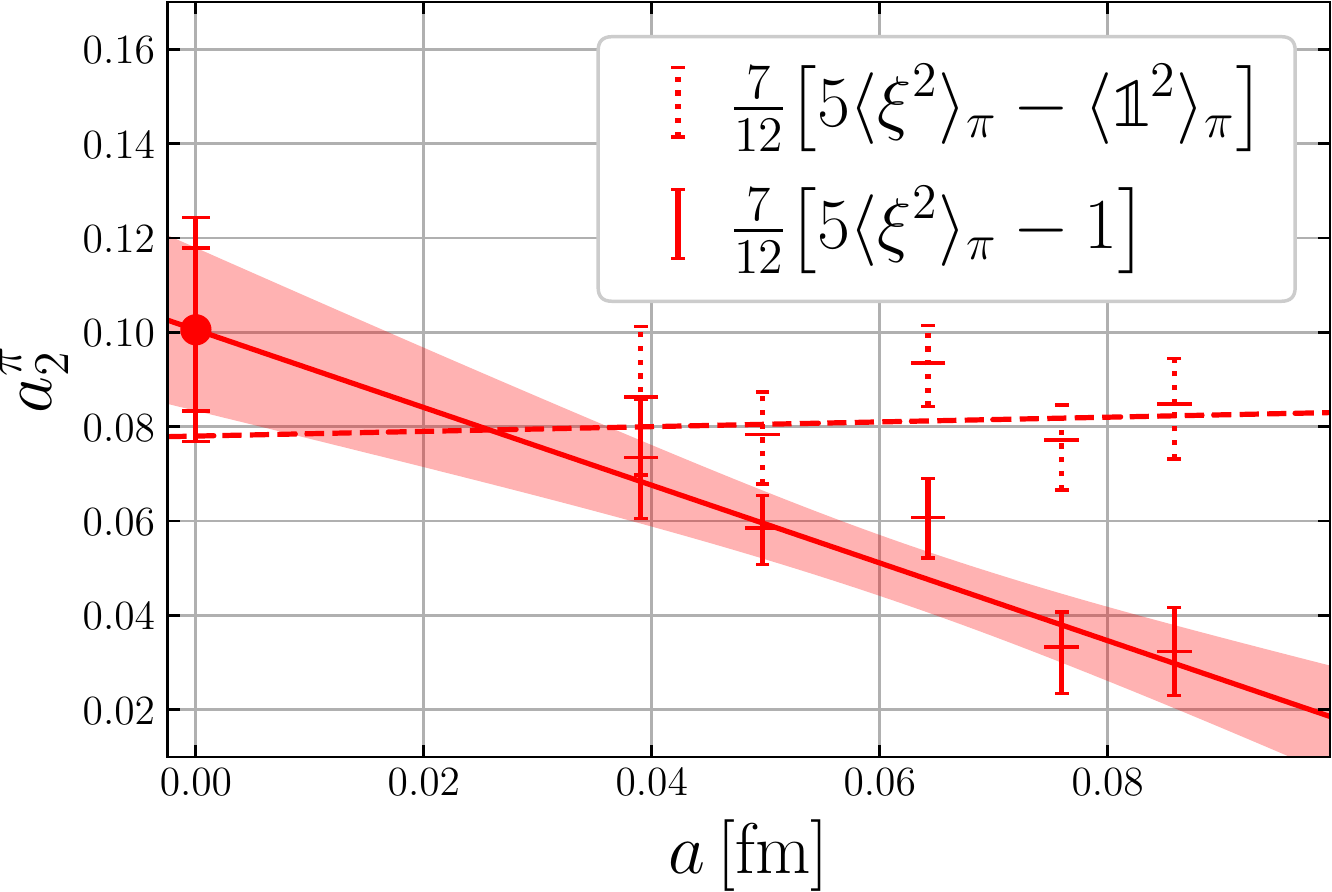}\hfill%
\includegraphics[height=\figheight]{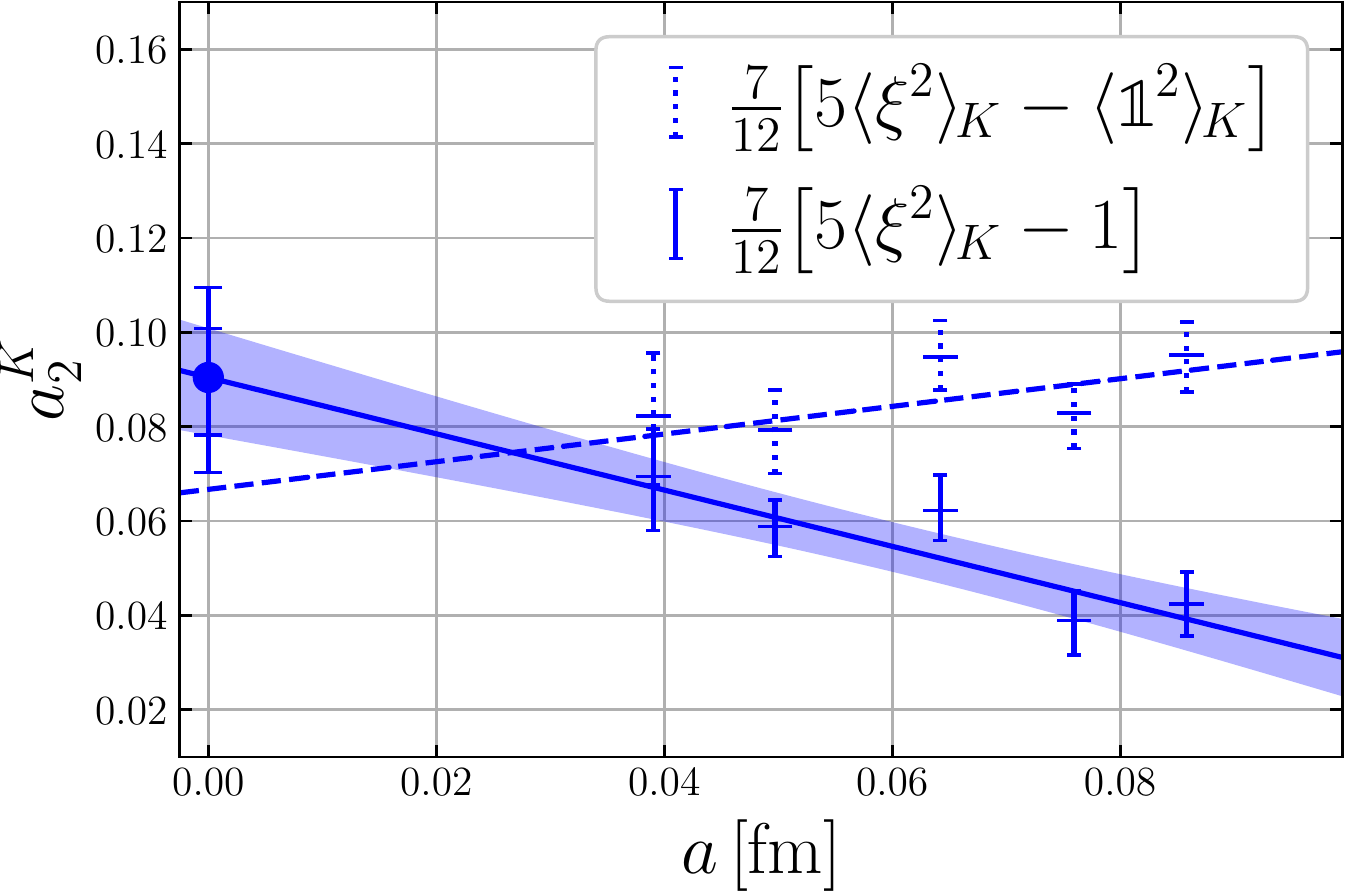}%
\caption{Illustration of a different approach to quantify the discretization effect uncertainty for~$a_2^M$. Instead of performing an alternative fit to the same data set (as in figure~\ref{fig_a_xi2}) one could perform the same fit to an alternative data set. The points shown have been obtained by translating all data along the fitted functions (keeping the lattice spacing fixed) and then averaging measurements with the same~$a$. In this picture, the solid lines and shaded bands represent the central values and statistical errors. The dashed lines correspond to the mean value of the fit to the alternative data points and the difference between solid and dashed lines could be used to estimate the systematic uncertainty due to the continuum extrapolation. (To avoid misunderstanding: the error estimates shown in this figure are not used in our determination of the moments of meson LCDAs.) For comparison we plot our final values for~$a_2^\pi$ and~$a_2^K$ as points at $a=0$, where the inner error bars are statistical only and the outer correspond to all errors added in quadrature.\label{fig_Braun}}%
\end{figure}%
\subsection{Extrapolation of the first LCDA moment}
\begin{figure}[t]%
\centering
\includegraphics[height=\figheight]{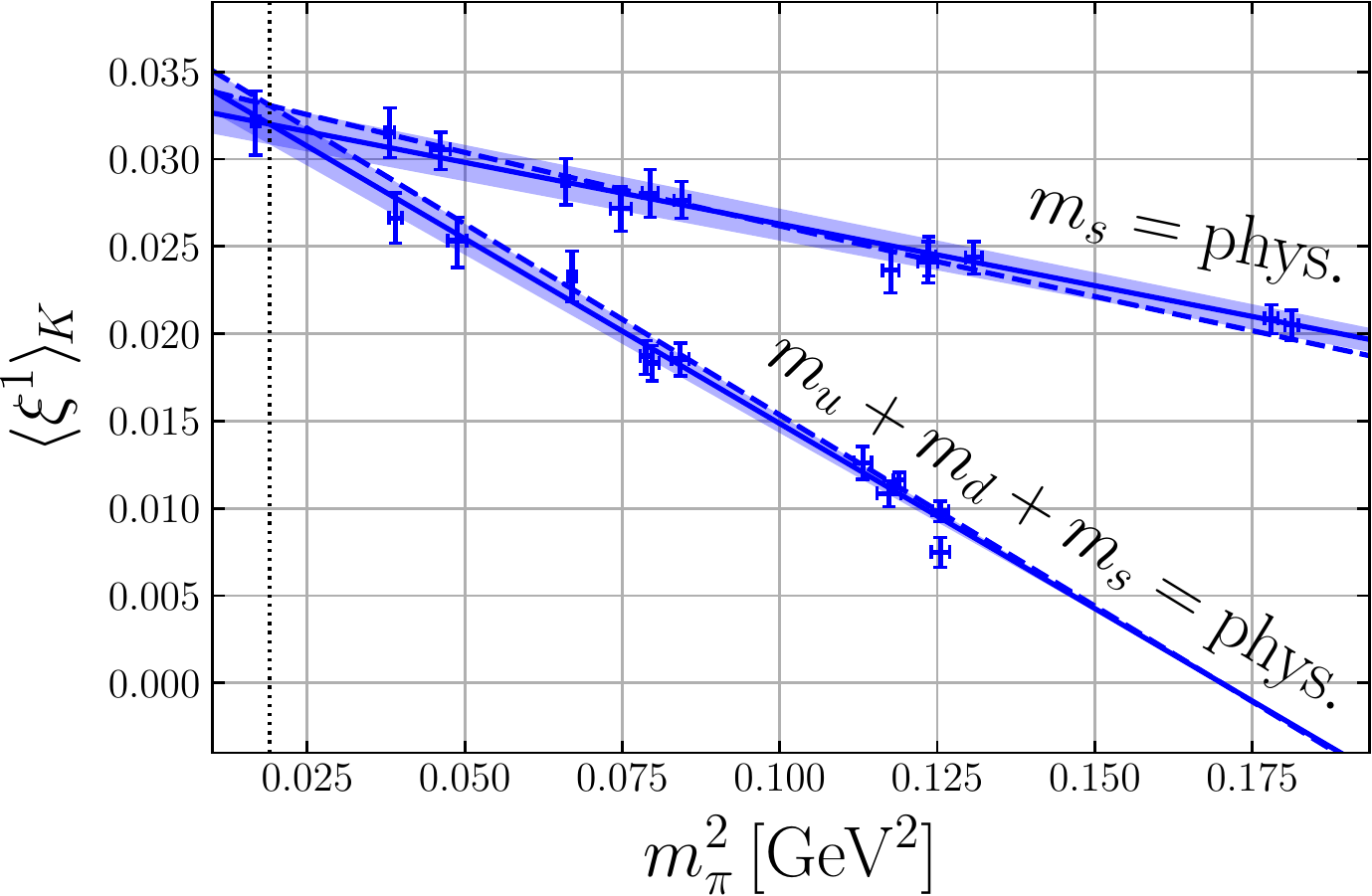}\hfill%
\includegraphics[height=\figheight]{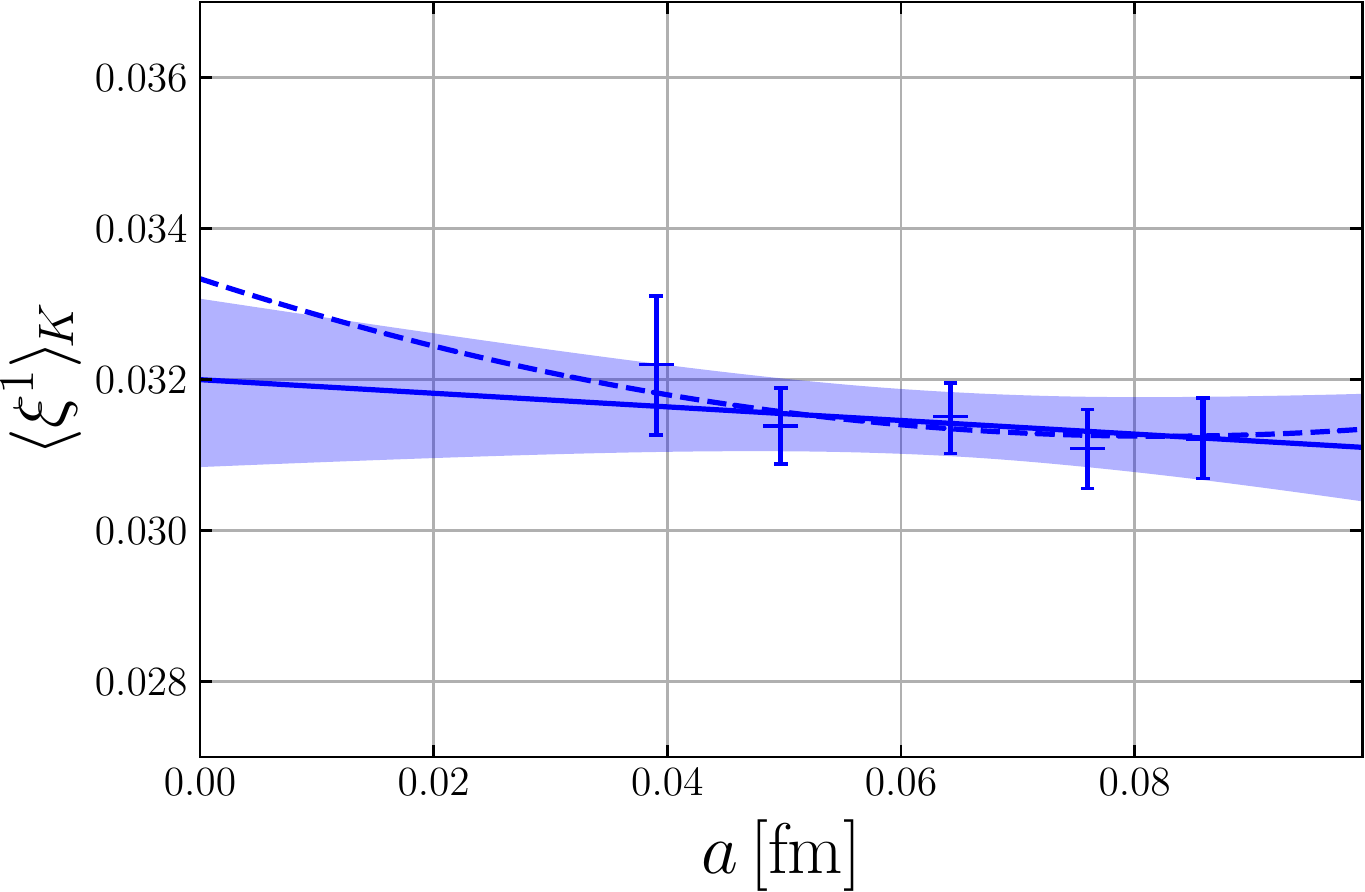}%
\caption{Left: The same as figure~\ref{fig_m_xi2}, but for the moment~$\langle\xi^1\rangle_{\!K}$; the two relevant trajectories have been condensed into one plot. Right: The same as figure~\ref{fig_a_xi2}, but for the moment~$\langle\xi^1\rangle_{\!K}$. The plots for the individual lattice spacings and trajectories can be found in figure~\ref{fig_xi1} in appendix~\ref{app_plots}.\label{fig_am_xi1}}%
\end{figure}

A combined continuum and chiral extrapolation of $\langle\xi^1\rangle_{\!K}$ is performed using eq.~\eqref{eq_extrapolate_xi1}, which automatically enforces the constraint that all odd moments have to vanish in the limit of exact flavor symmetry (which is also true for the lattice data). We therefore only have data points for lattices with nondegenerate quark masses, see figure~\ref{fig_am_xi1}~(left). The mass dependence in the continuum limit is determined by the single parameter $\delta\!A^{(1)}=\unit{0.141(23)}{\power{\giga\electronvolt}{-2}}$, see eq.~\eqref{eq_extrapolate_xi1}. Notably, we find only a very mild dependence of the first moment on the lattice spacing that is consistent with a flat behavior within errors, see figure~\ref{fig_am_xi1}~(right). The parametrization dependence is investigated, as above, by performing two alternative fits, each including a single additional parameter. These fits are indicated by dashed lines in the corresponding plots; one includes the term~$\delta\!\bar{A}^{(1)}\bar{m}^2\delta m^2$ for the mass dependence,\footnote{One could use a term~$\propto\delta m^4$ instead (this adds a single parameter in the case of the odd moments), which leads to a very similar estimate for the uncertainty. Using~$\bar{m}^4$ is however not allowed since the whole fit function must be proportional to~$\delta m^2$ due to symmetry.} the other one includes the term~$c_{0,2}^{(1)}a^2$ for the lattice spacing dependence.\par%
\begin{figure}[t]%
\centering
\includegraphics[width=.485\textwidth]{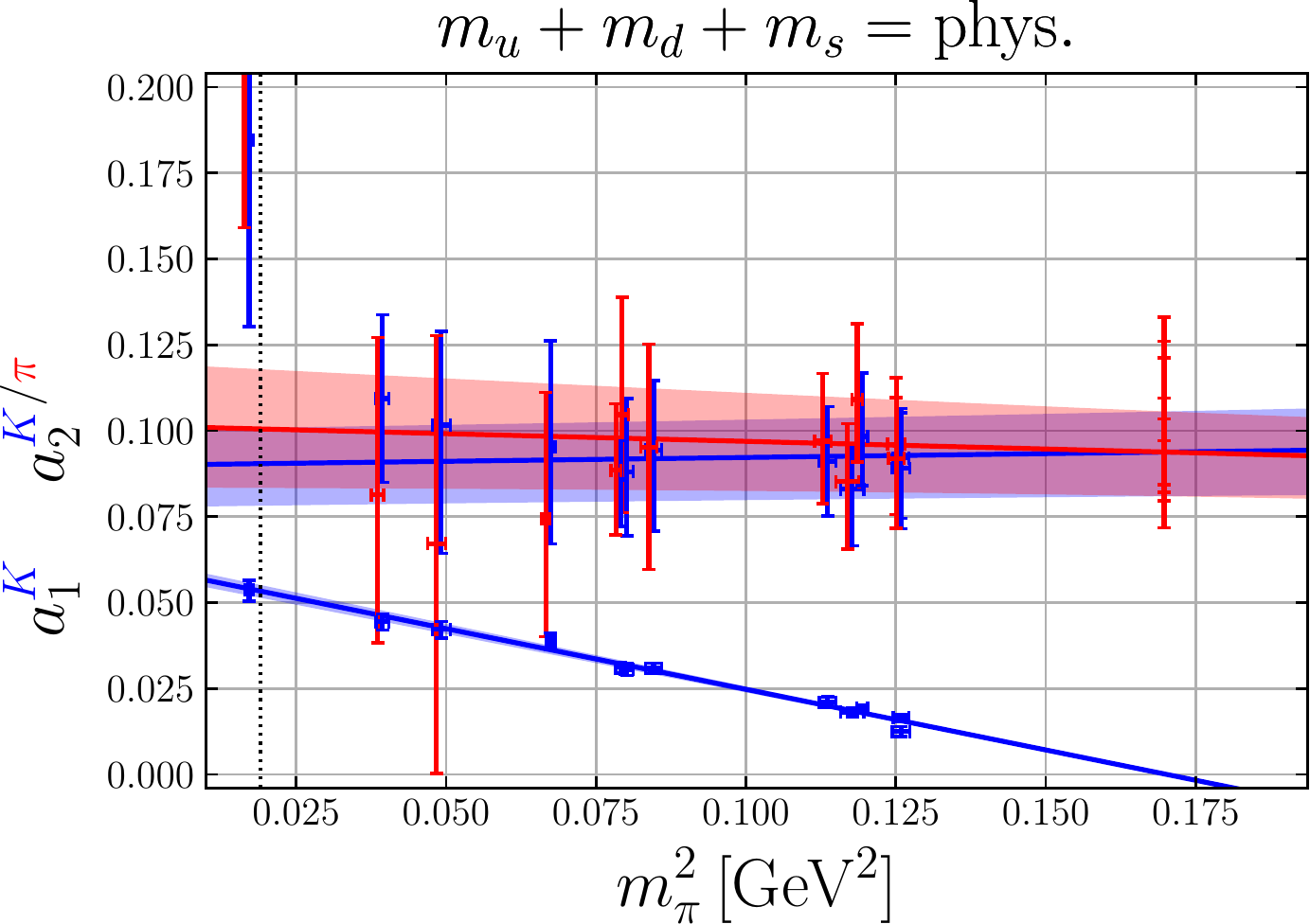}\hfill%
\includegraphics[width=.485\textwidth]{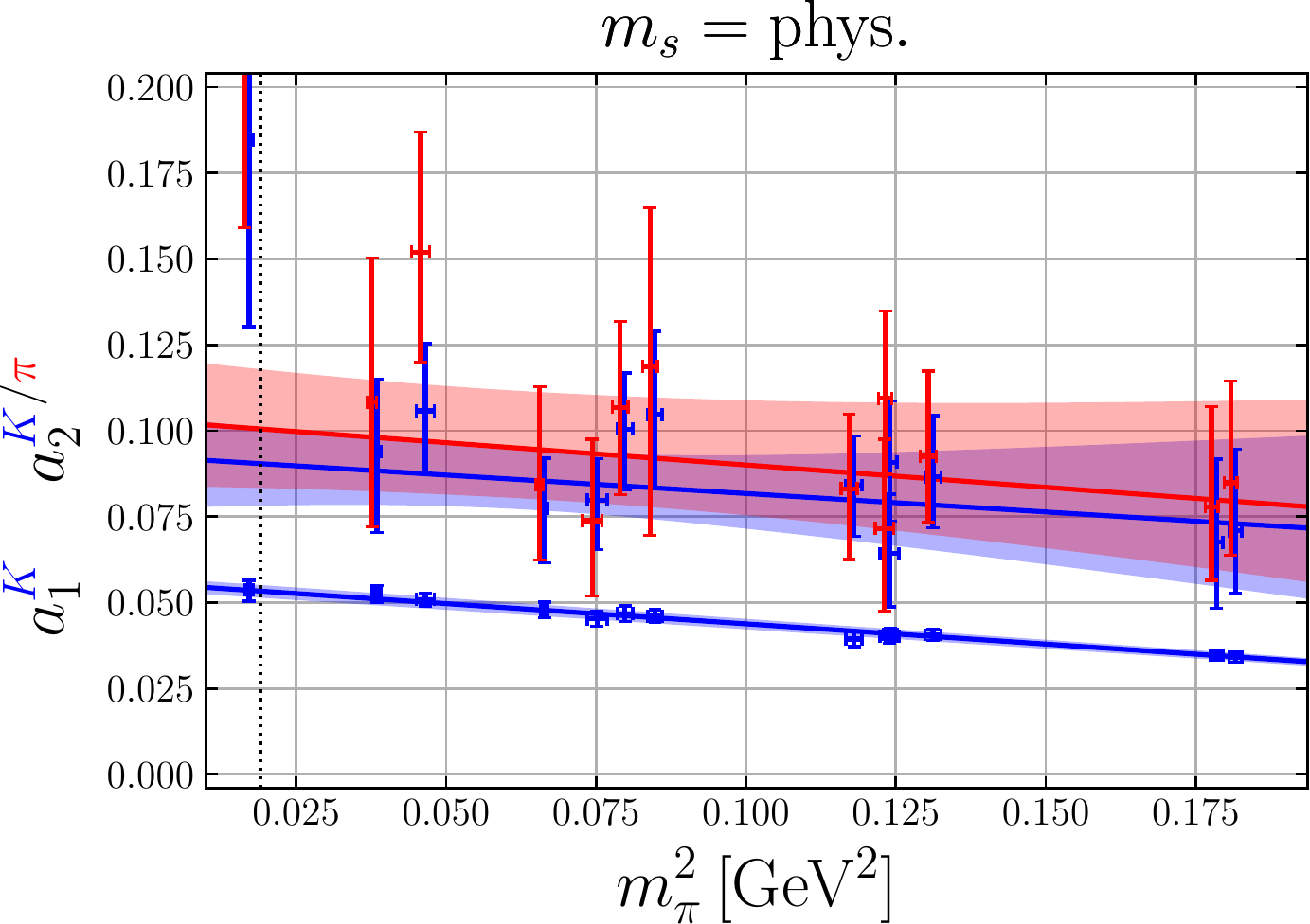}%
\caption{Summary plot for the first and second Gegenbauer moments of the pion~(red) and the kaon~(blue) in the continuum limit along two quark mass trajectories: fixed average quark mass~(left) and fixed strange quark mass~(right). These two trajectories intersect at the physical point (dotted vertical line). The error bands shown are statistical only.\label{fig_summary}}%
\end{figure}%
\subsection{Summary of the results}
\def\myrule{\rule[-2pt]{0sp}{12pt}}%
\begin{table}[t]%
\centering%
\caption{Continuum limit extrapolated values for the first two moments of the octet mesons. The results have been converted to the $\MSbar$ scheme at $\mu=\unit{2}{\giga\electronvolt}$ using intermediate $\text{RI}^\prime$ schemes and different loop orders in the perturbative matching. The statistical error given as sub- and superscript reflects the errors of the data after extrapolation. The numbers in parentheses give estimates of the systematic uncertainties due to the nonperturbative renormalization~($r$) as described in section~\ref{sec_renorm}, the continuum extrapolation~($a$), and the chiral extrapolation~($m$). As discussed in section~\ref{sec_lattices}, finite volume effects are negligible in our setting.\label{table_results}}%
\begin{widetable}{\textwidth}{crrE{.}{.}{-}{1.10}{{}^{+0}_{-0}()_r()_a()_m}E{.}{.}{-}{1.10}{{}^{+00}_{-00}()_r()_a()_m}}%
\toprule
\myrule$M$ & \multicolumn{1}{c}{$\text{RI}^\prime$} & \multicolumn{1}{c}{order} & \multicolumn{1}{c}{$\langle\xi^2\rangle_{\!M}$} & \multicolumn{1}{c}{$a_2^M$}\\
\midrule
\myrule$\pi$   & SMOM &      NNLO & 0.234^{+6}_{-6}(4)_r(4)_a(2)_m & 0.101^{+17}_{-17}(12)_r(10)_a(5)_m\\
\myrule$\pi$   & SMOM &       NLO & 0.227^{+6}_{-6}(5)_r(5)_a(2)_m & 0.078^{+18}_{-19}(16)_r(13)_a(5)_m\\
\myrule$K$     & SMOM &      NNLO & 0.231^{+4}_{-4}(4)_r(4)_a(1)_m & 0.090^{+10}_{-12}(11)_r(11)_a(4)_m\\
\myrule$K$     & SMOM &       NLO & 0.223^{+4}_{-5}(5)_r(5)_a(2)_m & 0.067^{+11}_{-13}(16)_r(14)_a(5)_m\\
\myrule$\etaa$ & SMOM &      NNLO & 0.230^{+4}_{-4}(4)_r(4)_a(1)_m & 0.087^{+10}_{-13}(11)_r(11)_a(4)_m\\
\myrule$\etaa$ & SMOM &       NLO & 0.222^{+4}_{-5}(6)_r(5)_a(2)_m & 0.063^{+11}_{-14}(16)_r(14)_a(5)_m\\
\midrule
\myrule$M$ & \multicolumn{1}{c}{$\text{RI}^\prime$} & \multicolumn{1}{c}{order} & \multicolumn{1}{c}{$\langle\xi^1\rangle_{\!M}$} & \multicolumn{1}{c}{$a_1^M$}\\
\midrule
\myrule$K$     & SMOM &      NNLO & 0.0320^{+11}_{-12}(3)_r(13)_a(11)_m & 0.0533^{+18}_{-19}(6)_r(22)_a(18)_m\\
\myrule$K$     & SMOM &       NLO & 0.0327^{+11}_{-12}(6)_r(14)_a(11)_m & 0.0545^{+18}_{-20}(9)_r(23)_a(18)_m\\
\myrule$K$     &  MOM & N${}^3$LO & 0.0315^{+11}_{-11}(1)_r(11)_a(10)_m & 0.0525^{+18}_{-19}(2)_r(19)_a(17)_m\\
\myrule$K$     &  MOM &      NNLO & 0.0319^{+11}_{-12}(1)_r(11)_a(10)_m & 0.0531^{+18}_{-19}(2)_r(18)_a(17)_m\\
\bottomrule
\end{widetable}%
\end{table}%
The mass dependence of the first two Gegenbauer moments $a_1^M=\tfrac{5}{3}\langle\xi^1\rangle_{\!M}$ and~$a_2^M=\tfrac{7}{12}\bigl[5\langle\xi^2\rangle_{\!M}-1\bigr]$ in the continuum limit is summarized in figure~\ref{fig_summary}. Our final results for the moments~$\langle\xi^1\rangle_{\!M}$ and~$\langle\xi^2\rangle_{\!M}$ as well as the corresponding Gegenbauer moments (in the continuum limit at the $\MSbar$~scale $\mu=\unit{2}{\giga\electronvolt}$) are collected in table~\ref{table_results}. It can be seen as a success of our strategy, i.e., generating ensembles on different quark mass trajectories while simultaneously reaching fine lattice spacings, that all the systematic uncertainties can be controlled and are of a similar or smaller size than the statistical accuracy. In analogy to the prevalent procedure used in determinations of parton distribution functions from experimental data, we quote separate results for the NLO (one-loop) and the NNLO (two-loop) analysis. Even though the results obtained using the SMOM scheme with NLO and NNLO matching almost agree within the given renormalization error, the central values still deviate considerably from each other so that a three-loop matching formula between the $\RI$ and $\MSbar$ schemes would be welcome. As to be expected, the systematic uncertainty due to renormalization decreases for increasing loop order. We quote our SMOM NNLO values as the final results in the abstract.\par%
\section{Discussion\label{sec_discussion}}%
\begin{table}[tp]%
\caption{The second moment of the pion LCDA at the $\MSbar$~scale~$\mu=\unit{2}{\giga\electronvolt}$. The CZ model fixes $a_2^\pi=2/3$ at the low scale $\mu\simeq\unit{500}{\mega\electronvolt}$; for a discussion of the extrapolation to higher scales see ref.~\cite{Bakulev:2002uc}. The abbreviations stand for: LQCD: lattice calculation; $N_f=2(+1)$: calculation using $N_f=2(+1)$ sea quarks; SW: nonperturbatively ${\mathcal O}(a)$ improved Sheikholeslami--Wohlert (i.e., Wilson-clover) fermion action; DWF: domain-wall fermions; QCDSR: QCD sum rules; NLC: nonlocal condensates; LCSR: light-cone sum rules; R: renormalon model for twist-$4$ corrections; DSE: Dyson--Schwinger equations with rainbow-ladder truncated (RL) or DCSB-improved (DB) kernels. The LCSR analysis is based on the experimental data from the CLEO~\cite{Gronberg:1997fj}, BaBar~\cite{Aubert:2009mc}, and Belle~\cite{Uehara:2012ag} collaborations. Among previous lattice studies only in ref.~\cite{Braun:2006dg} an attempt of a continuum limit extrapolation was made. The result of ref.~\cite{DelDebbio:2002mq} corresponds to $\mu=\unit{2.67}{\giga\electronvolt}$.\label{table_compare_P}}%
\begin{widetable}{\textwidth}{llll}%
\toprule
Method & \multicolumn{1}{c}{$\langle\xi^2\rangle_\pi$} & \multicolumn{1}{c}{$a_2^\pi$} & Reference\\
\midrule
LQCD, $N_f=2+1$, SW  & $0.234^{+6}_{-6}(4)(4)(2)$ & $0.101^{+17}_{-17}(12)(10)(5)$ & this article\\
LQCD, $N_f=2$, SW    & $0.2361(41)(39)$           & $0.1364(154)(145)$ & \cite{Braun:2015axa}\\
LQCD, $N_f=2+1$, DWF & $0.28(1)(2)$               & $0.233(29)(58)$ & \cite{Donnellan:2007xr,Arthur:2010xf}\\
LQCD, $N_f=2$, SW    & $0.269(39)$                & $0.201(114)$ & \cite{Braun:2006dg}\\
LQCD, $N_f=0$        & $0.280(49)^{+30}_{-13}$    & $0.233(143)^{+88}_{-38}$ & \cite{DelDebbio:2002mq}\\
\midrule
LO QCDSR  (CZ model)   & $0.334$             & $0.39$ & \cite{Chernyak:1981zz,Chernyak:1983ej}\\
\midrule
QCDSR                  & $0.26^{+5}_{-2}$    & $0.18^{+15}_{-6}$ & \cite{Khodjamirian:2004ga}\\
QCDSR                  & $0.265(21)$         & $0.19(6)$ & \cite{Ball:2006wn}\\
QCDSR, NLC (BMS model) & $0.251^{+18}_{-15}$ & $0.149^{+52}_{-43}$ & \cite{Mikhailov:1991pt,Bakulev:1998pf,Bakulev:2001pa,Mikhailov:2016klg}\\
\midrule
$F_{\pi\gamma\gamma^*}$ (CLEO), LCSR    & $0.245(10)$ & $0.13(3)$ & \cite{Schmedding:1999ap}\\
$F_{\pi\gamma\gamma^*}$ (CLEO), LCSR    & $0.275$     & $0.22$ & \cite{Bakulev:2002uc}\\
$F_{\pi\gamma\gamma^*}$ (CLEO), LCSR, R & $0.27$      & $0.19$ & \cite{Agaev:2005rc}\\
$F_{\pi\gamma\gamma^*}$ (BaBar), LCSR   & $0.233$     & $0.096$ & \cite{Agaev:2010aq}\\
$F_{\pi\gamma\gamma^*}$ (Belle), LCSR   & $0.223$     & $0.067$ & \cite{Agaev:2012tm}\\
\midrule
$F^\text{em}_\pi$, LCSR    & $0.258(34)(17)$ & $0.17(10)(5)$ & \cite{Braun:1999uj,Bijnens:2002mg}\\
$F^\text{em}_\pi$, LCSR, R & $0.248(7)$      & $0.14(2)$ & \cite{Agaev:2005gu}\\
\midrule
$F_{B\to\pi\ell\nu}$, LCSR & $0.245(45)$ & $0.13(13)$ & \cite{Ball:2005tb}\\
$F_{B\to\pi\ell\nu}$, LCSR & $0.238$     & $0.11$ & \cite{Duplancic:2008ix}\\
\midrule
DSE, RL & $0.280$ & $0.233$ & \cite{Chang:2013pq}\\
DSE, DB & $0.251$ & $0.149$ & \cite{Chang:2013pq}\\
\bottomrule
\end{widetable}%
\end{table}%
\begin{table}[tp]%
\caption{The first two Gegenbauer moments of the kaon LCDA at the $\MSbar$~scale $\mu=\unit{2}{\giga\electronvolt}$. The abbreviations have been explained in the caption of table~\ref{table_compare_P}.\label{table_compare_K}}%
\begin{widetable}{\textwidth}{llll}%
\toprule
Method & \multicolumn{1}{c}{$a_1^K$} & \multicolumn{1}{c}{$a_2^K$} & Reference\\
\midrule
LQCD, $N_f=2+1$, SW  & $0.0533^{+18}_{-19}(6)(22)(18)$ & $0.090^{+10}_{-12}(11)(11)(4)$ & this article\\
LQCD, $N_f=2+1$, DWF & $0.0600(17)(33)$                & $0.175(29)(58)$ & \cite{Donnellan:2007xr,Arthur:2010xf}\\
LQCD, $N_f=2$, SW    & $0.0453(9)(28)$                 & $0.175(18)(47)$ & \cite{Braun:2006dg}\\
\midrule
QCDSR & $0.04(2)$ & $0.18^{+15}_{-6}$ & \cite{Khodjamirian:2004ga}\\
QCDSR & $0.05(2)$ & $0.17(10)$ & \cite{Ball:2005vx, Ball:2006fz, Ball:2006wn}\\
QCDSR & $0.08(4)$ & --- & \cite{Chetyrkin:2007vm}\\
\midrule
DSE, RL & $0.183$ & $0.117$ & \cite{Shi:2014uwa}\\
DSE, DB & $0.067$ & $0.088$ & \cite{Shi:2014uwa}\\
\bottomrule
\end{widetable}%
\end{table}%
In table~\ref{table_compare_P} we compare our result for the second moment of the pion LCDA to values from the literature. Our number is compatible with the previous result~\cite{Braun:2015axa} obtained several years ago with~\mbox{$N_f=2$} clover fermions.\footnote{The result of ref.~\cite{Braun:2015axa} does not correspond to the continuum limit but to an average of data within a window of lattice spacings $a\approx\unit{0.06\text{--}0.08}{\femto\meter}$. Moreover, in this reference the values of $a_2$ and $\langle \xi^2\rangle$ are related via eq.~\eqref{eq_a2_otherlat}, where $\langle \mathbb{1}^2 \rangle\neq 1$ for $a>0$. Directly comparing our results to those of ref.~\cite{Braun:2015axa} at a finite lattice spacing may be misleading as in that simulation a different number of sea quarks and a different gluonic action were used.} The quality of the present data is much higher, enabling a controlled continuum extrapolation with quantifiable errors. Our result for $a_2^\pi$ is smaller by a factor of four in comparison to the original CZ calculation~\cite{Chernyak:1981zz,Chernyak:1983ej} evolved to the $\unit{2}{\giga\electronvolt}$ scale, but the difference to more recent QCD sum rule calculations is much smaller and in particular the sum rules involving nonlocal vacuum condensates~\cite{Mikhailov:1991pt,Bakulev:1998pf,Bakulev:2001pa,Mikhailov:2016klg} yield an estimate that is consistent with our results within the quoted error bar. The entries in table~\ref{table_compare_P} marked ``LCSR'' are obtained from experimental data in the factorization framework using LCSR-corrected coefficient functions to take into account the contributions of ``soft'' regions. It is interesting that new data from the BaBar~\cite{Aubert:2009mc} and Belle collaborations~\cite{Uehara:2012ag} generally support small values of the second moment, compatible with our result. Methods based on Dyson--Schwinger equations (DSE)~\cite{Chang:2013pq} suggest somewhat larger values.\par%
A similar comparison for the first two moments of the kaon is presented in table~\ref{table_compare_K}. Our result for the first moment is consistent with earlier lattice calculations as well as with results from QCD sum rules and is somewhat smaller compared to the DSE calculation in ref.~\cite{Shi:2014uwa}. Regarding the second moment of the kaon LCDA, our number is lower than ``old'' lattice estimates~\cite{Braun:2006dg,Donnellan:2007xr,Arthur:2010xf} but agrees remarkably well with the DSE prediction~\cite{Shi:2014uwa} based on the so-called DCSB-improved version of the truncation.\par%
As far as future calculations of the second moment of the pion and kaon LCDAs are concerned, the accuracy can be improved by increasing the statistics in particular for the ensembles at small lattice spacings and quark masses but also by adding additional simulation points. Also a three-loop calculation of the perturbative matching to the $\RI$ scheme is required to improve the overall accuracy.\par%
Regarding phenomenological applications, the first inverse moment%
\begin{align}%
 \frac13 \langle(1-x)^{-1}\rangle_{\!M} = \frac13 \int_0^1 \!\frac{\mathrm{d}x}{1-x}\, \phi_{M}(x) = 1 + a_1^M + a_2^M + a_3^M + \ldots\,,
\end{align}%
which is equal to the sum of all Gegenbauer coefficients, is of particular importance since this quantity enters at leading order in factorization theorems (see, e.g., ref.~\cite{Chernyak:1983ej}). Unfortunately, there is no known way to evaluate it directly on the lattice. As an illustration, we compare two phenomenologically acceptable models of the pion LCDA. The first model is the expansion in Gegenbauer polynomials truncated after~$n=2$ and the second model is based on a simple power-law parametrization:%
\begin{subequations}%
\begin{align}%
\phi^{(\mathrm{I})}(x) &= 6x(1-x)\bigl(1+a_1^{}C_1^{\smash[b]{3/2}}(\xi)+a_2^{}C_2^{\smash[b]{3/2}}(\xi)\bigr)\,, \label{eq_model_I} \\
\phi^{(\mathrm{II})}(x) &= \frac{\Gamma(2+\alpha^++\alpha^-)}{\Gamma(1+\alpha^+)\Gamma(1+\alpha^-)} \,x^{\alpha^+}\!(1-x)^{\alpha^-}\,.\label{eq_model_II}
\end{align}%
\end{subequations}%
Both formulae have two parameters, where for the pion, of course, $a_1^\pi=0$ and~\mbox{$\alpha^-_\pi=\alpha^+_\pi$}. We fix them such that our calculated values for~$a_1^M$ and~$a_2^M$ (in the SMOM scheme at two-loop order, cf.\ table~\ref{table_results}) are exactly reproduced also in the second model. Hence, both models have by construction the same value for the first two Gegenbauer coefficients, but differ in higher-order coefficients.\par%
\begin{figure}[t]%
\centering%
\includegraphics[width=0.485\textwidth]{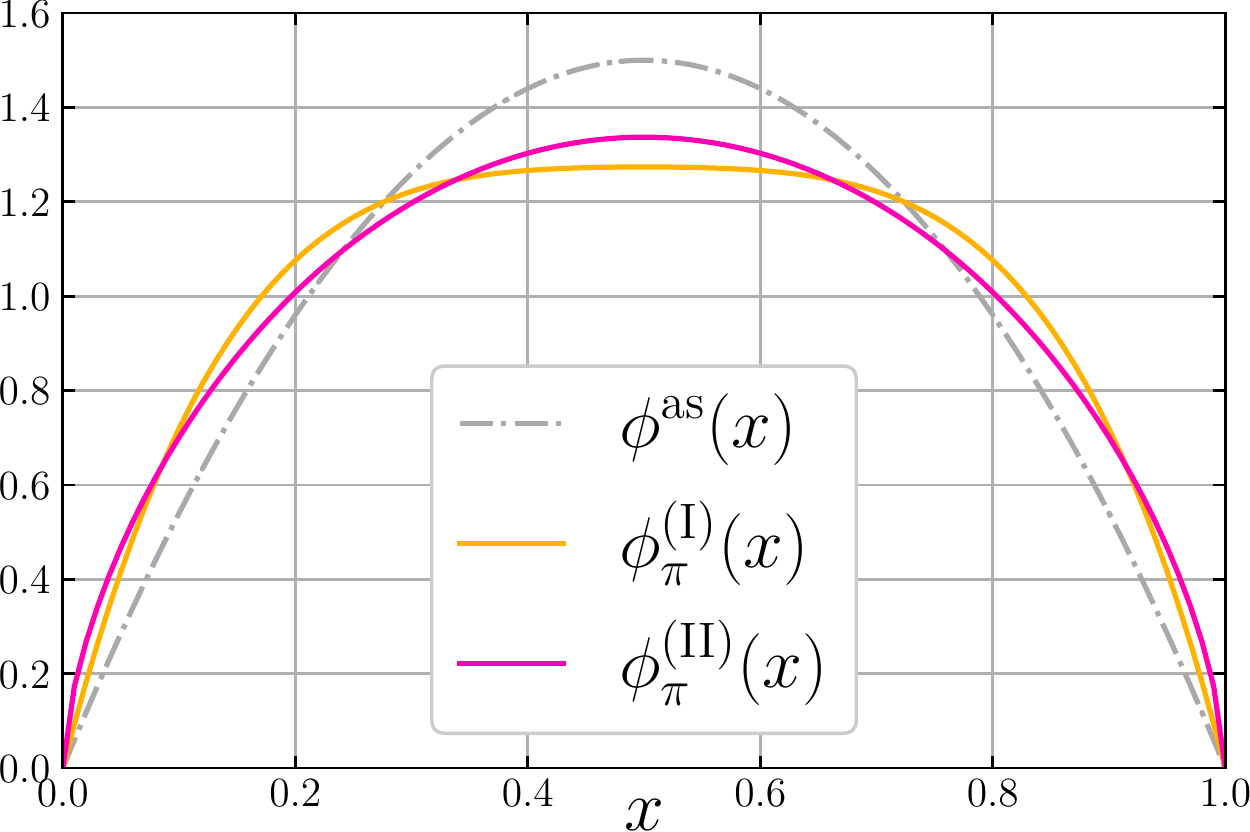}\hfill%
\includegraphics[width=0.485\textwidth]{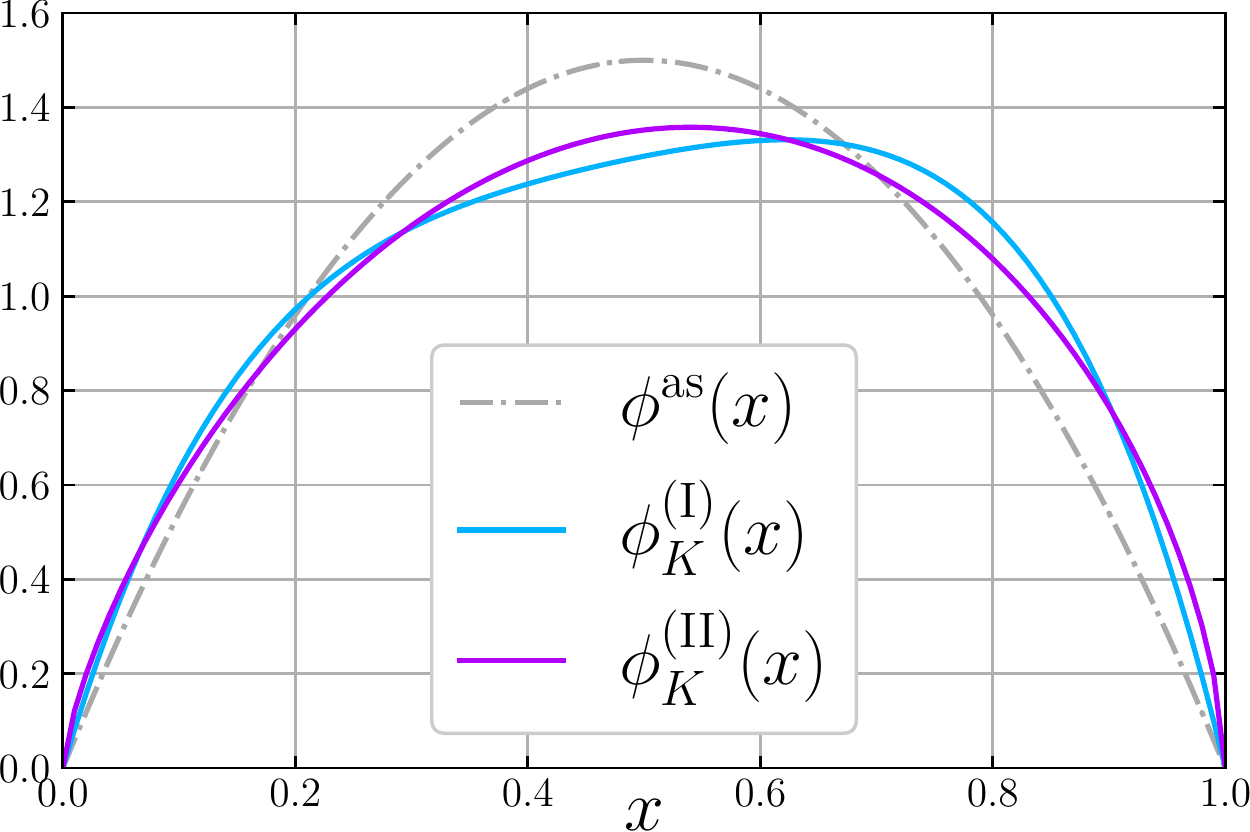}%
\caption{The truncated Gegenbauer expansion~\eqref{eq_model_I} and the power-law parametrization~\eqref{eq_model_II} at $\mu=\unit{2}{\giga\electronvolt}$ obtained using our results for~$a_1^M$ and~$a_2^M$. The left panel shows the resulting DAs for the pion, which are symmetric under $x\leftrightarrow1-x$, while the results for the kaon DA on the right panel are slightly skewed towards the strange quark due to flavor symmetry breaking. In all cases the deviation from the asymptotic shape is significant.\label{fig_DAs_PK}}%
\end{figure}%
The results are shown in figure~\ref{fig_DAs_PK}. Both models are somewhat ``flatter'' in comparison to the asymptotic LCDA shown by the gray curve, and in general do not seem to differ very much. The model dependence of the first inverse moment is, however, sizable. We obtain for the pion%
\begin{subequations}%
\begin{align}%
\langle(1-x)^{-1}\rangle_\pi^{(\mathrm{I})} &= 3.30^{+7}_{-7}\,, &
\langle(1-x)^{-1}\rangle_\pi^{(\mathrm{II})} &= 3.58^{+20}_{-17}\,,\subtag{2}
\end{align}%
\end{subequations}%
where the errors have been obtained by adding the individual errors of table~\ref{table_results} in quadrature. Both numbers are phenomenologically viable, in particular the second one is very close to $\langle(1-x)^{-1}\rangle_\pi=3.6$ (at $\mu=\unit{2}{\giga\electronvolt}$) from the model of ref.~\cite{Agaev:2012tm}, which provides a good description of the Belle data~\cite{Uehara:2012ag} for the $\pi\gamma\gamma^*$ form factor.\par%
The QCD description of form factors based on our models~(I) and~(II) will differ by as much as~$10\%$. The necessity to go beyond the second Gegenbauer moment is thus obvious. A brute-force extension of the present approach to operators with a larger number of derivatives does not seem to be viable even if the problem of the mixing with lower-dimensional operators is solved. Consider in particular the fourth moment, $\langle\xi^4\rangle_{\!M}=3/35+(8/35)a_2^M+(8/77)a_4^M$, for which we obtain in the two models%
\begin{subequations}%
\begin{align}%
\langle\xi^4\rangle^{(\mathrm{I})}_\pi &= 0.109^{+5}_{-5}\,, &
\langle\xi^4\rangle^{(\mathrm{II})}_\pi &= 0.112^{+7}_{-6}\,.\subtag{2}
\end{align}%
\end{subequations}%
One sees that even if both~$\langle\xi^2\rangle_\pi$ and~$\langle\xi^4\rangle_\pi$ were measured with $1\%$~precision on the lattice (which is already optimistic given our statistical error of $\sim 2.5\%$ on~$\langle\xi^2\rangle_\pi$), the value of~$a_4^\pi$ cannot be extracted reliably as it is overshadowed by the uncertainty in~$a_2^\pi$. Therefore, alternative methods should also be investigated.\par%
In the past few years exploratory studies appeared aiming at the extraction of the pion LCDA from lattice calculations of suitable Euclidean correlation functions in position space~\cite{Braun:2007wv,Zhang:2017bzy,Bali:2017gfr,Bali:2018spj}, see also refs.~\cite{Radyushkin:2017gjd,Chen:2017gck,Detmold:2018kwu}. After taking the continuum and other appropriate limits, these can be expressed in terms of LCDAs in the framework of QCD factorization within the continuum theory, in analogy to the extraction of parton distributions from fits to experimentally measured structure functions. In other words, the role of lattice QCD is in this case to provide a complementary set of observables from which the LCDAs can be extracted. In particular, in ref.~\cite{Bali:2018spj} it has been demonstrated that using the approach of ref.~\cite{Braun:2007wv}, the contributions of different Gegenbauer moments can be separated, at least in principle, by considering the correlation functions at large ``Ioffe times''. These new techniques generally require hadron sources with very large momentum combined with good statistical accuracy and very fine lattices to control the corresponding discretization errors. Whether these position space methods or the moment method employed here will be more useful to constrain higher moments of LCDAs is at present unclear.\par%
\section{Addendum: Three-loop matching\label{sec_addendum}}
We have rerun our analysis (see~\cite{Bali:2020addendum}) using the newly available three-loop matching (N${}^3$LO) for the conversion from the $\RI$ to the $\MSbar$ scheme~\cite{Kniehl:2020sgo,Kniehl:2020nhw}. The results can be taken from table~\ref{table_results_inclSMOM3}, which is an updated version of table~\ref{table_results}. As one can see, the errors of the renormalization procedure are reduced considerably. The new N${}^3$LO results for $a_2^\pi$ and $a_2^K$ are slightly larger, but are still consistent with the previous
NNLO result within errors. Considering the first moment of the kaon distribution amplitude, where we can compare results using $\RI$ or $\RInoS$ as intermediate schemes, it is encouraging to see that the final results in the $\MSbar$ scheme agree perfectly, if one uses three-loop matching in both cases.\par%
\def\myrule{\rule[-2pt]{0sp}{12pt}}%
\begin{table}[t]%
\centering%
\caption{Updated version of table~\ref{table_results} including N${}^3$LO matching. Continuum limit extrapolated values for the first two moments of the octet mesons. The results have been converted to the $\MSbar$ scheme at $\mu=\unit{2}{\giga\electronvolt}$ using intermediate $\text{RI}^\prime$ schemes and different loop orders in the perturbative matching. The statistical error given as sub- and superscript reflects the errors of the data after extrapolation. The numbers in parentheses give estimates of the systematic uncertainties due to the nonperturbative renormalization~($r$)
as described in section~\ref{sec_renorm}, the continuum extrapolation~($a$), and the chiral extrapolation~($m$). As discussed in section~\ref{sec_lattices}, finite volume effects are negligible in our setting.\label{table_results_inclSMOM3}}%
\begin{widetable}{\textwidth}{crrE{.}{.}{-}{1.10}{{}^{+0}_{-0}()_r()_a()_m}E{.}{.}{-}{1.10}{{}^{+00}_{-00}()_r()_a()_m}}%
\toprule
\myrule$M$ & \multicolumn{1}{c}{$\text{RI}^\prime$} & \multicolumn{1}{c}{order} & \multicolumn{1}{c}{$\langle\xi^2\rangle_{\!M}$} & \multicolumn{1}{c}{$a_2^M$}\\
\midrule
\myrule$\pi$   & SMOM & N${}^3$LO & 0.240^{+6}_{-6}(2)_r(3)_a(2)_m & 0.116^{+16}_{-17}(4)_r(9)_a(5)_m\\
\myrule$\pi$   & SMOM &      NNLO & 0.234^{+6}_{-6}(4)_r(4)_a(2)_m & 0.101^{+17}_{-17}(12)_r(10)_a(5)_m\\
\myrule$\pi$   & SMOM &       NLO & 0.227^{+6}_{-6}(5)_r(5)_a(2)_m & 0.078^{+18}_{-19}(16)_r(13)_a(5)_m\\
\myrule$K$     & SMOM & N${}^3$LO & 0.236^{+3}_{-4}(1)_r(3)_a(1)_m & 0.106^{+10}_{-12}(4)_r(9)_a(4)_m\\
\myrule$K$     & SMOM &      NNLO & 0.231^{+4}_{-4}(4)_r(4)_a(1)_m & 0.090^{+10}_{-12}(11)_r(11)_a(4)_m\\
\myrule$K$     & SMOM &       NLO & 0.223^{+4}_{-5}(5)_r(5)_a(2)_m & 0.067^{+11}_{-13}(16)_r(14)_a(5)_m\\
\myrule$\etaa$ & SMOM & N${}^3$LO & 0.235^{+3}_{-4}(1)_r(3)_a(1)_m & 0.103^{+10}_{-13}(4)_r(9)_a(4)_m\\
\myrule$\etaa$ & SMOM &      NNLO & 0.230^{+4}_{-4}(4)_r(4)_a(1)_m & 0.087^{+10}_{-13}(11)_r(11)_a(4)_m\\
\myrule$\etaa$ & SMOM &       NLO & 0.222^{+4}_{-5}(6)_r(5)_a(2)_m & 0.063^{+11}_{-14}(16)_r(14)_a(5)_m\\
\midrule
\myrule$M$ & \multicolumn{1}{c}{$\text{RI}^\prime$} & \multicolumn{1}{c}{order} & \multicolumn{1}{c}{$\langle\xi^1\rangle_{\!M}$} & \multicolumn{1}{c}{$a_1^M$}\\
\midrule
\myrule$K$     & SMOM & N${}^3$LO & 0.0315^{+10}_{-11}(2)_r(12)_a(10)_m & 0.0525^{+17}_{-19}(3)_r(20)_a(17)_m\\
\myrule$K$     & SMOM &      NNLO & 0.0320^{+11}_{-12}(3)_r(13)_a(11)_m & 0.0533^{+18}_{-19}(6)_r(22)_a(18)_m\\
\myrule$K$     & SMOM &       NLO & 0.0327^{+11}_{-12}(6)_r(14)_a(11)_m & 0.0545^{+18}_{-20}(9)_r(23)_a(18)_m\\
\myrule$K$     &  MOM & N${}^3$LO & 0.0315^{+11}_{-11}(1)_r(11)_a(10)_m & 0.0525^{+18}_{-19}(2)_r(19)_a(17)_m\\
\myrule$K$     &  MOM &      NNLO & 0.0319^{+11}_{-12}(1)_r(11)_a(10)_m & 0.0531^{+18}_{-19}(2)_r(18)_a(17)_m\\
\bottomrule
\end{widetable}%
\end{table}%
Adding all errors in quadrature we obtain with three-loop matching%
\begin{align*}
  a_2^\pi &= 0.116^{+19}_{-20} \,, \\
  a_1^K &= 0.0525^{+31}_{-33} \,, \\
  a_2^K &= 0.106^{+15}_{-16} \,.
\end{align*}
We include updated figures using the new, slightly shifted values in appendix~\ref{app_updated_figures}.\par%
\FloatBarrier
\acknowledgments
We thank Sara Collins, Stefano Piemonte, Jakob Simeth, and Wolfgang S\"oldner for discussions, Benjamin Gl{\"a}{\ss}le, Daniel Richtmann, and Stefan Solbrig for technical support, and all our other CLS colleagues for the joint generation of gauge ensembles, the planning of the simulations, and the exchange of scientific ideas.
\par
This work was supported by Deutsche Forschungsgemeinschaft SFB/TRR\nobreakdash-55 and by the Polish NCN (grant no.\ UMO-2016/21/B/ST2/01492). The authors gratefully acknowledge the computing time granted by the John von Neumann Institute for Computing (NIC) and provided on the Booster partition of the supercomputer JURECA~\cite{jureca} at J\"ulich Supercomputing Centre (JSC, \url{http://www.fz-juelich.de/ias/jsc/}). The authors also gratefully acknowledge the Interdisciplinary Centre for Mathematical and Computational Modelling~(ICM) of the University of Warsaw for computer time on Okeanos (grant nos.\ GA67\nobreakdash-12, GA69\nobreakdash-20, GA71\nobreakdash-26), the PLGRID consortium for a computer time allocation on the Prometheus machine hosted by Cyfronet Krakow (grants hadronspectrum, nspt, \mbox{pionda}), PRACE (Partnership for Advanced Computing in Europe, \url{http://www.prace-ri.eu}) for awarding us access to the Marconi-KNL machine hosted by CINECA at Bologna, Italy, and the Leibniz Supercomputer Centre (LRZ, \url{https://www.lrz.de}) in Garching for access to the coolMUC3 cluster. Additional computations have been carried out on the Regensburg QPACE~2 computer~\cite{Arts:2015jia} and the QPACE~3 machine of SFB/TRR\nobreakdash-55. Some of the $m_\ell=m_s$ gauge ensembles used were generated by members of the Mainz group on the Wilson and Clover HPC Clusters of IKP Mainz. We acknowledge the Gauss Centre for Supercomputing (GCS) for providing computing time for GCS large-scale projects on the GCS share of the supercomputers SuperMUC at LRZ and JUQUEEN at JSC, where many of the ensembles used here were generated. GCS is the alliance of the three national supercomputing centers HLRS (Uni\-ver\-si\-t\"at Stutt\-gart), JSC (For\-schungs\-zen\-trum J\"u\-lich) and LRZ (Bay\-e\-ri\-sche Aka\-de\-mie der Wis\-sen\-schaf\-ten), funded by the German Federal Ministry of Education and Research (BMBF) and the German State Ministries for Research of Ba\-den-W{\"u}rt\-tem\-berg (MWK), Bay\-ern (StMWFK) and Nord\-rhein-West\-fa\-len~(MIWF).
\par
We used a modified version of the {\sc Chroma}~\cite{Edwards:2004sx} software package, along with the {\sc Lib\-Hadron\-Analysis} library and, depending on the target machine, either the multigrid DD-$\alpha$AMG solver~\cite{Frommer:2013fsa} implementation of refs.~\cite{Heybrock:2015kpy,Georg:2017zua} or the domain decomposition solver of {\sc openQCD}~\cite{Luscher:2012av} (\url{https://luscher.web.cern.ch/luscher/openQCD/}). Most gauge ensembles have been generated by CLS (\url{https://wiki-zeuthen.desy.de/CLS/}) using {\sc openQCD}. A few additional ensembles have been generated by RQCD on QPACE, employing the BQCD~code~\cite{Nakamura:2010qh}.\par%
\clearpage
\FloatBarrier
\appendix%
\section{Lattice ensembles and supplementary figures\label{app_ens}\label{app_plots}}%
Below we list the properties of the analyzed lattice ensembles for the three quark mass trajectories: $\operatorname{Tr}\mathcal{M}=\text{phys.}$ in table~\ref{table_TrM}, $m_s=\mathrm{phys.}$ in table~\ref{table_msc}, and $m_\ell=m_s$ in table~\ref{table_sym}. The latter also contains the ensembles that have been used solely for the determination of renormalization factors.\par
\begin{figure}[p]%
\centering%
\includegraphics[width=\textwidth]{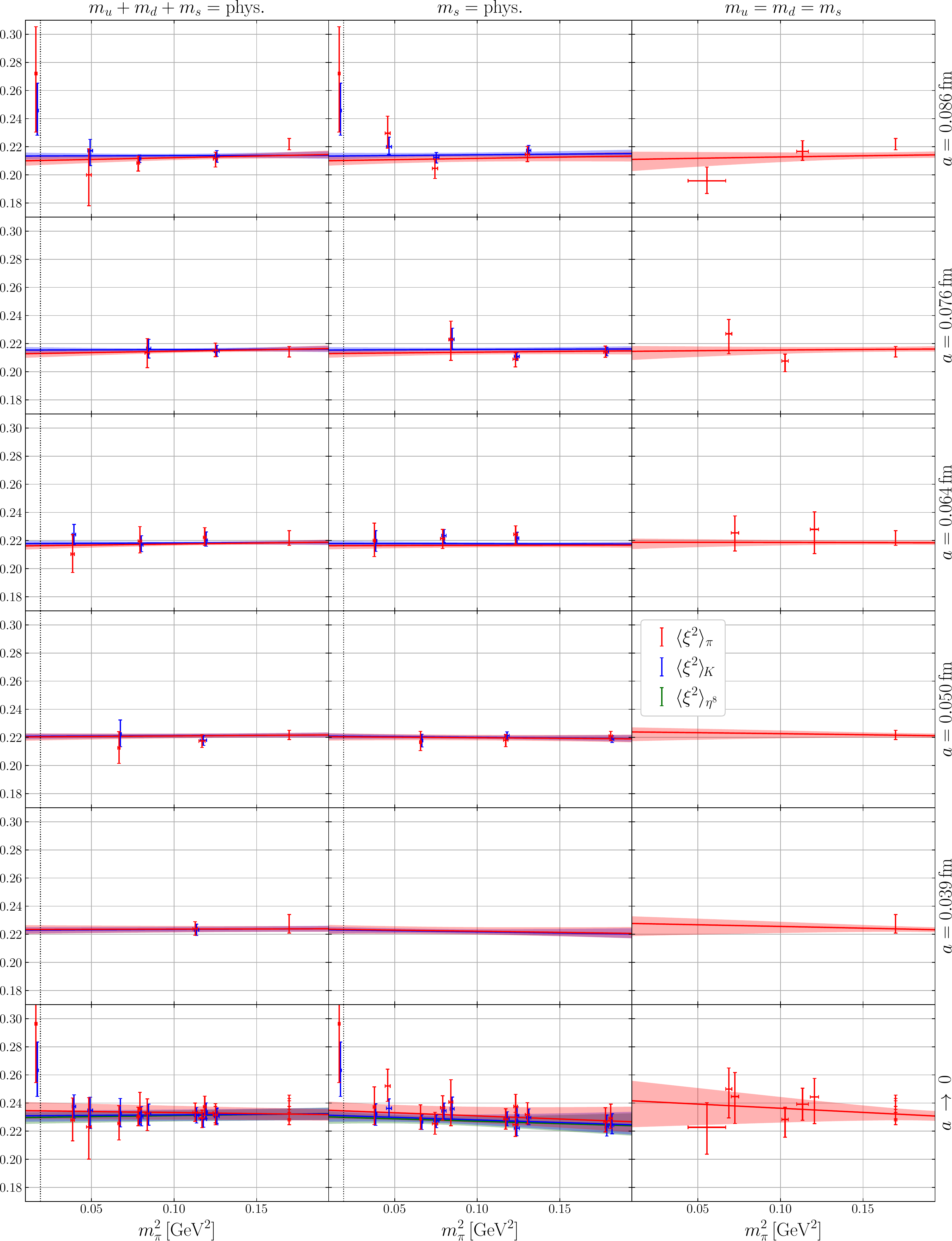}%
\caption{The pion mass dependence of the moments~$\langle\xi^2\rangle_{\!\M}$, defined in eq.~\subeqref{eq_renormalized_moments2}{a}, plotted (top to bottom) for all lattice spacings as well as in the continuum limit (where, for illustrative purposes, all points have been translated along the fitted function). The columns correspond to the lines of physical average quark mass~(left), physical strange quark mass~(middle), and symmetric quark masses~(right), cf.\ figure~\ref{fig_trajectories}. The dotted gray lines mark the physical meson masses.\label{fig_xi2}}%
\end{figure}

\begin{figure}[p]%
\centering%
\includegraphics[width=\textwidth]{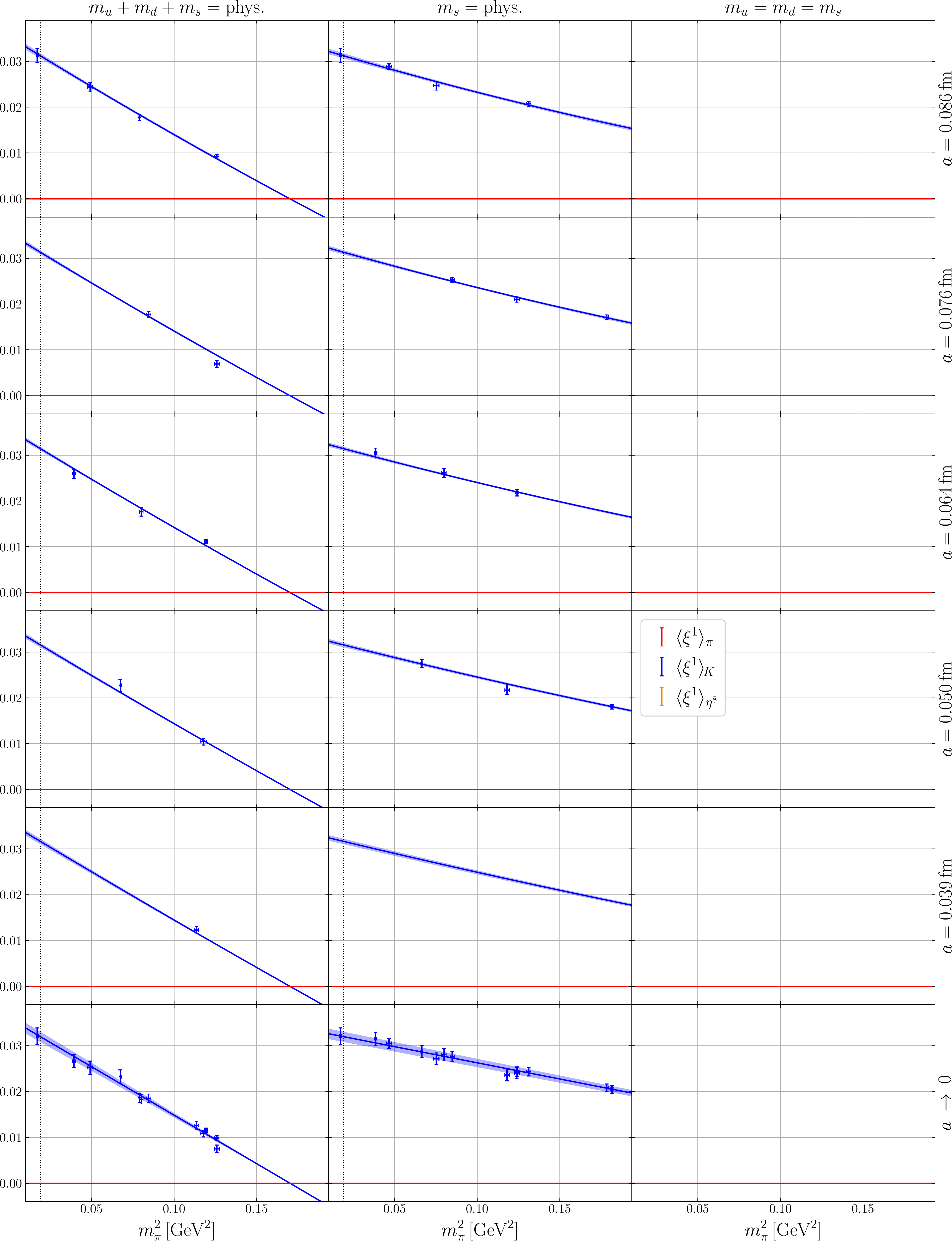}%
\caption{The pion mass dependence of the moments~$\langle\xi^1\rangle_{\!\M}$, defined in eq.~\subeqref{eq_renormalized_moments1}{a}, plotted for all lattice spacings as well as in the continuum limit (where, for illustrative purposes, all points have been translated along the fitted function). The columns correspond to the three quark mass trajectories, cf.\ figure~\ref{fig_trajectories}. The dotted gray lines mark the physical meson masses. Due to symmetry, this moment vanishes exactly for the $\pi$~and~$\etaa$ mesons as well as on the symmetric line.\label{fig_xi1}}%
\end{figure}

%
We also show the results of the global fit for the second moments~$\langle\xi^2\rangle_{\!M}$ in figure~\ref{fig_xi2} and for the first moments~$\langle\xi^1\rangle_{\!M}$ in figure~\ref{fig_xi1}. These are exactly the same fits that have been used to produce the more concise figures~\ref{fig_m_xi2},~\ref{fig_a_xi2},~\ref{fig_am_xi1}, and~\ref{fig_summary}. In contrast to the figures of the main text we resolve the dependence on all relevant variables simultaneously, i.e., we display the full mass dependence along the three individual trajectories for each of the five lattice spacings as well as in the continuum limit.\par%
\def\shrink{-.035\normalbaselineskip}%
\begin{table}[h]%
\centering%
\caption{List of the ensembles on the $\operatorname{Tr}\mathcal{M}=\text{phys.}$ trajectory. The inverse gauge coupling~$\beta$ determines the lattice spacing (cf.\ table~\ref{tab_spacings}), while the spatial and temporal extents fix the lattice geometry~$N_s^3\times N_t$. Boundary conditions in time direction are either periodic~(p) or open~(o). The light and strange hopping parameters, $\kappa_\ell$~and~$\kappa_s$, determine the corresponding quark masses; the resulting approximate meson masses~$m_\pi$ and~$m_K$ are given in units of~$\mega\electronvolt$, followed by the spatial lattice size in pion mass units. Finally, we give the number of gauge configurations used to measure the second moments.\label{table_TrM}}%
\begin{widetable}{\textwidth}{rccrcllcccr}%
\toprule
\multicolumn{1}{c}{Ens.} & $\beta$ & $N_s$ & \multicolumn{1}{c}{$N_t$} & bc & \multicolumn{1}{c}{$\kappa_\ell$} & \multicolumn{1}{c}{$\kappa_s$} & $m_\pi$ & $m_K$ & $m_\pi L$ & \multicolumn{1}{c}{conf.}\\
\midrule
D150    & $3.40$ & $64$ & $128$ & p & $0.137088         $ & $0.13610755       $ & $130$ & $481$ & $3.6$ & $ 566$\\[\shrink]
C101    & $3.40$ & $48$ & $ 96$ & o & $0.13703          $ & $0.136222041      $ & $221$ & $472$ & $4.6$ & $1547$\\[\shrink]
H105    & $3.40$ & $32$ & $ 96$ & o & $0.13697          $ & $0.13634079       $ & $281$ & $466$ & $3.9$ & $2022$\\[\shrink]
H102    & $3.40$ & $32$ & $ 96$ & o & $0.136865         $ & $0.136549339      $ & $354$ & $441$ & $4.9$ & $1997$\\[\shrink]
H101    & $3.40$ & $32$ & $ 96$ & o & $0.13675962       $ & $0.13675962       $ & $420$ & $420$ & $5.8$ & $2000$\\[\shrink]
\midrule
N401    & $3.46$ & $48$ & $128$ & o & $0.1370616        $ & $0.1365480771     $ & $290$ & $467$ & $5.4$ & $1088$\\[\shrink]
S400    & $3.46$ & $32$ & $128$ & o & $0.136984         $ & $0.136702387      $ & $354$ & $445$ & $4.4$ & $1740$\\[\shrink]
B450    & $3.46$ & $32$ & $ 64$ & p & $0.13689          $ & $0.13689          $ & $419$ & $419$ & $5.2$ & $1612$\\[\shrink]
\midrule
D200    & $3.55$ & $64$ & $128$ & o & $0.1372           $ & $0.136601748      $ & $197$ & $484$ & $4.1$ & $1169$\\[\shrink]
N200    & $3.55$ & $48$ & $128$ & o & $0.13714          $ & $0.13672086       $ & $282$ & $463$ & $4.4$ & $1409$\\[\shrink]
N203    & $3.55$ & $48$ & $128$ & o & $0.13708          $ & $0.136840284      $ & $345$ & $442$ & $5.4$ & $1496$\\[\shrink]
N202    & $3.55$ & $48$ & $128$ & o & $0.137            $ & $0.137            $ & $412$ & $412$ & $6.4$ & $ 881$\\[\shrink]
\midrule
J303    & $3.70$ & $64$ & $192$ & o & $0.137123         $ & $0.1367546608     $ & $259$ & $474$ & $4.2$ & $ 657$\\[\shrink]
N302    & $3.70$ & $48$ & $128$ & o & $0.137064         $ & $0.1368721791358  $ & $343$ & $450$ & $4.1$ & $1383$\\[\shrink]
N300    & $3.70$ & $48$ & $128$ & o & $0.137            $ & $0.137            $ & $421$ & $421$ & $5.1$ & $2027$\\[\shrink]
\midrule
J501    & $3.85$ & $64$ & $192$ & o & $0.1369032        $ & $0.136749715      $ & $336$ & $450$ & $4.3$ & $1532$\\[\shrink]
J500    & $3.85$ & $64$ & $192$ & o & $0.136852         $ & $0.136852         $ & $410$ & $410$ & $5.2$ & $ 843$\\[\shrink]
\bottomrule
\end{widetable}%
\end{table}%
\clearpage
\begin{table}[h]%
\centering%
\caption{The same as table~\ref{table_TrM}, but for the $m_s=\mathrm{phys.}$ trajectory.\label{table_msc}}%
\begin{widetable}{\textwidth}{rccrcllcccr}%
\toprule
\multicolumn{1}{c}{Ens.} & $\beta$ & $N_s$ & \multicolumn{1}{c}{$N_t$} & bc & \multicolumn{1}{c}{$\kappa_\ell$} & \multicolumn{1}{c}{$\kappa_s$} & $m_\pi$ & $m_K$ & $m_\pi L$ & \multicolumn{1}{c}{conf.}\\
\midrule
D150    & $3.40$ & $64$ & $128$ & p & $0.137088         $ & $0.13610755       $ & $130$ & $481$ & $3.6$ & $ 566$\\[\shrink]
C102    & $3.40$ & $48$ & $ 96$ & o & $0.13705084580022 $ & $0.13612906255557 $ & $215$ & $501$ & $4.5$ & $1500$\\[\shrink]
H106    & $3.40$ & $32$ & $ 96$ & o & $0.137015570024   $ & $0.136148704478   $ & $273$ & $517$ & $3.8$ & $1468$\\[\shrink]
H107    & $3.40$ & $32$ & $ 96$ & o & $0.13694566590798 $ & $0.136203165143476$ & $362$ & $546$ & $5.0$ & $1481$\\[\shrink]
\midrule
N450    & $3.46$ & $48$ & $128$ & p & $0.1370986        $ & $0.136352601      $ & $291$ & $531$ & $5.4$ & $1132$\\[\shrink]
B452    & $3.46$ & $32$ & $ 64$ & p & $0.1370455        $ & $0.136378044      $ & $351$ & $547$ & $4.3$ & $1944$\\[\shrink]
B451    & $3.46$ & $32$ & $ 64$ & p & $0.1369814        $ & $0.136408545      $ & $422$ & $575$ & $5.2$ & $2000$\\[\shrink]
\midrule
D201    & $3.55$ & $64$ & $128$ & o & $0.1372067        $ & $0.136546844      $ & $195$ & $501$ & $4.1$ & $1078$\\[\shrink]
N201    & $3.55$ & $48$ & $128$ & o & $0.13715968       $ & $0.136561319      $ & $282$ & $524$ & $4.4$ & $1070$\\[\shrink]
N204    & $3.55$ & $48$ & $128$ & o & $0.137112         $ & $0.136575049      $ & $352$ & $546$ & $5.5$ & $1500$\\[\shrink]
\midrule
J304    & $3.70$ & $64$ & $192$ & o & $0.13713          $ & $0.1366569203     $ & $257$ & $522$ & $4.1$ & $1408$\\[\shrink]
N304    & $3.70$ & $48$ & $128$ & o & $0.137079325093654$ & $0.136665430105663$ & $343$ & $551$ & $4.1$ & $1482$\\[\shrink]
N305    & $3.70$ & $48$ & $128$ & o & $0.137025         $ & $0.136676119      $ & $426$ & $583$ & $5.1$ & $2001$\\[\shrink]
\bottomrule
\end{widetable}%
\end{table}%
\begin{table}[h]%
\newcommand*{\ted}{\text{---}}%
\centering%
\caption{The same as table~\ref{table_TrM}, but for the $m_\ell=m_s$ trajectory. Renormalization factors are determined from the lattices marked by an asterisk. The number of configurations refers to those used for the measurement of the second moments, i.e., ensembles with $\ted^*$ are only used for renormalization.\label{table_sym}}%
\begin{widetable}{\textwidth}{rccrcllcccr}%
\toprule
\multicolumn{1}{c}{Ens.} & $\beta$ & $N_s$ & \multicolumn{1}{c}{$N_t$} & bc & \multicolumn{1}{c}{$\kappa_\ell$} & \multicolumn{1}{c}{$\kappa_s$} & $m_\pi$ & $m_K$ & $m_\pi L$ & \multicolumn{1}{c}{conf.}\\
\midrule
rqcd017 & $3.40$ & $32$ & $ 32$ & p & $0.136865         $ & $0.136865         $ & $236$ & $236$ & $3.3$ & $1799^{\mathrlap{*}}$\\[\shrink]
rqcd021 & $3.40$ & $32$ & $ 32$ & p & $0.136813         $ & $0.136813         $ & $337$ & $337$ & $4.7$ & $1541^{\mathrlap{*}}$\\[\shrink]
H101    & $3.40$ & $32$ & $ 96$ & o & $0.13675962       $ & $0.13675962       $ & $420$ & $420$ & $5.8$ & $2000$\\[\shrink]
rqcd016 & $3.40$ & $32$ & $ 32$ & p & $0.13675962       $ & $0.13675962       $ & $425$ & $425$ & $5.9$ & $\ted^{\mathrlap{*}}$\\[\shrink]
rqcd019 & $3.40$ & $32$ & $ 32$ & p & $0.1366           $ & $0.1366           $ & $611$ & $611$ & $8.5$ & $\ted^{\mathrlap{*}}$\\[\shrink]
\midrule
X450    & $3.46$ & $48$ & $ 64$ & p & $0.136994         $ & $0.136994         $ & $263$ & $263$ & $4.9$ & $ 398^{\mathrlap{*}}$\\[\shrink]
rqcd030 & $3.46$ & $32$ & $ 64$ & p & $0.1369587        $ & $0.1369587        $ & $321$ & $321$ & $4.0$ & $1224^{\mathrlap{*}}$\\[\shrink]
B450    & $3.46$ & $32$ & $ 64$ & p & $0.13689          $ & $0.13689          $ & $419$ & $419$ & $5.2$ & $1612^{\mathrlap{*}}$\\[\shrink]
rqcd029 & $3.46$ & $32$ & $ 64$ & p & $0.1366           $ & $0.1366           $ & $708$ & $708$ & $8.7$ & $\ted^{\mathrlap{*}}$\\[\shrink]
\midrule
X251    & $3.55$ & $48$ & $ 64$ & p & $0.1371           $ & $0.1371           $ & $270$ & $270$ & $4.2$ & $ 432^{\mathrlap{*}}$\\[\shrink]
X250    & $3.55$ & $48$ & $ 64$ & p & $0.13705          $ & $0.13705          $ & $348$ & $348$ & $5.4$ & $ 345^{\mathrlap{*}}$\\[\shrink]
N202    & $3.55$ & $48$ & $128$ & o & $0.137            $ & $0.137            $ & $412$ & $412$ & $6.4$ & $ 881$\\[\shrink]
rqcd025 & $3.55$ & $32$ & $ 64$ & p & $0.137            $ & $0.137            $ & $411$ & $411$ & $4.3$ & $\ted^{\mathrlap{*}}$\\[\shrink]
B250    & $3.55$ & $32$ & $ 64$ & p & $0.1367           $ & $0.1367           $ & $708$ & $708$ & $7.4$ & $\ted^{\mathrlap{*}}$\\[\shrink]
\midrule
N300    & $3.70$ & $48$ & $128$ & o & $0.137            $ & $0.137            $ & $421$ & $421$ & $5.1$ & $2027^{\mathrlap{*}}$\\[\shrink]
N303    & $3.70$ & $48$ & $128$ & o & $0.1368           $ & $0.1368           $ & $641$ & $641$ & $7.8$ & $\ted^{\mathrlap{*}}$\\[\shrink]
\midrule
J500    & $3.85$ & $64$ & $192$ & o & $0.136852         $ & $0.136852         $ & $410$ & $410$ & $5.2$ & $ 843^{\mathrlap{*}}$\\[\shrink]
N500    & $3.85$ & $48$ & $128$ & o & $0.13672514       $ & $0.13672514       $ & $599$ & $599$ & $5.7$ & $\ted^{\mathrlap{*}}$\\[\shrink]
\bottomrule
\end{widetable}%
\end{table}%
\FloatBarrier
\section{Updated figures\label{app_updated_figures}}
For completeness, we include figures~\ref{fig_m_xi2SMOM3}--\ref{fig_DAs_PKSMOM3}, employing the three-loop matching~\cite{Kniehl:2020sgo,Kniehl:2020nhw} between the $\RI$ and $\MSbar$ schemes. In figures~\ref{fig_m_xi2}--\ref{fig_DAs_PK} this was only carried out at the two-loop level. Note that this appendix is only contained in the present arXiv version.
\def\figheight{4.813cm}%
\begin{figure}[h]%
\vspace{1.28\baselineskip}
\centering
\includegraphics[width=\textwidth]{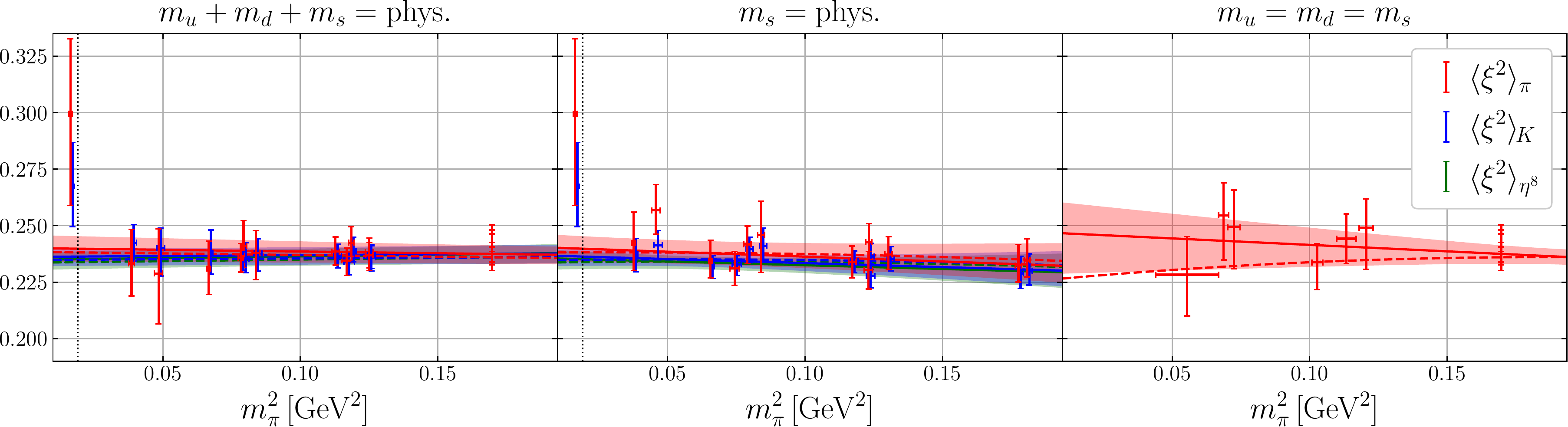}%
\caption{Updated version of figure~\ref{fig_m_xi2}.\label{fig_m_xi2SMOM3}\\[1.28\baselineskip]}%
% \end{figure}%
% %
% %
% \def\figheight{4.813cm}%
% \begin{figure}[tp]%
\centering
\includegraphics[height=\figheight]{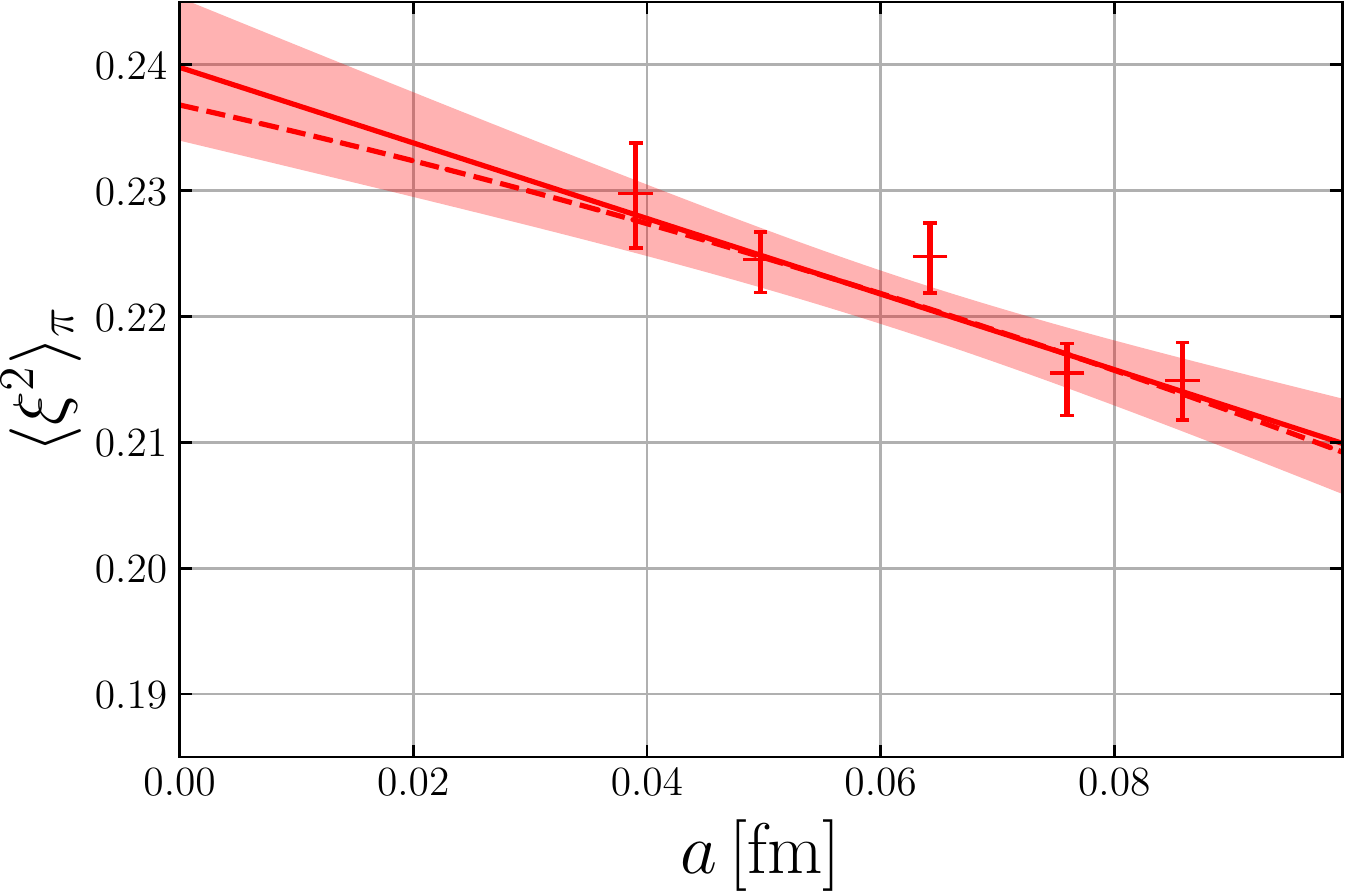}\hfill%
\includegraphics[height=\figheight]{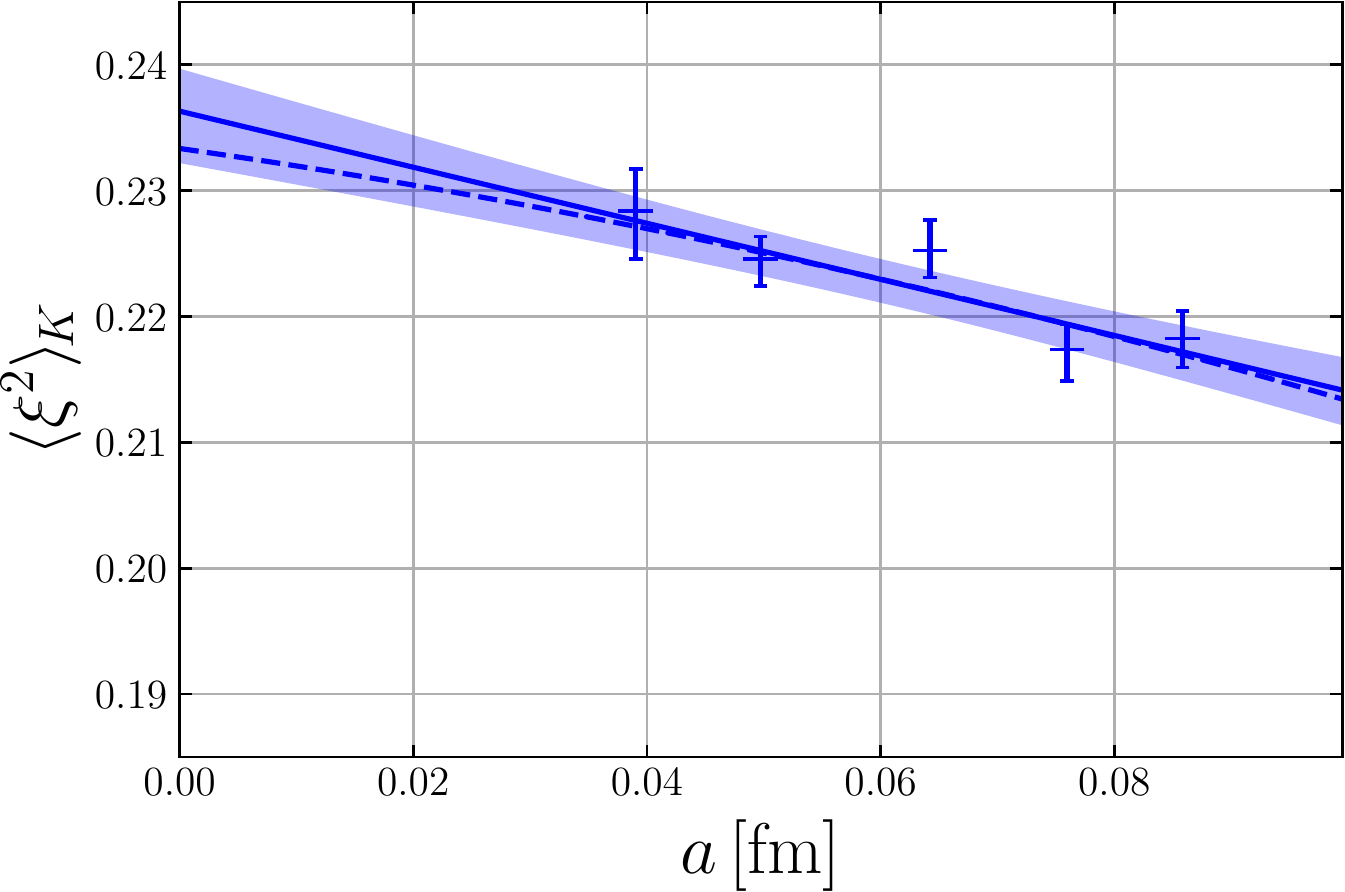}%
\caption{Updated version of figure~\ref{fig_a_xi2}.\label{fig_a_xi2SMOM3}\\[1.28\baselineskip]}%
% \end{figure}%
% %
% %
% \def\figheight{4.813cm}%
% \begin{figure}[tp]%
\centering
\includegraphics[height=\figheight]{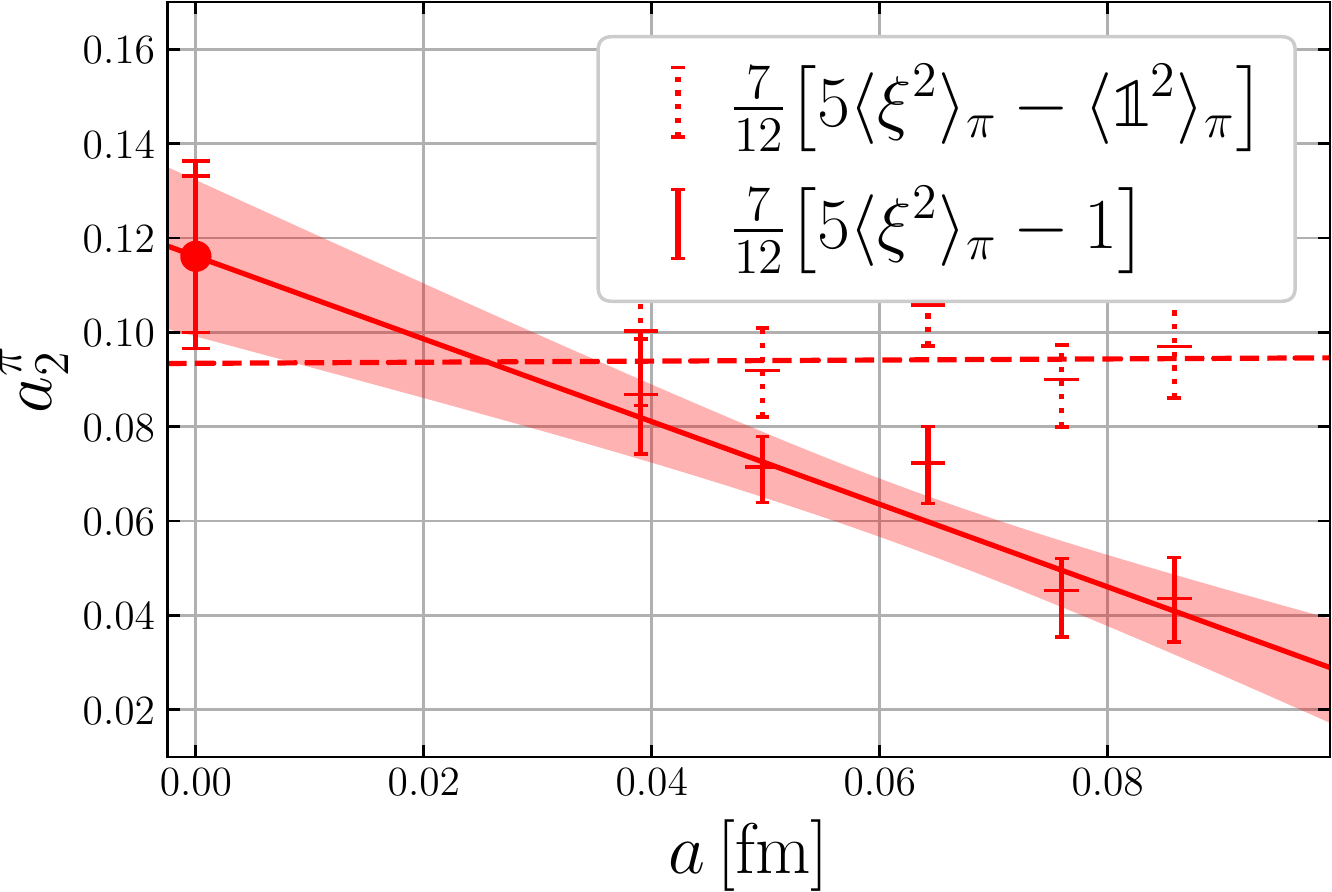}\hfill%
\includegraphics[height=\figheight]{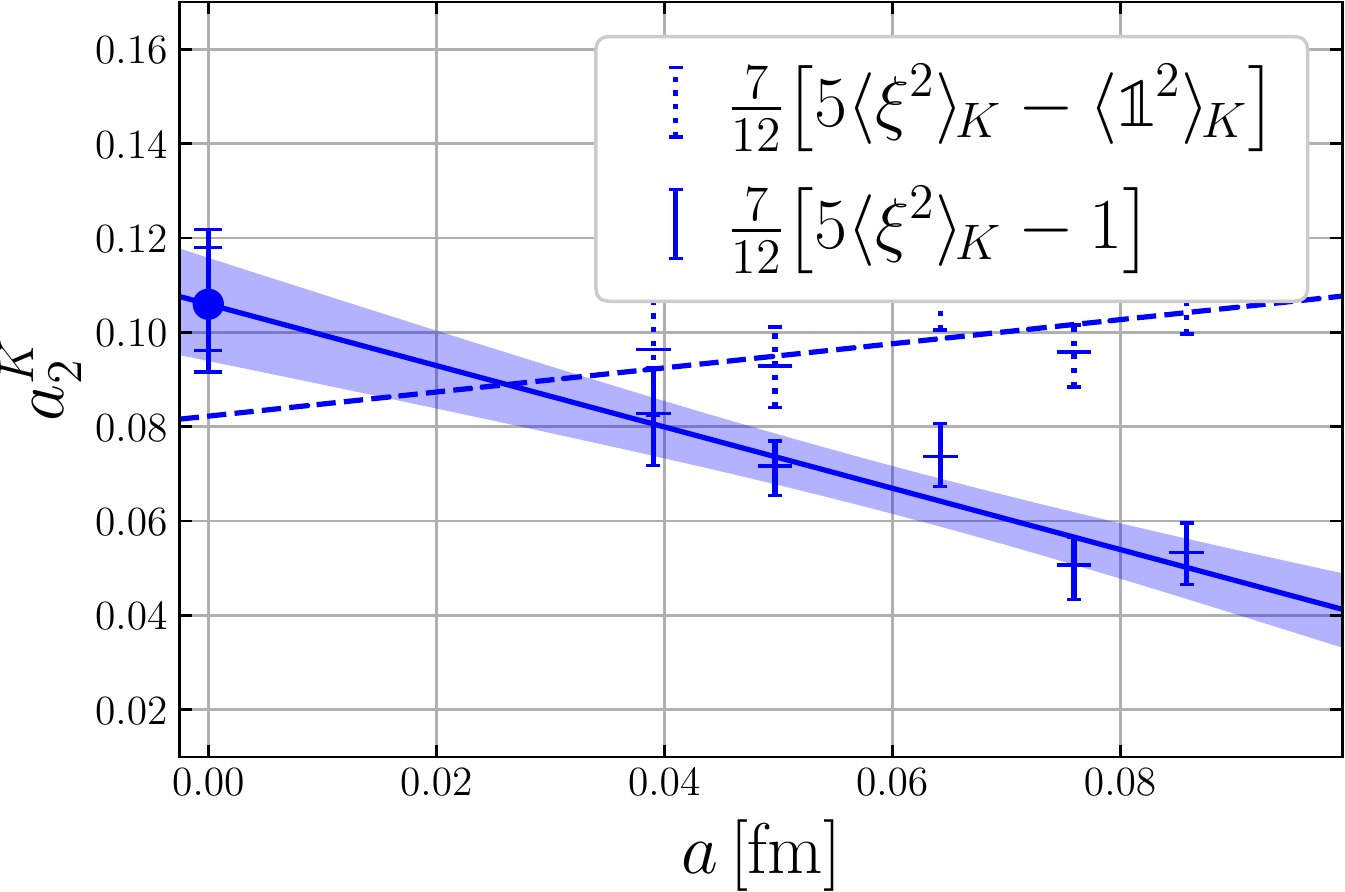}%
\caption{Updated version of figure~\ref{fig_Braun}.\label{fig_BraunSMOM3}}%
\end{figure}%
\begin{figure}[tp]%
\centering
\includegraphics[height=\figheight]{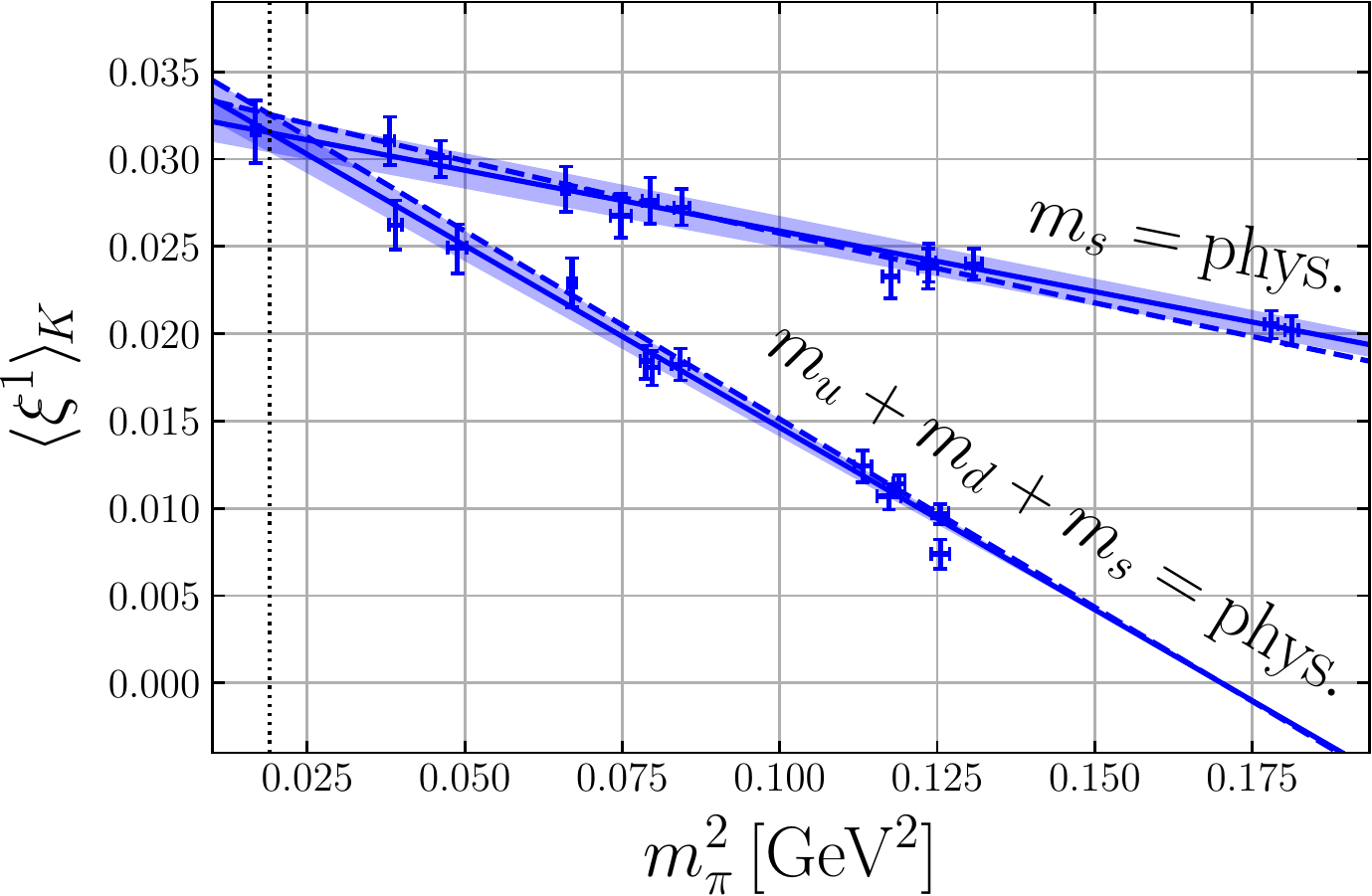}\hfill%
\includegraphics[height=\figheight]{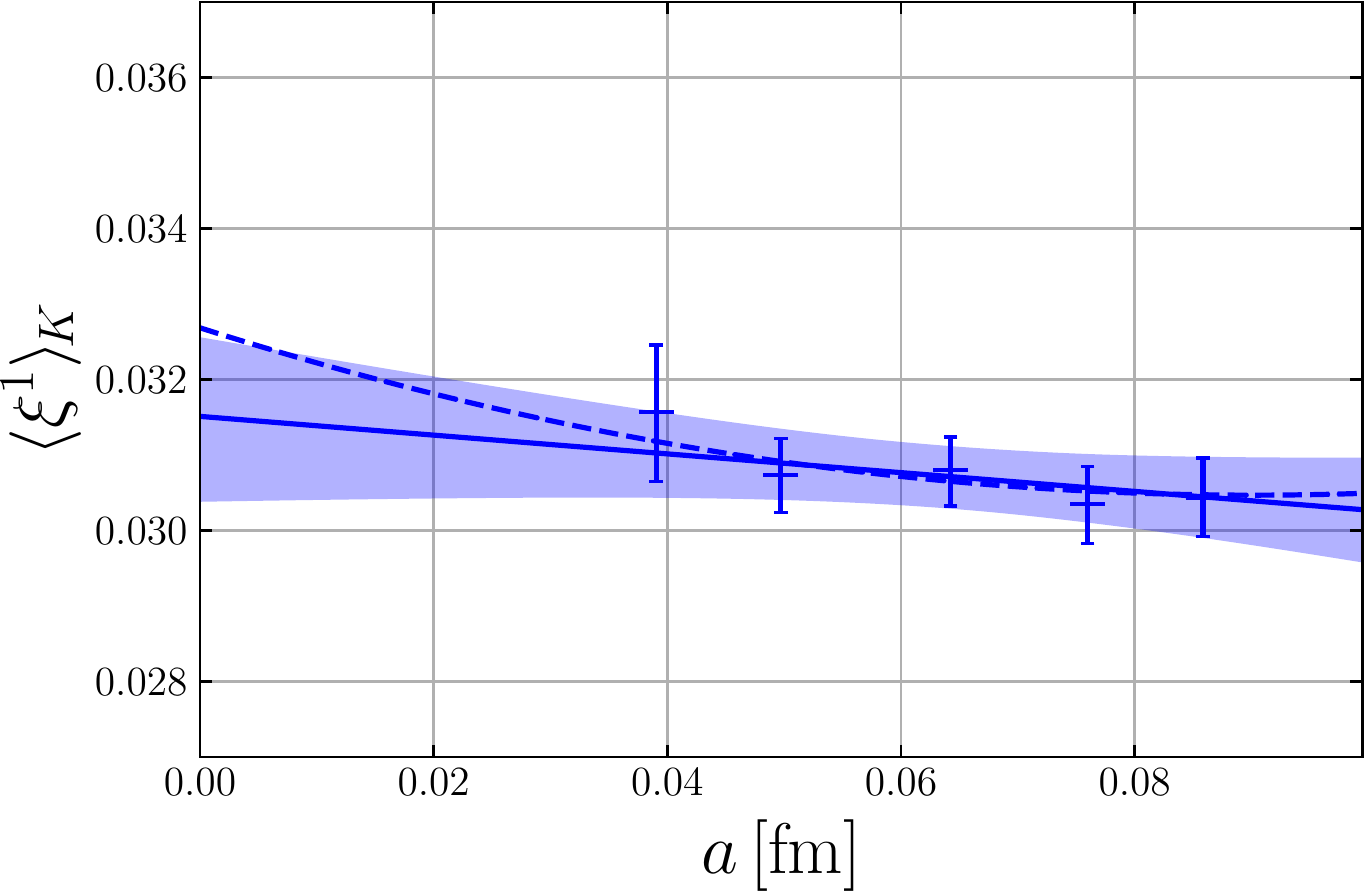}%
\caption{Updated version of figure~\ref{fig_am_xi1}.\label{fig_am_xi1SMOM3}\\[1.28\baselineskip]}%
% \end{figure}%
% %
% \begin{figure}[tp]%
\centering
\includegraphics[width=.485\textwidth]{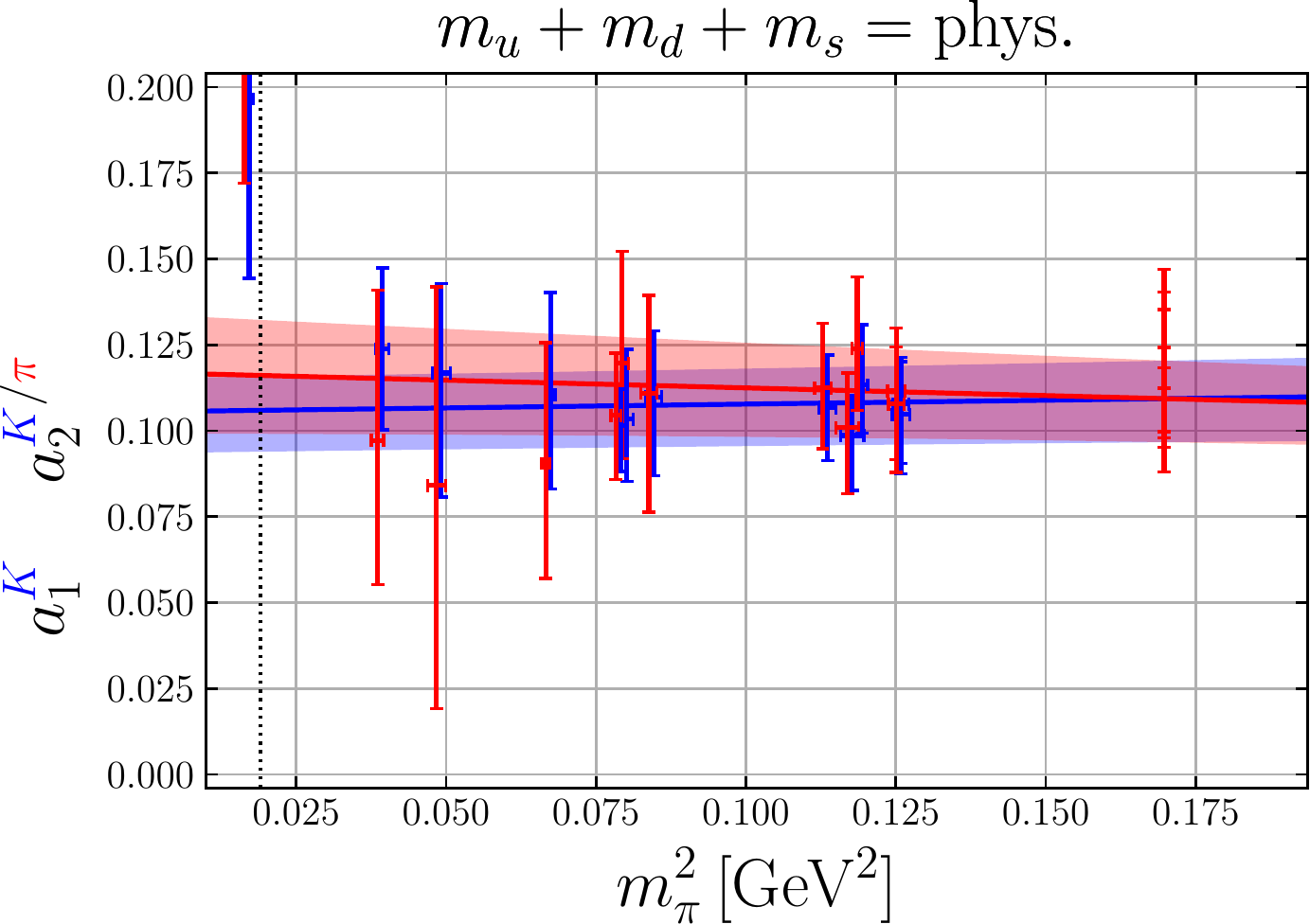}\hfill%
\includegraphics[width=.485\textwidth]{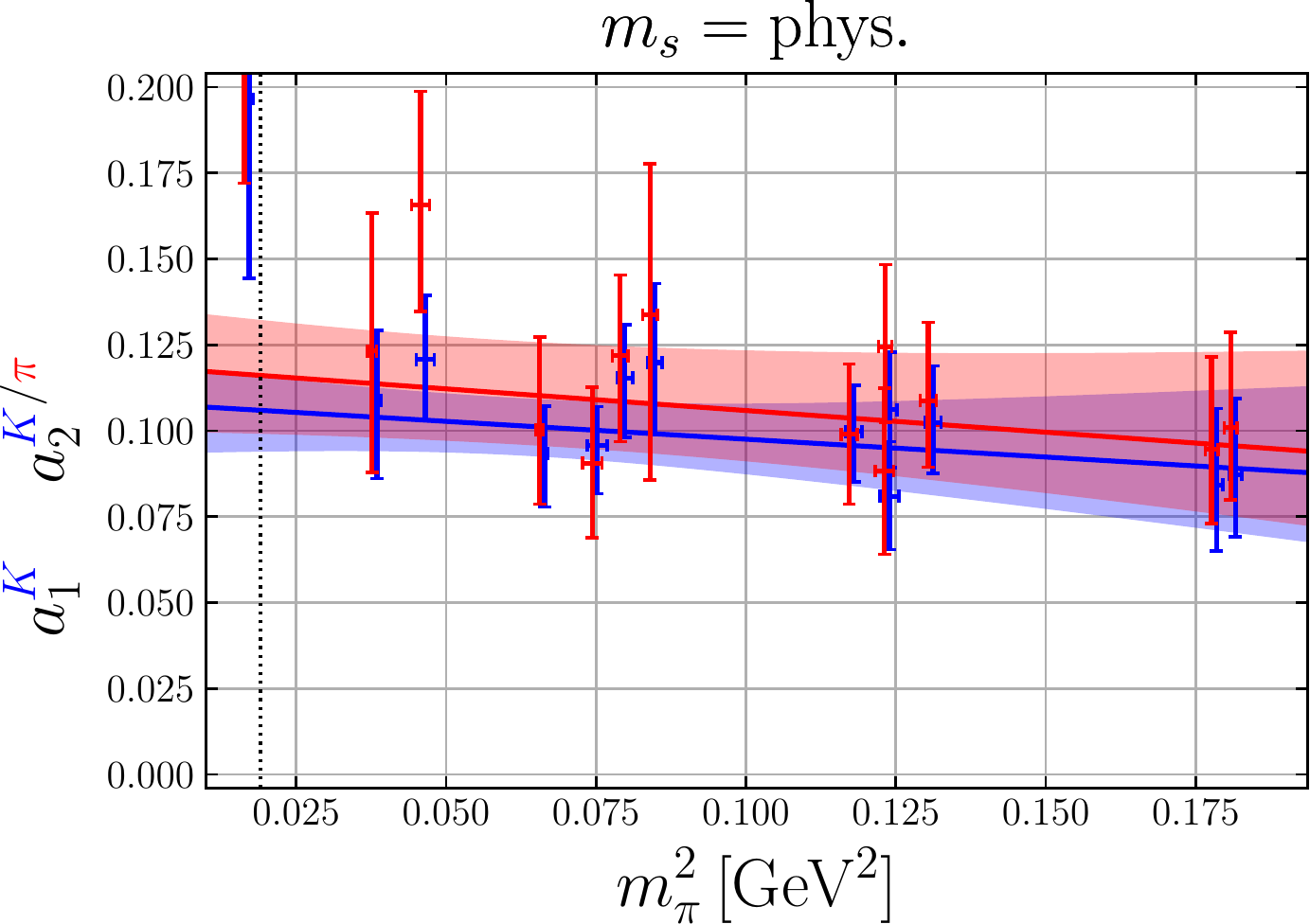}%
\caption{Updated version of figure~\ref{fig_summary}.\label{fig_summarySMOM3}\\[1.28\baselineskip]}%
% \end{figure}%
% %
% \begin{figure}[tp]%
\centering%
\includegraphics[width=0.485\textwidth]{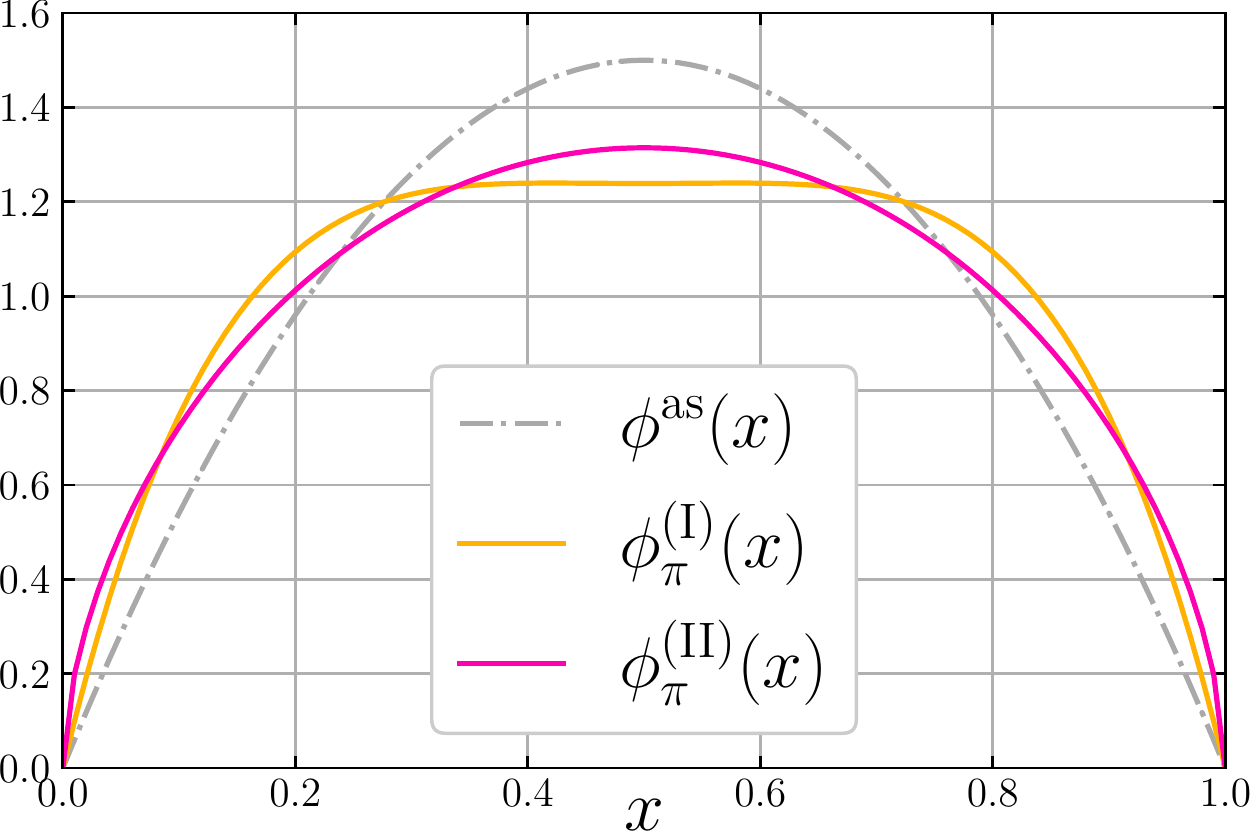}\hfill%
\includegraphics[width=0.485\textwidth]{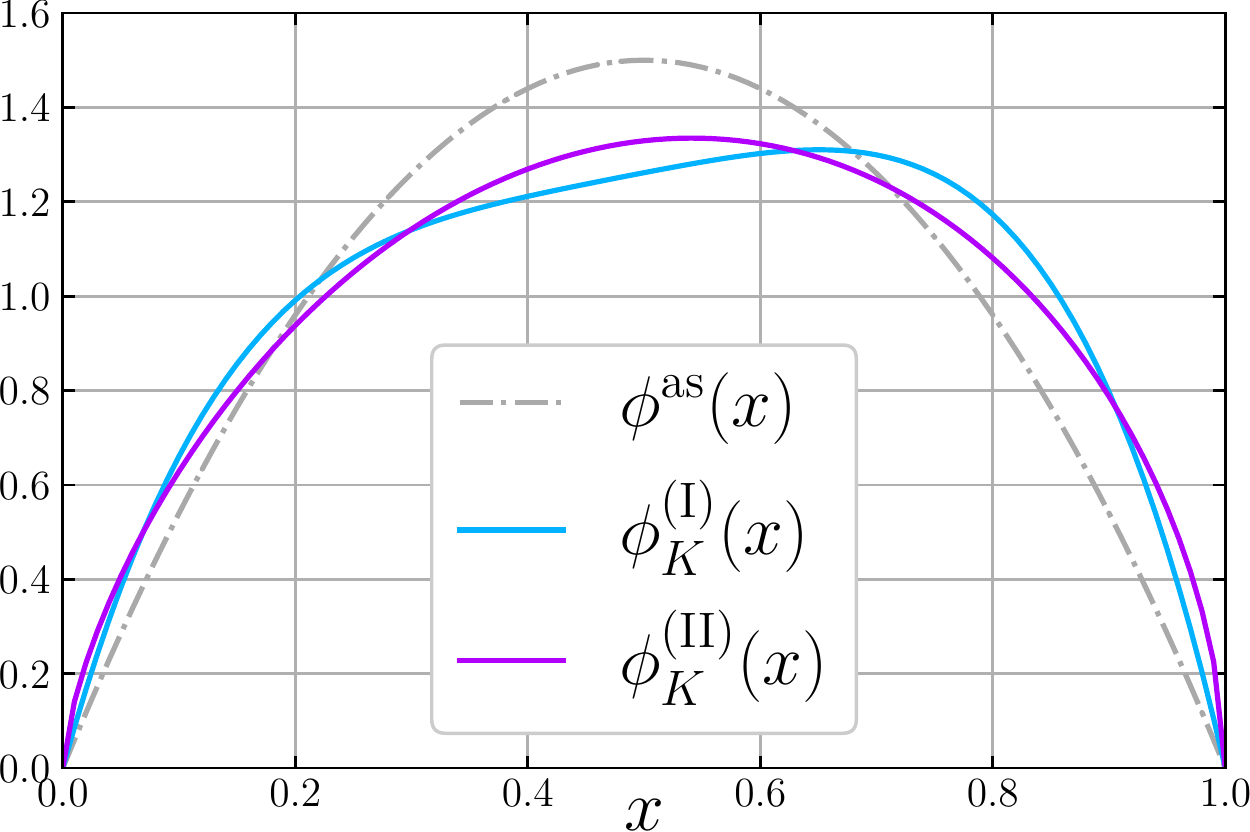}%
\caption{Updated version of figure~\ref{fig_DAs_PK}.\label{fig_DAs_PKSMOM3}}%
\end{figure}%
\FloatBarrier
\providecommand{\href}[2]{#2}\begingroup\raggedright\endgroup

\end{document}